\documentclass[aps,pra,twocolumn,superscriptaddress]{revtex4-2}
\usepackage{graphicx}    
\usepackage{bbold}
\usepackage{dcolumn}    
\usepackage{amssymb}   
\usepackage{amsmath}    
\usepackage{amsthm}
\usepackage{commath}   
\usepackage{subfigure}    
\usepackage{braket}
\usepackage{natbib}
\usepackage{float}
\usepackage{color}
\usepackage{siunitx}
\usepackage[colorlinks=true, allcolors=blue]{hyperref}
\usepackage{hyphenat}
\usepackage[T1]{fontenc}
\usepackage[normalem]{ulem}

\usepackage{MnSymbol}
\usepackage{tabularx}
\usepackage{array}
\usepackage{dsfont}
\usepackage{physics}
\usepackage{empheq}
\usepackage{makecell}
\usepackage{mathtools}
\usepackage{multirow}
\usepackage{titletoc}

\newcommand{\thx}{\thanks{Equal contribution.}}

\newcommand{\pme}{\affiliation{Pritzker School of Molecular Engineering, University of Chicago, Chicago, Illinois 60637, USA}}
\newcommand{\physics}{\affiliation{Department of Physics, University of Chicago, Chicago, IL 60637, USA}}

\newcommand{\JFI}{\affiliation{James Franck Institute, University of Chicago, Chicago, IL 60637, USA}}

\newcommand{\QNext}{\affiliation{Q-NEXT, Argonne National Laboratory, Lemont, Illinois 60439, USA}}

\newcommand{\argonne}{\affiliation{Materials Science Division, Argonne National Laboratory, Lemont, Illinois 60439, USA}}

\graphicspath{{Figures/}}

\titlecontents{section}[0pt]
  {\small}
  {\contentslabel{2.2em}}
  {\hspace*{-2.2em}}     
  {\titlerule*[.5pc]{.}\contentspage}

\titlecontents{subsection}[2.2em]
  {\small}
  {\contentslabel{3em}}
  {}
  {\titlerule*[.5pc]{.}\contentspage}

\begin{document}

\title{Pump-Free Microwave-Optical Bell Pair Generation\\for Teleportation-Based Quantum Transduction}

\author{Fangxin Li}\email{fangxinli@uchicago.edu}\thx\physics

\author{Jaesung Heo}\email{jheo@uchicago.edu}\thx\pme

\author{Zhaoyou Wang}\pme

\author{Benjamin Pingault}\pme\QNext\argonne

\author{Xingyu Gao}\pme\QNext

\author{Tengyang Ruan}\pme

\author{Anjun Chu}\pme

\author{David D. Awschalom}\physics\pme\QNext\argonne

\author{Andrew N. Cleland}\pme

\author{Andrew P. Higginbotham}\physics\JFI

\author{Alexander A. High}\pme

\author{Liang Jiang}\email{liangjiang@uchicago.edu}\pme

\date{\today}

\begin{abstract}
The coherent conversion between microwave and optical photons, known as quantum transduction, is critical for connecting superconducting processors to optical networks. Existing methods are limited by complications associated with optical pumping. We propose a pump-free microwave-optical Bell-pair source for teleportation-based transduction. Using a spin or atomic system resonantly coupled to optical and microwave cavities, the scheme generates loss-robust heralded Bell pairs. Across color centers, atomic ensembles, and phonon-mediated systems, this assembly achieves kilohertz-range heralding rates with high fidelity.
\end{abstract}

\maketitle

\emph{Introduction}\textemdash 
A quantum internet could enable secure communication, distributed quantum computing, and enhanced sensing by connecting heterogeneous quantum systems over long distances \cite{Kimble2008}. A key challenge in realizing such networks is quantum transduction, the coherent conversion of quantum states between microwave and optical frequencies. Quantum transduction provides an interface between quantum processors and optical communication channels \cite{Kimble2008,Cirac1997, Bochmann2013, Lauk2020, Han2021}. Recent advancements have achieved internal conversion efficiencies approaching unity. However, they still have not reached the necessary external efficiency combined with large bandwidths and sub-quantum noise floors~\cite{Lauk2020,Han2021,sekine_microwave--optical_2026}. One of the key limitations is the strong optical pump, which is inevitable in many cases due to the intrinsically weak light-matter couplings during the conversion process~\cite{Hafezi2012, Hisatomi2016, Gard2017, Fan2018, Vogt2019, Rueda2019, Zhong2020, Lauk2020, sekine_microwave--optical_2026, Mirhosseini2020, Holzgrafe2020, Krastanov2021, Han2021, Wu2021, Zhong2022, Sahu2023, Rochman2023, Meesala24, Meesala2024PRX, Xie2025}. The optical pump can degrade transduction performance in many ways: absorbed pump photons can create effective thermal baths and additional damping \cite{Meenehan2014,Meenehan2015,Higginbotham2018,MacCabe2020,Ren2020, forsch_microwave--optics_2020, arnold_converting_2020, Meesala2024PRX,Zhong2024}, and induce microwave loss and noise in superconducting devices through quasiparticle generation and heating \cite{hease_bidirectional_2020,stockill_ultra-low-noise_2022,Rochman2023,Arnold2025}. Pump photons can also contaminate the output signal, requiring filtering that could introduce severe loss of the detected signal \cite{sahu_quantum-enabled_2022,Rochman2023,Meesala2024PRX,Xie2025}. Therefore, a microwave-optical quantum interconnect protocol that does not rely on an optical pump offers significant advantages toward realizing high-fidelity quantum transduction devices.

\begin{figure}[t!]
\centering
\includegraphics[width=0.95\columnwidth]{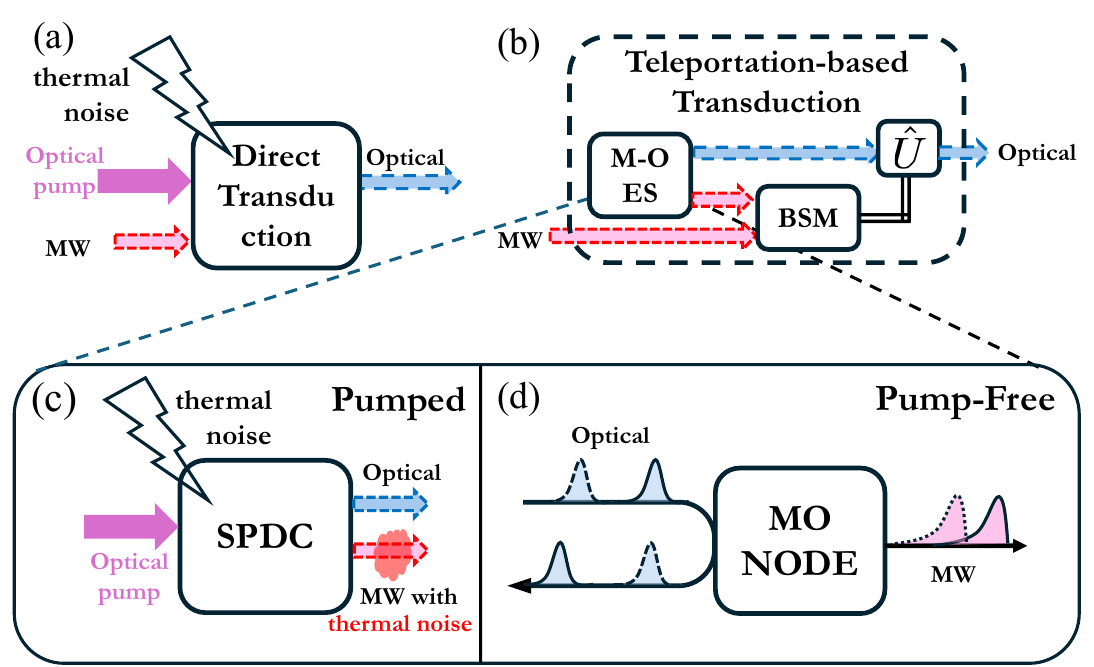}
\caption{(a) Direct quantum transduction using up-conversion is affected by thermal noise due to the optical pump. (b) Teleportation-based quantum transduction. A microwave state is teleported to an optical state by consuming an M-O Bell pair. An optical state can also be teleported to a microwave state (not illustrated here). In the panel, a Bell-state measurement (BSM) is jointly performed on the microwave input and the microwave photon from an M-O entanglement source (M-O ES). The measurement result is classically communicated (double-line) to perform the corresponding unitary $\hat{U}$ on the optical qubit. (c) M-O entanglement source using spontaneous parametric down conversion (SPDC). The process suffers from pump-induced thermal noise. (d) The pump-free M-O entanglement source. A reflected optical photon is entangled with an emitted microwave photon through the M-O no-optical-drive entangler (MO-NODE). \label{Fig1}}
\end{figure}

\begin{figure}[t!]
\centering
\includegraphics[width=0.95\columnwidth]{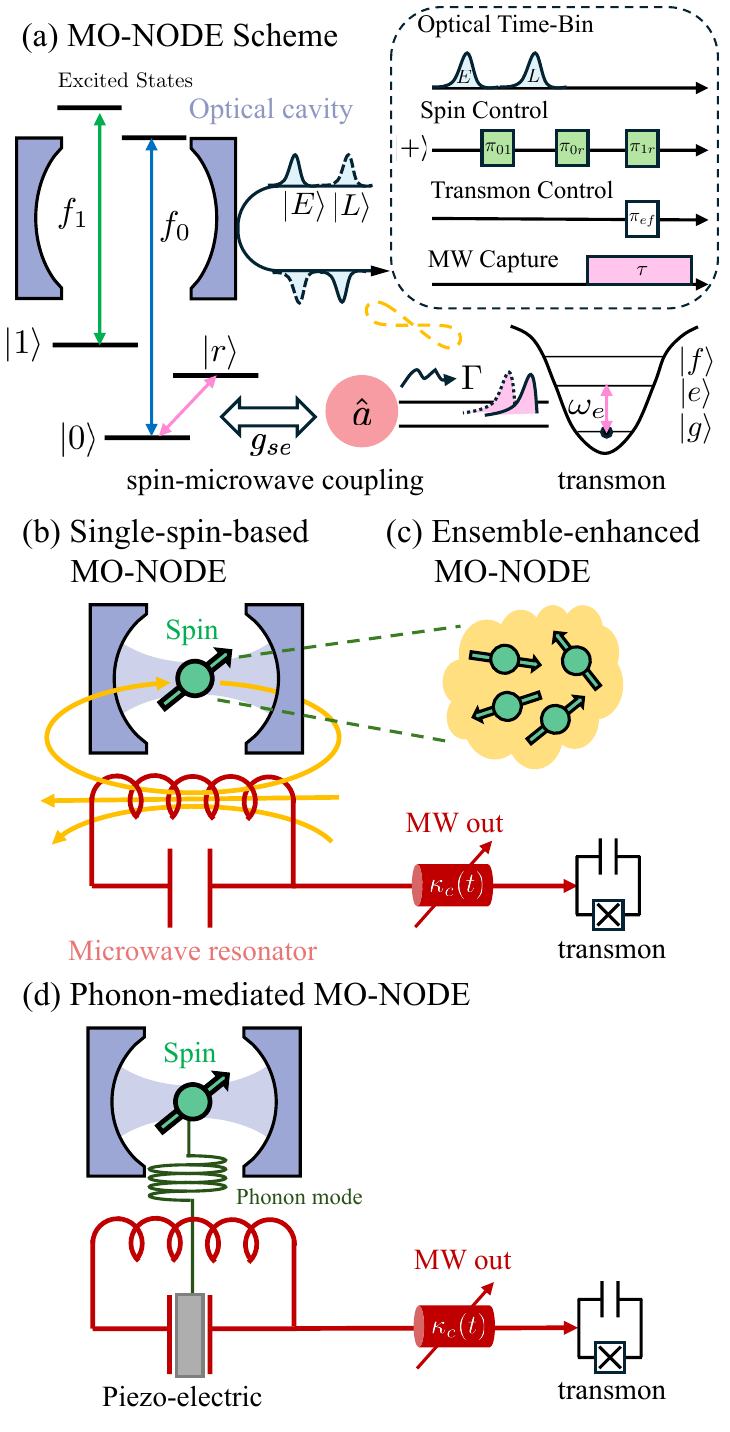}
\caption{(a) Illustration of the MO-NODE with relevant states and parameters. The $\ket{0}$, $\ket{1}$ and $\ket{r}$ states are in a ground state manifold, which can be split by external field (magnetic or strain). The microwave readout transition $0-r$ is resonantly coupled either directly or effectively to a microwave resonator mode $\hat{a}$. The optical transition to the excited state is coupled to an optical cavity.
A transmon with three lowest energy levels $\ket{g}$, $\ket{e}$, and $\ket{f}$ is used to herald microwave emission. The inset shows timed control sequences of the scheme. The $\pi_{0r}$ represents a $\pi$-rotation in the $\ket{0},\ket{r}$ subspace of the three-level system. For a physical implementation, (b) a single spin is coupled directly to an optical cavity and a microwave resonator.
(c) The spin-microwave coupling is enhanced by the spin ensemble. (d) The spin-microwave coupling is mediated by the phonon, which couples through spin-phonon and piezo-electric interactions.\label{Fig2}}
\end{figure}

In this work, we propose an optical-pump-free protocol for generating microwave-optical (M-O) Bell pairs, termed MO-NODE (M-O No-Optical-Drive Entangler). These Bell pairs can serve as an entanglement resource for teleportation-based M-O quantum transduction. The conventional quantum transduction approach, shown in Fig.~\ref{Fig1}(a), utilizes frequency conversion, where the optical pump supplies the energy difference. It suffers from the stringent requirement of high conversion efficiency, bandwidth and low added noise~\cite{Lauk2020, Han2021,sekine_microwave--optical_2026}. The teleportation-based transduction, shown in Fig.~\ref{Fig1}(b), bypasses the efficiency constraint by first generating M-O entanglement, and completing the quantum state transfer across frequency through teleportation~\cite{Zhong2020, Wu2021,shi_overcoming_2024,caleffi_quantum_2026}. The transduction capacity of this approach depends on the rate and fidelity of the M-O entanglement source~\cite{wang_quantum_2022,Zhong2022}. Currently, the entanglement source, shown in Fig.~\ref{Fig1}(c), is based on spontaneous parametric down-conversion (SPDC) \cite{Zhong2020,Krastanov2021,Meesala24,Meesala2024PRX} which also suffers from the pump-induced limitations. To avoid the strong pump undermining the entanglement source fidelity, the pump power is severely limited, leading to a low generation rate~\cite{Meesala24,Meesala2024PRX}. MO-NODE circumvents the use of an optical pump completely by operating in the single-optical-photon regime during the entanglement generation. It uses a spin system with both optical and microwave transitions, resonantly coupled to optical and microwave cavities. By reflecting a single optical photon off our MO-NODE, shown in Fig.~\ref{Fig1}(d), with timed microwave control pulses, the reflected photon and the released microwave signal constitute a time-bin encoded M-O Bell pair. With the pump-free mechanism, the MO-NODE promises to be a reliable source for teleportation-based transduction. Moreover, as crucial resources in hybrid platform quantum communication, such Bell pairs can also be used in a DLCZ-type protocol to entangle remote superconducting circuits via entanglement swapping \cite{Duan2001, Krastanov2021,zhong2020A}.

The MO-NODE starts by generating spin-optical-photon entanglement, which requires high optical cooperativity~\cite{PhysRevLett.92.127902,PhysRevX.4.031022,Nguyen2019PRB,PhysRevApplied.22.044013}, and is completed by mapping the spin state in the microwave manifold onto a microwave photon through Purcell-enhanced decay with a rate $\Gamma$, as shown in Fig.~\ref{Fig2}(a). A key requirement is to retrieve the spin state encoded in the microwave manifold before decoherence. This requires the coupling rate between the microwave mode and the spin system to be larger than the intrinsic decoherence rate of the system. To this end, we consider three solutions illustrated in Fig.~\ref{Fig2}(b)-(d): 1) a single spin system (e.g. color center in diamond~\cite{Sipahigil2016integrated,Nguyen2019PRB,Nguyen2019PRL,hanson2019,Rosenthal2023,Guo2023,harris2025high}) with an integrated microwave-optical resonator design that achieves large single-spin-microwave coupling~\cite{PFMO_device_design_2025}, 2) a spin ensemble system (e.g. an ultracold atom ensemble~\cite{Verd2009,Bernon2013,Hattermann2017,Chen2022,Wilde2025}), where the spin-microwave interaction is collectively enhanced, and 3) a spin-phonon system~\cite{Meesala2018,joe_purcell-enhanced_2026}, where the interaction with microwave is mediated by the phonon~\cite{arrangoiz-arriola_coupling_2018, Meesala2024PRX}.
Our approach, in principle, can be generalized to other atom-like systems with comparable level structure and coupling strength. Combining state-of-the-art experimental parameters and simulations, we expect our protocol to generate heralded M-O Bell pairs beyond kilohertz-scale rates while maintaining high fidelity, paving the way for high-fidelity information protocols between different quantum modalities in the future.

\emph{Single-spin-based MO-NODE}\textemdash We begin by describing the protocol based on single-spin-microwave coupling, which captures the essential idea of MO-NODE.

The protocol is realized in two steps. First, we generate spin-photon entanglement using spin-dependent cavity reflectivity~\cite{PhysRevLett.92.127902, PhysRevX.4.031022,Nguyen2019PRB,PhysRevApplied.22.044013}, where distinct transition frequencies ($f_0$ and $f_1$ in Fig.~\ref{Fig2}(a)) make the $\ket{0}$ state highly reflective and $\ket{1}$ non-reflective. By initializing a photon in the early ($\ket{E}$) and late ($\ket{L}$) time-bin superposition $(\ket{E}+\ket{L})/\sqrt{2}$ and a spin in $\ket{+} = (\ket{0}+\ket{1})/\sqrt{2}$, and applying a spin $\pi$-pulse between the time-bins (Fig.~\ref{Fig2}(a) inset), we create the reflected Bell state $\ket{\Phi}_{\text{Photon,Spin}}=(\ket{E} \ket{1} + \ket{L}\ket{0})/\sqrt{2}$.

A simple cavity QED model shows that in the ideal critical-coupling limit, the protocol approaches unit Bell-state fidelity, with success probability $P_{\text{opt}}=0.5C^2/(C+1)^2$. In this formula, $C=\frac{g_o^2}{\kappa_o\gamma_o}$ is the spin-cavity cooperativity, where $g_o$ is the coupling between the relevant transition and the optical cavity; $\kappa_o$ is the loss rate of the optical cavity, and $\gamma_o$ is the decay rate of the excitation state. The prefactor $0.5$ is due to the non-reflective spin state, which discards photons half of the time (see Supplemental Material Sec.~\ref{sec:schemes}). A series of experiments has demonstrated the protocol \cite{Nguyen2019PRB, Nguyen2019PRL, Stas2022, Knaut2024, Wei2025} with the best-achievable Bell state fidelity exceeding $0.97$ \cite{Wei2025}. Successful optical cavity reflection can be heralded by nondestructive photon detection \cite{Niemietz2021}.

Next, we map the spin state to a microwave photon by coupling the transition between one of the spin states and a readout level $\ket{r}$ to a superconducting microwave resonator mode. The coupling is modeled using the Jaynes-Cummings Hamiltonian:
$
     H_\mathrm{int} = \hbar g_{se} (\hat{a} \sigma_+ + \hat{a}^\dagger \sigma_-), 
$
where $g_{se}$ is an effective single-photon microwave coupling, $\hat{a}$ ($\hat{a}^\dagger$) is the annihilation (creation) operator for photons in the microwave resonator mode, and $\sigma_+=\ket{r}\bra{0}$ and $\sigma_-=\ket{0}\bra{r}$ are operators associated with the readout transition shown in Fig.~\ref{Fig2}(a). The microwave resonator mode has a decay rate $\kappa_{e}$. Spin state retrieval is in the Purcell regime, where the resonator-enhanced spin relaxation is the dominant process, with a rate $\Gamma_0 = 4g_{se}^2/\kappa_{e}$ for $g_{se} \lesssim \kappa_{e}$. As $\kappa_e$ is reduced toward $g_{se}$, the system crosses over from the Purcell regime to coherent spin–resonator exchange, with the fastest rate remaining of order $\Gamma_0\sim g_{se}$. 

To convert a spin qubit state $\alpha\ket{0}+\beta\ket{1}$ into a microwave time-bin state, we utilize Purcell-enhancement by tuning the resonator in resonance with the readout level and suppressing the decay of the off-resonant $\ket{1}$ state. Two time-bins are created by sequentially transferring the population from the $\ket{0}$ and the $\ket{1}$ states to the readout level $\ket{r}$ using microwave control pulses, as shown in the inset of Fig.~\ref{Fig2}(a), and allowing the level $\ket{r}$ to decay through the cavity. This procedure creates entanglement between the optical and microwave photons.

To herald the microwave photon, a three-level transmon captures the time-bin microwave photons via a high-efficiency pitch-and-catch sequence \cite{wenner_catching_2014, axline_-demand_2018, campagne-ibarcq_deterministic_2018, kurpiers_deterministic_2018,PhysRevApplied.12.044067}. We map the time-bins to the transmon by resonantly capturing the first time-bin microwave photon and shelving it in $|f\rangle$, which frees the $g$–$e$ transition to capture the second time-bin (Fig.~\ref{Fig2}(a)). A lossless transfer maps the time-bins to the $\{|e\rangle, |f\rangle\}$ subspace, yielding the Bell state $\ket{\Psi}_{\text{Photon,Transmon}}=(\ket{E} \ket{e} + \ket{L}\ket{f})/\sqrt{2}$.

The microwave emission probability is $p_\text{mw}=1-e^{-\Gamma \tau/2}$ over a detection time $\tau$, with Purcell rate of $\Gamma = \Gamma_0$. Because photon loss leaves the transmon in $|g\rangle$, we herald successful transfer by measuring the transmon to verify it is not in $|g\rangle$, without distinguishing between $|e\rangle$ and $|f\rangle$ \cite{jerger_realization_2016}. Accounting for a measurement error $\epsilon$ (failing to distinguish $|g\rangle$ from excited states), the overall heralding click probability is $P_{\text{mw}} = p_\text{mw}(1-\epsilon)+\epsilon(1-p_\text{mw})$, assuming a detection time shorter than the transmon relaxation time $\tau \ll T_{1,t}$ (see Supplemental Material Sec.~\ref{figmerit}).

The intrinsic heralding rate $R_{\text{herald}} = \mathcal{P}/T_r$ is defined as the number of heralding events per second, where $\mathcal{P} = P_{\text{opt}}P_{\text{mw}}$ is the joint heralding click probability and $T_r$ is the cycling duration of the Bell pair generation scheme. $T_r$ consists of a variable detection time $\tau$ and a fixed execution time $T_{\text{exe}}$ which includes both reset and gate time. In explicit form, the heralding rate can be written as 
\begin{equation}
    R_{\text{herald}} = 0.5 \cdot \frac{C^2}{(1+C)^2} \cdot \frac{(1-\epsilon)(1-e^{-\Gamma  \tau/2})+\epsilon \cdot e^{-\Gamma \tau/2}}{\tau + T_{\text{exe}}}.
\end{equation}
This expression entails an optimal detection time $\tau$ for a maximal rate. The heralding rate first increases with $\tau$ as the emission probability increases, and then decreases as the repetition rate is reduced, eventually going to $0$ as $\tau \rightarrow \infty$. This trade-off leads to a maximum heralding rate $\sim \frac{1}{2}\frac{C^2}{(1+C)^2}\frac{1}{2/\Gamma+T_{\mathrm{exe}}}$ at a time set by $\Gamma$ and $T_{\text{exe}}$. 

Conditioning on a heralding event, we restrict the transmon to the logical subspace $\{\ket{e},\ket{f}\}$. The fidelity of microwave retrieval depends on the spin dephasing time $T_{2,\text{s}}$ and the transmon relaxation time $T_{1,\text{t}}$. The dephasing process occurs throughout the time window $\tau$, while the relaxation of the transmon occurs after the shelving. The error process can be modeled by a dephasing channel combined with an amplitude damping channel in the subspace $\ket{f}\rightarrow \ket{e}$ (see Supplemental Material Sec.~\ref{figmerit}). Assuming the spin-optical entanglement protocol produces unit-fidelity spin-photon Bell pairs, the fidelity of the heralded M-O Bell pair is given by
\begin{equation}\label{mwfidelity}
\mathcal{F}_{\text{herald}}=\frac{F_M}{4}(1+e^{-\tau/T_{1,\text{t}}}+2e^{-\tau/2T_{1,\text{t}}}e^{-\tau/T_{2,\text{s}}}),
\end{equation}
where the pre-factor $F_M = \frac{(1-\epsilon)p_\text{mw}}{P_{\text{mw}}}$ is the proportion of correct heralding, taking into account the transmon measurement error. (see Supplemental Material Sec.~\ref{figmerit}). Such measurement error could lead to false heralding events even when there is no photon (dark counts), thereby degrading the overall heralding fidelity. 

The fidelity is also maximized at an optimal detection time set by the trade-off between microwave-photon emission and decoherence. Increasing the detection time initially improves the fidelity by increasing the emission probability, whereas at longer times, spin decoherence and transmon decay reduce it.

\begin{figure*}[t!]
\centering
\includegraphics[width=\textwidth]{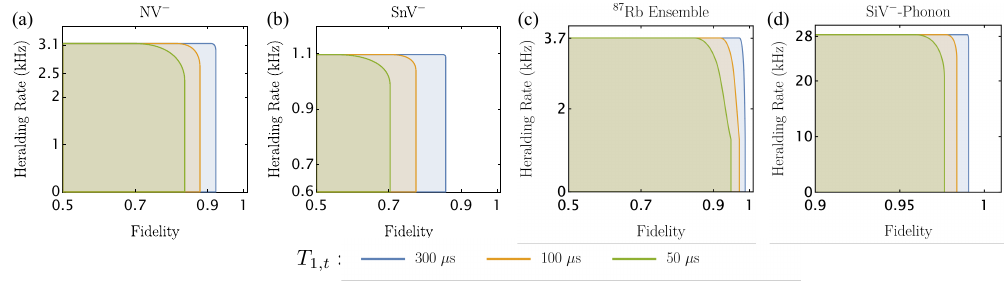}
\caption{Figures of merit for our MO-NODE across different platforms and transmon lifetimes $T_{1,t}$ with fixed $\epsilon=0.01$. \label{Fig3}}
\end{figure*}

\begin{table}
\begin{tabular}{|c|c|c|c|c|}
\hline \textbf{MO-NODE} & \multicolumn{2}{c|}{\textbf{Single Spin}} & \textbf{Ensemble} & \textbf{Phonon}\\
\hline Spin & NV$^-$ & ${}^{117}\text{SnV}^-$ & ${}^{87}\text{Rb}$ & SiV-Phonon\\
\hline $C$ & 9.7 \cite{PFMO_device_design_2025} & 31 \cite{PFMO_device_design_2025} & 3~\cite{Chen2022} & 31~\cite{joe_purcell-enhanced_2026}\\
\hline $\Gamma/2\pi$ (kHz) & 3.8 \cite{PFMO_device_design_2025} & 0.9 \cite{PFMO_device_design_2025} & 40~\cite{Verd2009} & 178\\
\hline $T_{2,s}$ (ms) & 1~\cite{rondin_magnetometry_2014} & 2.5~\cite{harris2025high} & 1000~\cite{Bohi2009,Bernon2013} & 1~\cite{sukachev_silicon-vacancy_2017}\\
\hline $T_{\text{exe}}$ ($\mu$s) & 10 & 10 & 50 & 10\\
\hline
\end{tabular}
\caption{Key parameters of different spin systems considered for realizing our protocol. The $\Gamma$ of the SiV is calculated in the main text from literature values. The $T_{2,s}$ values are dynamical decoupling values with one or two pulses, taken respectively from the cited work. The $T_{\text{exe}}$ includes microwave spin control time and transmon reset time. The control pulse time for color centers in diamond ranges from tens of nanoseconds to microseconds \cite{pingault_coherent_2017,fuchs_gigahertz_2009,harris2025high,dolde_high-fidelity_2014,karapatzakis_microwave_2024,Rosenthal2023,Guo2023,doherty_nitrogen-vacancy_2013}, while that of neutral atoms is $\sim 4.1$ µs \cite{Bohi2009}. Active transmon reset can be performed on a submicrosecond timescale. \cite{magnard_fast_2018,egger_pulsed_2018,zhou_rapid_2021}. These parameters are used to calculate the figures of merit for each implementation.
}\label{parameter_fom}
\end{table}

To understand the performance of our scheme, we consider NV$^-$ and ${}^{117}\mathrm{SnV}^-$ centers in diamond as examples. The key challenge for single diamond color centers is to realize strong microwave coupling between the color center and a microwave resonator while maintaining high cooperativity $C$ with an optical cavity despite metal-induced loss. We address this challenge with a hybrid device design described in our companion paper \cite{PFMO_device_design_2025}. Our design achieves a microwave coupling rate in the kHz range while maintaining optical cooperativity above unity, $C\gg1$ \cite{PFMO_device_design_2025} (also see Supplemental Material Sec.~\ref{sec:color}).

Figs.~\ref{Fig3}(a,b) show the achievable region in terms of the heralding rate and fidelity for NV$^-$ and ${}^{117}\mathrm{SnV}^-$ centers, using the key parameters listed in Table~\ref{parameter_fom}. The upper boundary of the achievable region is determined by the maximum heralding rate, which, as mentioned above, depends only on $C$, $\Gamma$, and $T_{\text{exe}}$. The rightmost boundary is set by the maximum fidelity. Since the detection time that optimizes the heralding rate is not necessarily the same as the one that optimizes the fidelity, there is no point in the plot that simultaneously achieves both the maximum fidelity and the maximum heralding rate. For a transmon lifetime of $T_{1,t}=300~\mu\mathrm{s}$, the NV$^-$ (${}^{117}\mathrm{SnV}^-$) center can achieve a maximum fidelity of $0.92$ ($0.86$) with a corresponding intrinsic heralding rate of $3.0$ ($1.1$)~$\mathrm{kHz}$.

It is worth commenting that if the spin lacks the three-level structure required for time-bins, we can instead coherently swap its two-level state to a microwave photon (or transmon) by dynamically tuning their resonance or coupling \cite{sillanpaa_coherent_2007,hofheinz_generation_2008,zeytinoglu_microwave-induced_2015}. This converts the entanglement as $\ket{\Phi}_{\text{Photon,Spin}}=(\ket{E} \ket{1} + \ket{L}\ket{0})/\sqrt{2} \longrightarrow\ket{\Psi}_{\text{Photon,mw}}=(\ket{E} \ket{1}_{\text{mw}} + \ket{L}\ket{0}_{\text{mw}})/\sqrt{2}$, using microwave Fock states. Operating in the coherent regime ($g_{se}\gg \kappa_e$), the swapping time is $t_{\text{swap}}=\pi/(2 g_{se})$. Although unheralded photon loss reduces expected fidelity compared to the pitch-and-catch scheme, strong spin-microwave coupling can still ensure a fast, high-fidelity protocol. 

\emph{Ensemble-enhanced MO-NODE}\textemdash 
One extension of the single-spin-based MO-NODE is to use a spin ensemble, as illustrated in Fig.~\ref{Fig2}(c). The key idea is to use a single collective excitation in the microwave manifold as the ensemble analog of the single spin. For an ensemble of $N$ spins, the collective coupling $G_m\sim \sqrt{N}g_{se}$ allows $\Gamma\sim G_m$ for $\kappa\gtrsim G_m$~\cite{Tavis1968, Plankensteiner2019}, thereby improving the microwave retrieval efficiency and hence the overall protocol performance. This concept requires at least four relevant levels in the microwave manifold.

In analogy with the single-spin-based MO-NODE, the two states participating in spin-photon entanglement are denoted by $\ket{0}$ and $\ket{1}$, and the microwave readout state by $\ket{r}$. The additional level is the lowest-energy state $\ket{g_a}$, which is distinct from $\ket{0}$. The microwave resonator is resonant with the $\ket{g_a}\leftrightarrow\ket{r}$ transition. We first consider the interaction between the selected spin states and the corresponding cavities as modeled in analogy with the Tavis-Cummings model~\cite{Tavis1968} in the ideal limit of no inhomogeneous broadening or mode mismatch. We define the single collective excitation state as $\ket{W_x}\equiv\left(\sqrt{N}\right)^{-1}\sum_n \ket{g_a g_a\cdots x_n \cdots g_a g_a}$, where $x\in\{0,1,r\}$, and $n$ labels the $n$-th spin in the ensemble.

From the ensemble initialized in $\ket{G}=\ket{g_a g_a \cdots g_a g_a}$, single excitation can be created by injecting a microwave photon emitted from a transmon coupled to the system \footnote{Rydberg blockade can be used in a neutral atom ensemble for the preparation of the single collective excitation \cite{Mei2009}. Since Rydberg blockade relies on optical driving, the single-excitation preparation step should be temporally separated from the entangling step to avoid pump-induced heating. After this preparation step, no optical drive is required, and the entangling process remains pump-free.}, preparing the ensemble in the $\ket{W_r}$ state. Global drives then coherently manipulate the collective states $\ket{W_x}$, enabling an entangling protocol analogous to the single-spin-based MO-NODE, with each $\ket{W_x}$ playing the role of $\ket{x}$ \footnote{Preparation of the single collective excitation can be heralded by recapturing the emitted microwave photon with the transmon. We apply $\pi_{g_a0}$ before the initially prepared excitation decays and apply the same pulse again after the decay, leaving the ensemble in the state $\ket{W_0}$ while emitting the microwave photon used for heralding.}. The final output is an M-O Bell state $\ket{\Psi}_{\mathrm{Photon,Transmon}}=(\ket{E}\ket{e}+\ket{L}\ket{f})/\sqrt{2}$.

The ensemble nature of our protocol requires us to consider both the mismatch between the microwave and optical (M-O) cavity-coupling profiles and the effects of inhomogeneous broadening \cite{Wesenberg2011,Kurucz2011}. During the microwave interaction, the spin-wave components interfere collectively, yielding an enhancement determined by their overlap with the microwave bright mode. By contrast, the spin-wave components do not interfere with one another during the optical interaction, which implies that there is no dark mode for the optical cavity provided that each coupling rate is not zero. Thus, there is no mode-matching requirement between M-O cavity coupling profiles. (see Supplemental Material Sec.~\ref{ensemble_bright_dark}).
The effects of optical inhomogeneous broadening are captured by the average of the single-emitter cavity responses weighted by their populations in the initial collective state, while the spin-wave inhomogeneity produces second-order leakage into the microwave dark mode during the finite optical interaction time. (see Supplemental Material Sec.~\ref{ensemble_supplementary}). For robust microwave-state retrieval, cavity protection \cite{Kurucz2011,Putz2014} and dynamical-decoupling techniques \cite{Julsgaard2013} may be employed. To further suppress inhomogeneous broadening, we consider a neutral-atom ensemble as a platform for the ensemble-based MO-NODE.

A state-of-the-art experiment has demonstrated the simultaneous coupling of an atomic ensemble to optical and millimeter-wave modes \cite{kumar_quantum-enabled_2023}. In that work, however, the millimeter-wave transition is encoded between two Rydberg states, making an optical pump unavoidable. Instead, we consider an ensemble of $^{87}\mathrm{Rb}$ atoms coupled through the ground-state manifold to a microwave resonator with $G_m/2\pi\sim40~\mathrm{kHz}$ \cite{Verd2009}, while optical coupling is provided by a traveling-wave cavity with $C\sim3$ \cite{Chen2022}. The relatively long microwave drive time \cite{Bohi2009}, comparable to the detection window, degrades the fidelity and imposes a minimum protocol duration, resulting in the lower-branch cutoff in Fig.~\ref{Fig3}(c) (see Supplemental Material Sec.~\ref{ensemble_fom}). Nevertheless, for $T_{1,t}=300~\mu\mathrm{s}$, the $^{87}\mathrm{Rb}$ implementation reaches a maximum heralding rate of $3.7~\mathrm{kHz}$ at a fidelity of $0.97$. The impact of experimental constraints, including the challenge of integrating optical and microwave components, is left for future work. 

\emph{Phonon-mediated MO-NODE}\textemdash 
Instead of directly coupling the spin to microwave photons, an alternative is to use phonons to mediate the interaction between the spin and the microwave photon, as shown in Fig.~\ref{Fig2}(d). A recent experiment has demonstrated single-spin-phonon coupling reaching $300$ kHz in a photonic crystal cavity with co-localized optical and phonon modes \cite{joe_purcell-enhanced_2026}. Moreover, strong phonon-microwave coupling has been demonstrated using piezo-electric interaction integrated with the crystal cavity \cite{arrangoiz-arriola_coupling_2018, Meesala24,Meesala2024PRX}. Therefore, a natural extension of the spin-microwave scheme uses the optomechanical cavity to first create spin-photon entanglement, as described in an earlier paragraph, and retrieve the spin state through an effective spin-microwave coupling induced by the phonon. The interaction Hamiltonian of the system is $H_{\text{int,ph}} =  \hbar g_{sp} (\hat{b} \sigma_+ + \hat{b}^\dagger \sigma_-) + \hbar g_{pe}(\hat{a} \hat{b}^\dagger + \hat{a}^\dagger\hat{b} )$, where $\hat{a}$, $\hat{b}$ and $\hat\sigma$ denote the microwave mode, phonon mode and the relevant spin transition. $g_{sp}$ is the spin-phonon coupling rate predicted to reach the MHz-scale in near-term experiments \cite{Meesala2018,joe_purcell-enhanced_2026}, and $g_{pe}$ is the piezo-electric coupling rate that has already been demonstrated to reach MHz-scale level~\cite{arrangoiz-arriola_coupling_2018, Meesala2024PRX}. 

In the phonon-mediated MO-NODE, the Purcell-enhanced spin decay takes on a new effective rate $\Gamma = \Gamma_{sp} = \frac{4g^2_{sp}}{\kappa_m + 4g^2_{pe}/\kappa_e}$, where $\kappa_m$ and $\kappa_e$ are the loss rate of the phonon and the microwave mode, respectively. This rate is calculated by adiabatically eliminating the lossy microwave mode (assuming $\kappa_e \gg g_{pe}$), which effectively  increases the phonon linewidth to $\Gamma_m =\kappa_m + 4g^2_{pe}/\kappa_e$. The spin then decays through the acoustic Purcell-effect when $\Gamma_m \gg g_{sp}$. Unlike direct spin-microwave coupling, the probability of retrieving the microwave signal now also depends on the piezo-electric cooperativity $C_{pe} = 4g_{pe}^2/(\kappa_m\kappa_e)$ in addition to the exponential profile $p_\text{mw}'= \frac{C_{pe}}{C_{pe}+1} (1-e^{-\Gamma_{sp}\tau/2})$ (see Supplemental Material Sec.~\ref{phononappendix}). Once the microwave photon is emitted from the piezo-electric device, the heralding probability and the fidelity follow exactly the same equation as the one derived for single-spin-based MO-NODE, with $P_{\text{mw}} = p_\text{mw}'(1-\epsilon)+\epsilon(1-p_\text{mw}')$. Using experimentally relevant parameters $(g_{sp},g_{pe},\kappa_m,\kappa_e)/2\pi =(0.3,2.4, 0.1, 12)$ MHz \cite{Meesala2024PRX, joe_purcell-enhanced_2026}, the adiabatic estimate predicts a spin decay rate $\Gamma_{sp}/2\pi=178~\mathrm{kHz}$, which is one to two orders of magnitude larger than that in the single-spin protocol. The phonon-mediated protocol can reach a maximum heralding rate of $28.3~\mathrm{kHz}$ with a fidelity of $0.99$ at $T_{1,t}=300~\mu\mathrm{s}$ as shown in Fig.~\ref{Fig3}(d). This gives the best performance among the MO-NODE implementations considered in this work \footnote{It is worth mentioning that the strong spin-phonon coupling ($g_{sp}/\kappa_m>1$) also makes the platform
suitable for the coherent swap scheme.  A simple estimation, assuming sequential swaps between the spin-phonon and the phonon-microwave, gives the swapping time $t_{\text{swap}}=\pi/(2g_{sp})+\pi/(2g_{pe})$. Using the same set of parameter values as in the above paragraph, except for a reduced microwave loss rate of $0.1$ MHz, the swap time takes roughly $1~\mu$s, and the microwave retrieval is realized with fidelity $0.7$. (see Supplemental Material). Additional schemes that maps the spin state to a bosonic mode state has been considered, for example, in \cite{PhysRevLett.105.220501}.}.

\emph{Conclusion}\textemdash
We propose a pump-free scheme for generating microwave-optical (M-O) Bell pairs, suitable for platforms with strong optical interfaces and efficient microwave-state retrieval. Analyzing implementations with a single spin system, an atom/spin ensemble, and a spin-phonon system, we show that they all achieve intrinsic kilohertz-scale heralding rates with high fidelity. Although previous SPDC-based proposals predicted comparable performance~\cite{Krastanov2021,Zhong2022}, experimental pump-power and cooling limitations, as well as filtering loss, have severely bottlenecked their actual rates~\cite{Meesala24,Meesala2024PRX}. Our pump-free mechanism circumvents these constraints, offering a robust path to high-fidelity M-O entanglement that could be extended to other systems in the future.

\begin{acknowledgments}

We acknowledge helpful discussions with William Munro, Hong Qiao, Yang Shen, Judas Strayer, and Christopher Wang. We acknowledge support from the ARO (W911NF-23-1-0077), ARO MURI (W911NF-21-1-0325), AFOSR (FA9550-20-1-0270, FA9550-22-1-0370) and AFOSR MURI (FA9550-21-1-0209, FA9550-23-1-0338), ONR (N000142512032) and ONR MURI (N000142612102), DARPA (HR0011-24-9-0361 and QuSeN HR0011-25-2-0035), NSF (ERC-1941583, OMA-2137642, OSI-2326767, CCF-2312755, OSI-2426975). This material is based upon work supported by the U.S. Department of Energy, Office of Science, National Quantum Information Science Research Centers and Advanced Scientific Computing Research (ASCR) program under contract number DE-AC02-06CH11357 as part of the InterQnet quantum networking project. This material is based upon work supported by the U.S. Department of Energy Office of Science National Quantum Information Science Research Centers (QISRC) as part of the Q-NEXT center. This work was completed with resources provided by the University of Chicago’s Research Computing Center. We acknowledge support from the Quantum Leap Challenge Institute for Hybrid Quantum Architectures and Networks (HQAN) (NSF OMA-2016136). We acknowledge help from the Sonnet Software Technical Support.  
\end{acknowledgments}

\bibliography{reference}

\begin{thebibliography}{106}%
\makeatletter
\providecommand \@ifxundefined [1]{%
 \@ifx{#1\undefined}
}%
\providecommand \@ifnum [1]{%
 \ifnum #1\expandafter \@firstoftwo
 \else \expandafter \@secondoftwo
 \fi
}%
\providecommand \@ifx [1]{%
 \ifx #1\expandafter \@firstoftwo
 \else \expandafter \@secondoftwo
 \fi
}%
\providecommand \natexlab [1]{#1}%
\providecommand \enquote  [1]{``#1''}%
\providecommand \bibnamefont  [1]{#1}%
\providecommand \bibfnamefont [1]{#1}%
\providecommand \citenamefont [1]{#1}%
\providecommand \href@noop [0]{\@secondoftwo}%
\providecommand \href [0]{\begingroup \@sanitize@url \@href}%
\providecommand \@href[1]{\@@startlink{#1}\@@href}%
\providecommand \@@href[1]{\endgroup#1\@@endlink}%
\providecommand \@sanitize@url [0]{\catcode `\\12\catcode `\$12\catcode `\&12\catcode `\#12\catcode `\^12\catcode `\_12\catcode `\%12\relax}%
\providecommand \@@startlink[1]{}%
\providecommand \@@endlink[0]{}%
\providecommand \url  [0]{\begingroup\@sanitize@url \@url }%
\providecommand \@url [1]{\endgroup\@href {#1}{\urlprefix }}%
\providecommand \urlprefix  [0]{URL }%
\providecommand \Eprint [0]{\href }%
\providecommand \doibase [0]{https://doi.org/}%
\providecommand \selectlanguage [0]{\@gobble}%
\providecommand \bibinfo  [0]{\@secondoftwo}%
\providecommand \bibfield  [0]{\@secondoftwo}%
\providecommand \translation [1]{[#1]}%
\providecommand \BibitemOpen [0]{}%
\providecommand \bibitemStop [0]{}%
\providecommand \bibitemNoStop [0]{.\EOS\space}%
\providecommand \EOS [0]{\spacefactor3000\relax}%
\providecommand \BibitemShut  [1]{\csname bibitem#1\endcsname}%
\let\auto@bib@innerbib\@empty
\bibitem [{\citenamefont {Kimble}(2008)}]{Kimble2008}%
  \BibitemOpen
  \bibfield  {author} {\bibinfo {author} {\bibfnamefont {H.~J.}\ \bibnamefont {Kimble}},\ }\bibfield  {title} {\bibinfo {title} {The quantum internet},\ }\href {https://doi.org/10.1038/nature07127} {\bibfield  {journal} {\bibinfo  {journal} {Nature}\ }\textbf {\bibinfo {volume} {453}},\ \bibinfo {pages} {1023} (\bibinfo {year} {2008})}\BibitemShut {NoStop}%
\bibitem [{\citenamefont {Cirac}\ \emph {et~al.}(1997)\citenamefont {Cirac}, \citenamefont {Zoller}, \citenamefont {Kimble},\ and\ \citenamefont {Mabuchi}}]{Cirac1997}%
  \BibitemOpen
  \bibfield  {author} {\bibinfo {author} {\bibfnamefont {J.~I.}\ \bibnamefont {Cirac}}, \bibinfo {author} {\bibfnamefont {P.}~\bibnamefont {Zoller}}, \bibinfo {author} {\bibfnamefont {H.~J.}\ \bibnamefont {Kimble}},\ and\ \bibinfo {author} {\bibfnamefont {H.}~\bibnamefont {Mabuchi}},\ }\bibfield  {title} {\bibinfo {title} {Quantum state transfer and entanglement distribution among distant nodes in a quantum network},\ }\href {https://doi.org/10.1103/PhysRevLett.78.3221} {\bibfield  {journal} {\bibinfo  {journal} {Phys. Rev. Lett.}\ }\textbf {\bibinfo {volume} {78}},\ \bibinfo {pages} {3221} (\bibinfo {year} {1997})}\BibitemShut {NoStop}%
\bibitem [{\citenamefont {Bochmann}\ \emph {et~al.}(2013)\citenamefont {Bochmann}, \citenamefont {Vainsencher}, \citenamefont {Awschalom},\ and\ \citenamefont {Cleland}}]{Bochmann2013}%
  \BibitemOpen
  \bibfield  {author} {\bibinfo {author} {\bibfnamefont {J.}~\bibnamefont {Bochmann}}, \bibinfo {author} {\bibfnamefont {A.}~\bibnamefont {Vainsencher}}, \bibinfo {author} {\bibfnamefont {D.~D.}\ \bibnamefont {Awschalom}},\ and\ \bibinfo {author} {\bibfnamefont {A.~N.}\ \bibnamefont {Cleland}},\ }\bibfield  {title} {\bibinfo {title} {Nanomechanical coupling between microwave and optical photons},\ }\href {https://doi.org/10.1038/nphys2748} {\bibfield  {journal} {\bibinfo  {journal} {Nature Physics}\ }\textbf {\bibinfo {volume} {9}},\ \bibinfo {pages} {712} (\bibinfo {year} {2013})}\BibitemShut {NoStop}%
\bibitem [{\citenamefont {Lauk}\ \emph {et~al.}(2020)\citenamefont {Lauk}, \citenamefont {Sinclair}, \citenamefont {Barzanjeh}, \citenamefont {Covey}, \citenamefont {Saffman}, \citenamefont {Spiropulu},\ and\ \citenamefont {Simon}}]{Lauk2020}%
  \BibitemOpen
  \bibfield  {author} {\bibinfo {author} {\bibfnamefont {N.}~\bibnamefont {Lauk}}, \bibinfo {author} {\bibfnamefont {N.}~\bibnamefont {Sinclair}}, \bibinfo {author} {\bibfnamefont {S.}~\bibnamefont {Barzanjeh}}, \bibinfo {author} {\bibfnamefont {J.~P.}\ \bibnamefont {Covey}}, \bibinfo {author} {\bibfnamefont {M.}~\bibnamefont {Saffman}}, \bibinfo {author} {\bibfnamefont {M.}~\bibnamefont {Spiropulu}},\ and\ \bibinfo {author} {\bibfnamefont {C.}~\bibnamefont {Simon}},\ }\bibfield  {title} {\bibinfo {title} {Perspectives on quantum transduction},\ }\href {https://doi.org/10.1088/2058-9565/ab788a} {\bibfield  {journal} {\bibinfo  {journal} {Quantum Science and Technology}\ }\textbf {\bibinfo {volume} {5}},\ \bibinfo {pages} {020501} (\bibinfo {year} {2020})}\BibitemShut {NoStop}%
\bibitem [{\citenamefont {Han}\ \emph {et~al.}(2021)\citenamefont {Han}, \citenamefont {Fu}, \citenamefont {Zou}, \citenamefont {Jiang},\ and\ \citenamefont {Tang}}]{Han2021}%
  \BibitemOpen
  \bibfield  {author} {\bibinfo {author} {\bibfnamefont {X.}~\bibnamefont {Han}}, \bibinfo {author} {\bibfnamefont {W.}~\bibnamefont {Fu}}, \bibinfo {author} {\bibfnamefont {C.-L.}\ \bibnamefont {Zou}}, \bibinfo {author} {\bibfnamefont {L.}~\bibnamefont {Jiang}},\ and\ \bibinfo {author} {\bibfnamefont {H.~X.}\ \bibnamefont {Tang}},\ }\bibfield  {title} {\bibinfo {title} {Microwave-optical quantum frequency conversion},\ }\href {https://doi.org/10.1364/OPTICA.425414} {\bibfield  {journal} {\bibinfo  {journal} {Optica}\ }\textbf {\bibinfo {volume} {8}},\ \bibinfo {pages} {1050} (\bibinfo {year} {2021})}\BibitemShut {NoStop}%
\bibitem [{\citenamefont {Sekine}\ \emph {et~al.}(2026)\citenamefont {Sekine}, \citenamefont {Murakami},\ and\ \citenamefont {Doi}}]{sekine_microwave--optical_2026}%
  \BibitemOpen
  \bibfield  {author} {\bibinfo {author} {\bibfnamefont {A.}~\bibnamefont {Sekine}}, \bibinfo {author} {\bibfnamefont {R.}~\bibnamefont {Murakami}},\ and\ \bibinfo {author} {\bibfnamefont {Y.}~\bibnamefont {Doi}},\ }\bibfield  {title} {\bibinfo {title} {Microwave-to-optical quantum transduction of photons for quantum interconnects},\ }\bibfield  {journal} {\bibinfo  {journal} {npj Nanophotonics}\ }\href {https://doi.org/10.1038/s44310-026-00130-8} {10.1038/s44310-026-00130-8} (\bibinfo {year} {2026})\BibitemShut {NoStop}%
\bibitem [{\citenamefont {Hafezi}\ \emph {et~al.}(2012)\citenamefont {Hafezi}, \citenamefont {Kim}, \citenamefont {Rolston}, \citenamefont {Orozco}, \citenamefont {Lev},\ and\ \citenamefont {Taylor}}]{Hafezi2012}%
  \BibitemOpen
  \bibfield  {author} {\bibinfo {author} {\bibfnamefont {M.}~\bibnamefont {Hafezi}}, \bibinfo {author} {\bibfnamefont {Z.}~\bibnamefont {Kim}}, \bibinfo {author} {\bibfnamefont {S.~L.}\ \bibnamefont {Rolston}}, \bibinfo {author} {\bibfnamefont {L.~A.}\ \bibnamefont {Orozco}}, \bibinfo {author} {\bibfnamefont {B.~L.}\ \bibnamefont {Lev}},\ and\ \bibinfo {author} {\bibfnamefont {J.~M.}\ \bibnamefont {Taylor}},\ }\bibfield  {title} {\bibinfo {title} {Atomic interface between microwave and optical photons},\ }\href {https://doi.org/10.1103/PhysRevA.85.020302} {\bibfield  {journal} {\bibinfo  {journal} {Phys. Rev. A}\ }\textbf {\bibinfo {volume} {85}},\ \bibinfo {pages} {020302} (\bibinfo {year} {2012})}\BibitemShut {NoStop}%
\bibitem [{\citenamefont {Hisatomi}\ \emph {et~al.}(2016)\citenamefont {Hisatomi}, \citenamefont {Osada}, \citenamefont {Tabuchi}, \citenamefont {Ishikawa}, \citenamefont {Noguchi}, \citenamefont {Yamazaki}, \citenamefont {Usami},\ and\ \citenamefont {Nakamura}}]{Hisatomi2016}%
  \BibitemOpen
  \bibfield  {author} {\bibinfo {author} {\bibfnamefont {R.}~\bibnamefont {Hisatomi}}, \bibinfo {author} {\bibfnamefont {A.}~\bibnamefont {Osada}}, \bibinfo {author} {\bibfnamefont {Y.}~\bibnamefont {Tabuchi}}, \bibinfo {author} {\bibfnamefont {T.}~\bibnamefont {Ishikawa}}, \bibinfo {author} {\bibfnamefont {A.}~\bibnamefont {Noguchi}}, \bibinfo {author} {\bibfnamefont {R.}~\bibnamefont {Yamazaki}}, \bibinfo {author} {\bibfnamefont {K.}~\bibnamefont {Usami}},\ and\ \bibinfo {author} {\bibfnamefont {Y.}~\bibnamefont {Nakamura}},\ }\bibfield  {title} {\bibinfo {title} {Bidirectional conversion between microwave and light via ferromagnetic magnons},\ }\href {https://doi.org/10.1103/PhysRevB.93.174427} {\bibfield  {journal} {\bibinfo  {journal} {Phys. Rev. B}\ }\textbf {\bibinfo {volume} {93}},\ \bibinfo {pages} {174427} (\bibinfo {year} {2016})}\BibitemShut {NoStop}%
\bibitem [{\citenamefont {Gard}\ \emph {et~al.}(2017)\citenamefont {Gard}, \citenamefont {Jacobs}, \citenamefont {McDermott},\ and\ \citenamefont {Saffman}}]{Gard2017}%
  \BibitemOpen
  \bibfield  {author} {\bibinfo {author} {\bibfnamefont {B.~T.}\ \bibnamefont {Gard}}, \bibinfo {author} {\bibfnamefont {K.}~\bibnamefont {Jacobs}}, \bibinfo {author} {\bibfnamefont {R.}~\bibnamefont {McDermott}},\ and\ \bibinfo {author} {\bibfnamefont {M.}~\bibnamefont {Saffman}},\ }\bibfield  {title} {\bibinfo {title} {Microwave-to-optical frequency conversion using a cesium atom coupled to a superconducting resonator},\ }\href {https://doi.org/10.1103/PhysRevA.96.013833} {\bibfield  {journal} {\bibinfo  {journal} {Phys. Rev. A}\ }\textbf {\bibinfo {volume} {96}},\ \bibinfo {pages} {013833} (\bibinfo {year} {2017})}\BibitemShut {NoStop}%
\bibitem [{\citenamefont {Fan}\ \emph {et~al.}(2018)\citenamefont {Fan}, \citenamefont {Zou}, \citenamefont {Cheng}, \citenamefont {Guo}, \citenamefont {Han}, \citenamefont {Gong}, \citenamefont {Wang},\ and\ \citenamefont {Tang}}]{Fan2018}%
  \BibitemOpen
  \bibfield  {author} {\bibinfo {author} {\bibfnamefont {L.}~\bibnamefont {Fan}}, \bibinfo {author} {\bibfnamefont {C.-L.}\ \bibnamefont {Zou}}, \bibinfo {author} {\bibfnamefont {R.}~\bibnamefont {Cheng}}, \bibinfo {author} {\bibfnamefont {X.}~\bibnamefont {Guo}}, \bibinfo {author} {\bibfnamefont {X.}~\bibnamefont {Han}}, \bibinfo {author} {\bibfnamefont {Z.}~\bibnamefont {Gong}}, \bibinfo {author} {\bibfnamefont {S.}~\bibnamefont {Wang}},\ and\ \bibinfo {author} {\bibfnamefont {H.~X.}\ \bibnamefont {Tang}},\ }\bibfield  {title} {\bibinfo {title} {Superconducting cavity electro-optics: A platform for coherent photon conversion between superconducting and photonic circuits},\ }\href {https://doi.org/10.1126/sciadv.aar4994} {\bibfield  {journal} {\bibinfo  {journal} {Science Advances}\ }\textbf {\bibinfo {volume} {4}},\ \bibinfo {pages} {eaar4994} (\bibinfo {year} {2018})}\BibitemShut {NoStop}%
\bibitem [{\citenamefont {Vogt}\ \emph {et~al.}(2019)\citenamefont {Vogt}, \citenamefont {Gross}, \citenamefont {Han}, \citenamefont {Pal}, \citenamefont {Lam}, \citenamefont {Kiffner},\ and\ \citenamefont {Li}}]{Vogt2019}%
  \BibitemOpen
  \bibfield  {author} {\bibinfo {author} {\bibfnamefont {T.}~\bibnamefont {Vogt}}, \bibinfo {author} {\bibfnamefont {C.}~\bibnamefont {Gross}}, \bibinfo {author} {\bibfnamefont {J.}~\bibnamefont {Han}}, \bibinfo {author} {\bibfnamefont {S.~B.}\ \bibnamefont {Pal}}, \bibinfo {author} {\bibfnamefont {M.}~\bibnamefont {Lam}}, \bibinfo {author} {\bibfnamefont {M.}~\bibnamefont {Kiffner}},\ and\ \bibinfo {author} {\bibfnamefont {W.}~\bibnamefont {Li}},\ }\bibfield  {title} {\bibinfo {title} {Efficient microwave-to-optical conversion using rydberg atoms},\ }\href {https://doi.org/10.1103/PhysRevA.99.023832} {\bibfield  {journal} {\bibinfo  {journal} {Phys. Rev. A}\ }\textbf {\bibinfo {volume} {99}},\ \bibinfo {pages} {023832} (\bibinfo {year} {2019})}\BibitemShut {NoStop}%
\bibitem [{\citenamefont {Rueda}\ \emph {et~al.}(2019)\citenamefont {Rueda}, \citenamefont {Hease}, \citenamefont {Barzanjeh},\ and\ \citenamefont {Fink}}]{Rueda2019}%
  \BibitemOpen
  \bibfield  {author} {\bibinfo {author} {\bibfnamefont {A.}~\bibnamefont {Rueda}}, \bibinfo {author} {\bibfnamefont {W.}~\bibnamefont {Hease}}, \bibinfo {author} {\bibfnamefont {S.}~\bibnamefont {Barzanjeh}},\ and\ \bibinfo {author} {\bibfnamefont {J.~M.}\ \bibnamefont {Fink}},\ }\bibfield  {title} {\bibinfo {title} {Electro-optic entanglement source for microwave to telecom quantum state transfer},\ }\href {https://doi.org/10.1038/s41534-019-0220-5} {\bibfield  {journal} {\bibinfo  {journal} {npj Quantum Information}\ }\textbf {\bibinfo {volume} {5}},\ \bibinfo {pages} {108} (\bibinfo {year} {2019})}\BibitemShut {NoStop}%
\bibitem [{\citenamefont {Zhong}\ \emph {et~al.}(2020{\natexlab{a}})\citenamefont {Zhong}, \citenamefont {Wang}, \citenamefont {Zou}, \citenamefont {Zhang}, \citenamefont {Han}, \citenamefont {Fu}, \citenamefont {Xu}, \citenamefont {Shankar}, \citenamefont {Devoret}, \citenamefont {Tang},\ and\ \citenamefont {Jiang}}]{Zhong2020}%
  \BibitemOpen
  \bibfield  {author} {\bibinfo {author} {\bibfnamefont {C.}~\bibnamefont {Zhong}}, \bibinfo {author} {\bibfnamefont {Z.}~\bibnamefont {Wang}}, \bibinfo {author} {\bibfnamefont {C.}~\bibnamefont {Zou}}, \bibinfo {author} {\bibfnamefont {M.}~\bibnamefont {Zhang}}, \bibinfo {author} {\bibfnamefont {X.}~\bibnamefont {Han}}, \bibinfo {author} {\bibfnamefont {W.}~\bibnamefont {Fu}}, \bibinfo {author} {\bibfnamefont {M.}~\bibnamefont {Xu}}, \bibinfo {author} {\bibfnamefont {S.}~\bibnamefont {Shankar}}, \bibinfo {author} {\bibfnamefont {M.~H.}\ \bibnamefont {Devoret}}, \bibinfo {author} {\bibfnamefont {H.~X.}\ \bibnamefont {Tang}},\ and\ \bibinfo {author} {\bibfnamefont {L.}~\bibnamefont {Jiang}},\ }\bibfield  {title} {\bibinfo {title} {Proposal for heralded generation and detection of entangled microwave--optical-photon pairs},\ }\href {https://doi.org/10.1103/PhysRevLett.124.010511} {\bibfield  {journal} {\bibinfo  {journal} {Phys. Rev. Lett.}\ }\textbf {\bibinfo {volume} {124}},\ \bibinfo {pages} {010511}
  (\bibinfo {year} {2020}{\natexlab{a}})}\BibitemShut {NoStop}%
\bibitem [{\citenamefont {Mirhosseini}\ \emph {et~al.}(2020)\citenamefont {Mirhosseini}, \citenamefont {Sipahigil}, \citenamefont {Kalaee},\ and\ \citenamefont {Painter}}]{Mirhosseini2020}%
  \BibitemOpen
  \bibfield  {author} {\bibinfo {author} {\bibfnamefont {M.}~\bibnamefont {Mirhosseini}}, \bibinfo {author} {\bibfnamefont {A.}~\bibnamefont {Sipahigil}}, \bibinfo {author} {\bibfnamefont {M.}~\bibnamefont {Kalaee}},\ and\ \bibinfo {author} {\bibfnamefont {O.}~\bibnamefont {Painter}},\ }\bibfield  {title} {\bibinfo {title} {Superconducting qubit to optical photon transduction},\ }\href {https://doi.org/10.1038/s41586-020-3038-6} {\bibfield  {journal} {\bibinfo  {journal} {Nature}\ }\textbf {\bibinfo {volume} {588}},\ \bibinfo {pages} {599} (\bibinfo {year} {2020})}\BibitemShut {NoStop}%
\bibitem [{\citenamefont {Holzgrafe}\ \emph {et~al.}(2020)\citenamefont {Holzgrafe}, \citenamefont {Sinclair}, \citenamefont {Zhu}, \citenamefont {Shams-Ansari}, \citenamefont {Colangelo}, \citenamefont {Hu}, \citenamefont {Zhang}, \citenamefont {Berggren},\ and\ \citenamefont {Lon\v{c}ar}}]{Holzgrafe2020}%
  \BibitemOpen
  \bibfield  {author} {\bibinfo {author} {\bibfnamefont {J.}~\bibnamefont {Holzgrafe}}, \bibinfo {author} {\bibfnamefont {N.}~\bibnamefont {Sinclair}}, \bibinfo {author} {\bibfnamefont {D.}~\bibnamefont {Zhu}}, \bibinfo {author} {\bibfnamefont {A.}~\bibnamefont {Shams-Ansari}}, \bibinfo {author} {\bibfnamefont {M.}~\bibnamefont {Colangelo}}, \bibinfo {author} {\bibfnamefont {Y.}~\bibnamefont {Hu}}, \bibinfo {author} {\bibfnamefont {M.}~\bibnamefont {Zhang}}, \bibinfo {author} {\bibfnamefont {K.~K.}\ \bibnamefont {Berggren}},\ and\ \bibinfo {author} {\bibfnamefont {M.}~\bibnamefont {Lon\v{c}ar}},\ }\bibfield  {title} {\bibinfo {title} {Cavity electro-optics in thin-film lithium niobate for efficient microwave-to-optical transduction},\ }\href {https://doi.org/10.1364/OPTICA.397513} {\bibfield  {journal} {\bibinfo  {journal} {Optica}\ }\textbf {\bibinfo {volume} {7}},\ \bibinfo {pages} {1714} (\bibinfo {year} {2020})}\BibitemShut {NoStop}%
\bibitem [{\citenamefont {Krastanov}\ \emph {et~al.}(2021)\citenamefont {Krastanov}, \citenamefont {Raniwala}, \citenamefont {Holzgrafe}, \citenamefont {Jacobs}, \citenamefont {Lon\ifmmode~\check{c}\else \v{c}\fi{}ar}, \citenamefont {Reagor},\ and\ \citenamefont {Englund}}]{Krastanov2021}%
  \BibitemOpen
  \bibfield  {author} {\bibinfo {author} {\bibfnamefont {S.}~\bibnamefont {Krastanov}}, \bibinfo {author} {\bibfnamefont {H.}~\bibnamefont {Raniwala}}, \bibinfo {author} {\bibfnamefont {J.}~\bibnamefont {Holzgrafe}}, \bibinfo {author} {\bibfnamefont {K.}~\bibnamefont {Jacobs}}, \bibinfo {author} {\bibfnamefont {M.}~\bibnamefont {Lon\ifmmode~\check{c}\else \v{c}\fi{}ar}}, \bibinfo {author} {\bibfnamefont {M.~J.}\ \bibnamefont {Reagor}},\ and\ \bibinfo {author} {\bibfnamefont {D.~R.}\ \bibnamefont {Englund}},\ }\bibfield  {title} {\bibinfo {title} {Optically heralded entanglement of superconducting systems in quantum networks},\ }\href {https://doi.org/10.1103/PhysRevLett.127.040503} {\bibfield  {journal} {\bibinfo  {journal} {Phys. Rev. Lett.}\ }\textbf {\bibinfo {volume} {127}},\ \bibinfo {pages} {040503} (\bibinfo {year} {2021})}\BibitemShut {NoStop}%
\bibitem [{\citenamefont {Wu}\ \emph {et~al.}(2021)\citenamefont {Wu}, \citenamefont {Cui}, \citenamefont {Fan},\ and\ \citenamefont {Zhuang}}]{Wu2021}%
  \BibitemOpen
  \bibfield  {author} {\bibinfo {author} {\bibfnamefont {J.}~\bibnamefont {Wu}}, \bibinfo {author} {\bibfnamefont {C.}~\bibnamefont {Cui}}, \bibinfo {author} {\bibfnamefont {L.}~\bibnamefont {Fan}},\ and\ \bibinfo {author} {\bibfnamefont {Q.}~\bibnamefont {Zhuang}},\ }\bibfield  {title} {\bibinfo {title} {Deterministic microwave-optical transduction based on quantum teleportation},\ }\href {https://doi.org/10.1103/PhysRevApplied.16.064044} {\bibfield  {journal} {\bibinfo  {journal} {Phys. Rev. Appl.}\ }\textbf {\bibinfo {volume} {16}},\ \bibinfo {pages} {064044} (\bibinfo {year} {2021})}\BibitemShut {NoStop}%
\bibitem [{\citenamefont {Zhong}\ \emph {et~al.}(2022)\citenamefont {Zhong}, \citenamefont {Han},\ and\ \citenamefont {Jiang}}]{Zhong2022}%
  \BibitemOpen
  \bibfield  {author} {\bibinfo {author} {\bibfnamefont {C.}~\bibnamefont {Zhong}}, \bibinfo {author} {\bibfnamefont {X.}~\bibnamefont {Han}},\ and\ \bibinfo {author} {\bibfnamefont {L.}~\bibnamefont {Jiang}},\ }\bibfield  {title} {\bibinfo {title} {Microwave and optical entanglement for quantum transduction with electro-optomechanics},\ }\href {https://doi.org/10.1103/PhysRevApplied.18.054061} {\bibfield  {journal} {\bibinfo  {journal} {Phys. Rev. Appl.}\ }\textbf {\bibinfo {volume} {18}},\ \bibinfo {pages} {054061} (\bibinfo {year} {2022})}\BibitemShut {NoStop}%
\bibitem [{\citenamefont {Sahu}\ \emph {et~al.}(2023)\citenamefont {Sahu}, \citenamefont {Qiu}, \citenamefont {Hease}, \citenamefont {Arnold}, \citenamefont {Minoguchi}, \citenamefont {Rabl},\ and\ \citenamefont {Fink}}]{Sahu2023}%
  \BibitemOpen
  \bibfield  {author} {\bibinfo {author} {\bibfnamefont {R.}~\bibnamefont {Sahu}}, \bibinfo {author} {\bibfnamefont {L.}~\bibnamefont {Qiu}}, \bibinfo {author} {\bibfnamefont {W.}~\bibnamefont {Hease}}, \bibinfo {author} {\bibfnamefont {G.}~\bibnamefont {Arnold}}, \bibinfo {author} {\bibfnamefont {Y.}~\bibnamefont {Minoguchi}}, \bibinfo {author} {\bibfnamefont {P.}~\bibnamefont {Rabl}},\ and\ \bibinfo {author} {\bibfnamefont {J.~M.}\ \bibnamefont {Fink}},\ }\bibfield  {title} {\bibinfo {title} {Entangling microwaves with light},\ }\href {https://doi.org/10.1126/science.adg3812} {\bibfield  {journal} {\bibinfo  {journal} {Science}\ }\textbf {\bibinfo {volume} {380}},\ \bibinfo {pages} {718} (\bibinfo {year} {2023})}\BibitemShut {NoStop}%
\bibitem [{\citenamefont {Rochman}\ \emph {et~al.}(2023)\citenamefont {Rochman}, \citenamefont {Xie}, \citenamefont {Bartholomew}, \citenamefont {Schwab},\ and\ \citenamefont {Faraon}}]{Rochman2023}%
  \BibitemOpen
  \bibfield  {author} {\bibinfo {author} {\bibfnamefont {J.}~\bibnamefont {Rochman}}, \bibinfo {author} {\bibfnamefont {T.}~\bibnamefont {Xie}}, \bibinfo {author} {\bibfnamefont {J.~G.}\ \bibnamefont {Bartholomew}}, \bibinfo {author} {\bibfnamefont {K.~C.}\ \bibnamefont {Schwab}},\ and\ \bibinfo {author} {\bibfnamefont {A.}~\bibnamefont {Faraon}},\ }\bibfield  {title} {\bibinfo {title} {Microwave-to-optical transduction with erbium ions coupled to planar photonic and superconducting resonators},\ }\href {https://doi.org/10.1038/s41467-023-36799-0} {\bibfield  {journal} {\bibinfo  {journal} {Nature Communications}\ }\textbf {\bibinfo {volume} {14}},\ \bibinfo {pages} {1153} (\bibinfo {year} {2023})}\BibitemShut {NoStop}%
\bibitem [{\citenamefont {Meesala}\ \emph {et~al.}(2024{\natexlab{a}})\citenamefont {Meesala}, \citenamefont {Wood}, \citenamefont {Lake}, \citenamefont {Chiappina}, \citenamefont {Zhong}, \citenamefont {Beyer}, \citenamefont {Shaw}, \citenamefont {Jiang},\ and\ \citenamefont {Painter}}]{Meesala24}%
  \BibitemOpen
  \bibfield  {author} {\bibinfo {author} {\bibfnamefont {S.}~\bibnamefont {Meesala}}, \bibinfo {author} {\bibfnamefont {S.}~\bibnamefont {Wood}}, \bibinfo {author} {\bibfnamefont {D.}~\bibnamefont {Lake}}, \bibinfo {author} {\bibfnamefont {P.}~\bibnamefont {Chiappina}}, \bibinfo {author} {\bibfnamefont {C.}~\bibnamefont {Zhong}}, \bibinfo {author} {\bibfnamefont {A.~D.}\ \bibnamefont {Beyer}}, \bibinfo {author} {\bibfnamefont {M.~D.}\ \bibnamefont {Shaw}}, \bibinfo {author} {\bibfnamefont {L.}~\bibnamefont {Jiang}},\ and\ \bibinfo {author} {\bibfnamefont {O.}~\bibnamefont {Painter}},\ }\bibfield  {title} {\bibinfo {title} {Non-classical microwave–optical photon pair generation with a chip-scale transducer},\ }\href {https://doi.org/10.1038/s41567-024-02409-z} {\bibfield  {journal} {\bibinfo  {journal} {Nature Physics}\ }\textbf {\bibinfo {volume} {20}},\ \bibinfo {pages} {871–877} (\bibinfo {year} {2024}{\natexlab{a}})}\BibitemShut {NoStop}%
\bibitem [{\citenamefont {Meesala}\ \emph {et~al.}(2024{\natexlab{b}})\citenamefont {Meesala}, \citenamefont {Lake}, \citenamefont {Wood}, \citenamefont {Chiappina}, \citenamefont {Zhong}, \citenamefont {Beyer}, \citenamefont {Shaw}, \citenamefont {Jiang},\ and\ \citenamefont {Painter}}]{Meesala2024PRX}%
  \BibitemOpen
  \bibfield  {author} {\bibinfo {author} {\bibfnamefont {S.}~\bibnamefont {Meesala}}, \bibinfo {author} {\bibfnamefont {D.}~\bibnamefont {Lake}}, \bibinfo {author} {\bibfnamefont {S.}~\bibnamefont {Wood}}, \bibinfo {author} {\bibfnamefont {P.}~\bibnamefont {Chiappina}}, \bibinfo {author} {\bibfnamefont {C.}~\bibnamefont {Zhong}}, \bibinfo {author} {\bibfnamefont {A.~D.}\ \bibnamefont {Beyer}}, \bibinfo {author} {\bibfnamefont {M.~D.}\ \bibnamefont {Shaw}}, \bibinfo {author} {\bibfnamefont {L.}~\bibnamefont {Jiang}},\ and\ \bibinfo {author} {\bibfnamefont {O.}~\bibnamefont {Painter}},\ }\bibfield  {title} {\bibinfo {title} {Quantum entanglement between optical and microwave photonic qubits},\ }\href {https://doi.org/10.1103/PhysRevX.14.031055} {\bibfield  {journal} {\bibinfo  {journal} {Phys. Rev. X}\ }\textbf {\bibinfo {volume} {14}},\ \bibinfo {pages} {031055} (\bibinfo {year} {2024}{\natexlab{b}})}\BibitemShut {NoStop}%
\bibitem [{\citenamefont {Xie}\ \emph {et~al.}(2025)\citenamefont {Xie}, \citenamefont {Fukumori}, \citenamefont {Li},\ and\ \citenamefont {Faraon}}]{Xie2025}%
  \BibitemOpen
  \bibfield  {author} {\bibinfo {author} {\bibfnamefont {T.}~\bibnamefont {Xie}}, \bibinfo {author} {\bibfnamefont {R.}~\bibnamefont {Fukumori}}, \bibinfo {author} {\bibfnamefont {J.}~\bibnamefont {Li}},\ and\ \bibinfo {author} {\bibfnamefont {A.}~\bibnamefont {Faraon}},\ }\bibfield  {title} {\bibinfo {title} {Scalable microwave-to-optical transducers at the single-photon level with spins},\ }\href {https://doi.org/10.1038/s41567-025-02884-y} {\bibfield  {journal} {\bibinfo  {journal} {Nature Physics}\ }\textbf {\bibinfo {volume} {21}},\ \bibinfo {pages} {931} (\bibinfo {year} {2025})}\BibitemShut {NoStop}%
\bibitem [{\citenamefont {Meenehan}\ \emph {et~al.}(2014)\citenamefont {Meenehan}, \citenamefont {Cohen}, \citenamefont {Gr\"oblacher}, \citenamefont {Hill}, \citenamefont {Safavi-Naeini}, \citenamefont {Aspelmeyer},\ and\ \citenamefont {Painter}}]{Meenehan2014}%
  \BibitemOpen
  \bibfield  {author} {\bibinfo {author} {\bibfnamefont {S.~M.}\ \bibnamefont {Meenehan}}, \bibinfo {author} {\bibfnamefont {J.~D.}\ \bibnamefont {Cohen}}, \bibinfo {author} {\bibfnamefont {S.}~\bibnamefont {Gr\"oblacher}}, \bibinfo {author} {\bibfnamefont {J.~T.}\ \bibnamefont {Hill}}, \bibinfo {author} {\bibfnamefont {A.~H.}\ \bibnamefont {Safavi-Naeini}}, \bibinfo {author} {\bibfnamefont {M.}~\bibnamefont {Aspelmeyer}},\ and\ \bibinfo {author} {\bibfnamefont {O.}~\bibnamefont {Painter}},\ }\bibfield  {title} {\bibinfo {title} {Silicon optomechanical crystal resonator at millikelvin temperatures},\ }\href {https://doi.org/10.1103/PhysRevA.90.011803} {\bibfield  {journal} {\bibinfo  {journal} {Phys. Rev. A}\ }\textbf {\bibinfo {volume} {90}},\ \bibinfo {pages} {011803} (\bibinfo {year} {2014})}\BibitemShut {NoStop}%
\bibitem [{\citenamefont {Meenehan}\ \emph {et~al.}(2015)\citenamefont {Meenehan}, \citenamefont {Cohen}, \citenamefont {MacCabe}, \citenamefont {Marsili}, \citenamefont {Shaw},\ and\ \citenamefont {Painter}}]{Meenehan2015}%
  \BibitemOpen
  \bibfield  {author} {\bibinfo {author} {\bibfnamefont {S.~M.}\ \bibnamefont {Meenehan}}, \bibinfo {author} {\bibfnamefont {J.~D.}\ \bibnamefont {Cohen}}, \bibinfo {author} {\bibfnamefont {G.~S.}\ \bibnamefont {MacCabe}}, \bibinfo {author} {\bibfnamefont {F.}~\bibnamefont {Marsili}}, \bibinfo {author} {\bibfnamefont {M.~D.}\ \bibnamefont {Shaw}},\ and\ \bibinfo {author} {\bibfnamefont {O.}~\bibnamefont {Painter}},\ }\bibfield  {title} {\bibinfo {title} {Pulsed excitation dynamics of an optomechanical crystal resonator near its quantum ground state of motion},\ }\href {https://doi.org/10.1103/PhysRevX.5.041002} {\bibfield  {journal} {\bibinfo  {journal} {Phys. Rev. X}\ }\textbf {\bibinfo {volume} {5}},\ \bibinfo {pages} {041002} (\bibinfo {year} {2015})}\BibitemShut {NoStop}%
\bibitem [{\citenamefont {Higginbotham}\ \emph {et~al.}(2018)\citenamefont {Higginbotham}, \citenamefont {Burns}, \citenamefont {Urmey}, \citenamefont {Peterson}, \citenamefont {Kampel}, \citenamefont {Brubaker}, \citenamefont {Smith}, \citenamefont {Lehnert},\ and\ \citenamefont {Regal}}]{Higginbotham2018}%
  \BibitemOpen
  \bibfield  {author} {\bibinfo {author} {\bibfnamefont {A.~P.}\ \bibnamefont {Higginbotham}}, \bibinfo {author} {\bibfnamefont {P.~S.}\ \bibnamefont {Burns}}, \bibinfo {author} {\bibfnamefont {M.~D.}\ \bibnamefont {Urmey}}, \bibinfo {author} {\bibfnamefont {R.~W.}\ \bibnamefont {Peterson}}, \bibinfo {author} {\bibfnamefont {N.~S.}\ \bibnamefont {Kampel}}, \bibinfo {author} {\bibfnamefont {B.~M.}\ \bibnamefont {Brubaker}}, \bibinfo {author} {\bibfnamefont {G.}~\bibnamefont {Smith}}, \bibinfo {author} {\bibfnamefont {K.~W.}\ \bibnamefont {Lehnert}},\ and\ \bibinfo {author} {\bibfnamefont {C.~A.}\ \bibnamefont {Regal}},\ }\bibfield  {title} {\bibinfo {title} {Harnessing electro-optic correlations in an efficient mechanical converter},\ }\href {https://doi.org/10.1038/s41567-018-0210-0} {\bibfield  {journal} {\bibinfo  {journal} {Nature Physics}\ }\textbf {\bibinfo {volume} {14}},\ \bibinfo {pages} {1038} (\bibinfo {year} {2018})}\BibitemShut {NoStop}%
\bibitem [{\citenamefont {MacCabe}\ \emph {et~al.}(2020)\citenamefont {MacCabe}, \citenamefont {Ren}, \citenamefont {Luo}, \citenamefont {Cohen}, \citenamefont {Zhou}, \citenamefont {Sipahigil}, \citenamefont {Mirhosseini},\ and\ \citenamefont {Painter}}]{MacCabe2020}%
  \BibitemOpen
  \bibfield  {author} {\bibinfo {author} {\bibfnamefont {G.~S.}\ \bibnamefont {MacCabe}}, \bibinfo {author} {\bibfnamefont {H.}~\bibnamefont {Ren}}, \bibinfo {author} {\bibfnamefont {J.}~\bibnamefont {Luo}}, \bibinfo {author} {\bibfnamefont {J.~D.}\ \bibnamefont {Cohen}}, \bibinfo {author} {\bibfnamefont {H.}~\bibnamefont {Zhou}}, \bibinfo {author} {\bibfnamefont {A.}~\bibnamefont {Sipahigil}}, \bibinfo {author} {\bibfnamefont {M.}~\bibnamefont {Mirhosseini}},\ and\ \bibinfo {author} {\bibfnamefont {O.}~\bibnamefont {Painter}},\ }\bibfield  {title} {\bibinfo {title} {Nano-acoustic resonator with ultralong phonon lifetime},\ }\href {https://doi.org/10.1126/science.abc7312} {\bibfield  {journal} {\bibinfo  {journal} {Science}\ }\textbf {\bibinfo {volume} {370}},\ \bibinfo {pages} {840} (\bibinfo {year} {2020})}\BibitemShut {NoStop}%
\bibitem [{\citenamefont {Ren}\ \emph {et~al.}(2020)\citenamefont {Ren}, \citenamefont {Matheny}, \citenamefont {MacCabe}, \citenamefont {Luo}, \citenamefont {Pfeifer}, \citenamefont {Mirhosseini},\ and\ \citenamefont {Painter}}]{Ren2020}%
  \BibitemOpen
  \bibfield  {author} {\bibinfo {author} {\bibfnamefont {H.}~\bibnamefont {Ren}}, \bibinfo {author} {\bibfnamefont {M.~H.}\ \bibnamefont {Matheny}}, \bibinfo {author} {\bibfnamefont {G.~S.}\ \bibnamefont {MacCabe}}, \bibinfo {author} {\bibfnamefont {J.}~\bibnamefont {Luo}}, \bibinfo {author} {\bibfnamefont {H.}~\bibnamefont {Pfeifer}}, \bibinfo {author} {\bibfnamefont {M.}~\bibnamefont {Mirhosseini}},\ and\ \bibinfo {author} {\bibfnamefont {O.}~\bibnamefont {Painter}},\ }\bibfield  {title} {\bibinfo {title} {Two-dimensional optomechanical crystal cavity with high quantum cooperativity},\ }\href {https://doi.org/10.1038/s41467-020-17182-9} {\bibfield  {journal} {\bibinfo  {journal} {Nature Communications}\ }\textbf {\bibinfo {volume} {11}},\ \bibinfo {pages} {3373} (\bibinfo {year} {2020})}\BibitemShut {NoStop}%
\bibitem [{\citenamefont {Forsch}\ \emph {et~al.}(2020)\citenamefont {Forsch}, \citenamefont {Stockill}, \citenamefont {Wallucks}, \citenamefont {Marinković}, \citenamefont {Gärtner}, \citenamefont {Norte}, \citenamefont {van Otten}, \citenamefont {Fiore}, \citenamefont {Srinivasan},\ and\ \citenamefont {Gröblacher}}]{forsch_microwave--optics_2020}%
  \BibitemOpen
  \bibfield  {author} {\bibinfo {author} {\bibfnamefont {M.}~\bibnamefont {Forsch}}, \bibinfo {author} {\bibfnamefont {R.}~\bibnamefont {Stockill}}, \bibinfo {author} {\bibfnamefont {A.}~\bibnamefont {Wallucks}}, \bibinfo {author} {\bibfnamefont {I.}~\bibnamefont {Marinković}}, \bibinfo {author} {\bibfnamefont {C.}~\bibnamefont {Gärtner}}, \bibinfo {author} {\bibfnamefont {R.~A.}\ \bibnamefont {Norte}}, \bibinfo {author} {\bibfnamefont {F.}~\bibnamefont {van Otten}}, \bibinfo {author} {\bibfnamefont {A.}~\bibnamefont {Fiore}}, \bibinfo {author} {\bibfnamefont {K.}~\bibnamefont {Srinivasan}},\ and\ \bibinfo {author} {\bibfnamefont {S.}~\bibnamefont {Gröblacher}},\ }\bibfield  {title} {\bibinfo {title} {Microwave-to-optics conversion using a mechanical oscillator in its quantum ground state},\ }\href {https://doi.org/10.1038/s41567-019-0673-7} {\bibfield  {journal} {\bibinfo  {journal} {Nature Physics}\ }\textbf {\bibinfo {volume} {16}},\ \bibinfo {pages} {69} (\bibinfo {year} {2020})}\BibitemShut {NoStop}%
\bibitem [{\citenamefont {Arnold}\ \emph {et~al.}(2020)\citenamefont {Arnold}, \citenamefont {Wulf}, \citenamefont {Barzanjeh}, \citenamefont {Redchenko}, \citenamefont {Rueda}, \citenamefont {Hease}, \citenamefont {Hassani},\ and\ \citenamefont {Fink}}]{arnold_converting_2020}%
  \BibitemOpen
  \bibfield  {author} {\bibinfo {author} {\bibfnamefont {G.}~\bibnamefont {Arnold}}, \bibinfo {author} {\bibfnamefont {M.}~\bibnamefont {Wulf}}, \bibinfo {author} {\bibfnamefont {S.}~\bibnamefont {Barzanjeh}}, \bibinfo {author} {\bibfnamefont {E.~S.}\ \bibnamefont {Redchenko}}, \bibinfo {author} {\bibfnamefont {A.}~\bibnamefont {Rueda}}, \bibinfo {author} {\bibfnamefont {W.~J.}\ \bibnamefont {Hease}}, \bibinfo {author} {\bibfnamefont {F.}~\bibnamefont {Hassani}},\ and\ \bibinfo {author} {\bibfnamefont {J.~M.}\ \bibnamefont {Fink}},\ }\bibfield  {title} {\bibinfo {title} {Converting microwave and telecom photons with a silicon photonic nanomechanical interface},\ }\href {https://doi.org/10.1038/s41467-020-18269-z} {\bibfield  {journal} {\bibinfo  {journal} {Nature Communications}\ }\textbf {\bibinfo {volume} {11}},\ \bibinfo {pages} {4460} (\bibinfo {year} {2020})}\BibitemShut {NoStop}%
\bibitem [{\citenamefont {Zhong}\ \emph {et~al.}(2024)\citenamefont {Zhong}, \citenamefont {Li}, \citenamefont {Meesala}, \citenamefont {Wood}, \citenamefont {Lake}, \citenamefont {Painter},\ and\ \citenamefont {Jiang}}]{Zhong2024}%
  \BibitemOpen
  \bibfield  {author} {\bibinfo {author} {\bibfnamefont {C.}~\bibnamefont {Zhong}}, \bibinfo {author} {\bibfnamefont {F.}~\bibnamefont {Li}}, \bibinfo {author} {\bibfnamefont {S.}~\bibnamefont {Meesala}}, \bibinfo {author} {\bibfnamefont {S.}~\bibnamefont {Wood}}, \bibinfo {author} {\bibfnamefont {D.}~\bibnamefont {Lake}}, \bibinfo {author} {\bibfnamefont {O.}~\bibnamefont {Painter}},\ and\ \bibinfo {author} {\bibfnamefont {L.}~\bibnamefont {Jiang}},\ }\bibfield  {title} {\bibinfo {title} {Microwave-optical entanglement from pulse-pumped electro-optomechanics},\ }\href {https://doi.org/10.1103/PhysRevApplied.22.064047} {\bibfield  {journal} {\bibinfo  {journal} {Phys. Rev. Appl.}\ }\textbf {\bibinfo {volume} {22}},\ \bibinfo {pages} {064047} (\bibinfo {year} {2024})}\BibitemShut {NoStop}%
\bibitem [{\citenamefont {Hease}\ \emph {et~al.}(2020)\citenamefont {Hease}, \citenamefont {Rueda}, \citenamefont {Sahu}, \citenamefont {Wulf}, \citenamefont {Arnold}, \citenamefont {Schwefel},\ and\ \citenamefont {Fink}}]{hease_bidirectional_2020}%
  \BibitemOpen
  \bibfield  {author} {\bibinfo {author} {\bibfnamefont {W.}~\bibnamefont {Hease}}, \bibinfo {author} {\bibfnamefont {A.}~\bibnamefont {Rueda}}, \bibinfo {author} {\bibfnamefont {R.}~\bibnamefont {Sahu}}, \bibinfo {author} {\bibfnamefont {M.}~\bibnamefont {Wulf}}, \bibinfo {author} {\bibfnamefont {G.}~\bibnamefont {Arnold}}, \bibinfo {author} {\bibfnamefont {H.~G.}\ \bibnamefont {Schwefel}},\ and\ \bibinfo {author} {\bibfnamefont {J.~M.}\ \bibnamefont {Fink}},\ }\bibfield  {title} {\bibinfo {title} {Bidirectional {Electro}-{Optic} {Wavelength} {Conversion} in the {Quantum} {Ground} {State}},\ }\href {https://doi.org/10.1103/PRXQuantum.1.020315} {\bibfield  {journal} {\bibinfo  {journal} {PRX Quantum}\ }\textbf {\bibinfo {volume} {1}},\ \bibinfo {pages} {020315} (\bibinfo {year} {2020})}\BibitemShut {NoStop}%
\bibitem [{\citenamefont {Stockill}\ \emph {et~al.}(2022)\citenamefont {Stockill}, \citenamefont {Forsch}, \citenamefont {Hijazi}, \citenamefont {Beaudoin}, \citenamefont {Pantzas}, \citenamefont {Sagnes}, \citenamefont {Braive},\ and\ \citenamefont {Gröblacher}}]{stockill_ultra-low-noise_2022}%
  \BibitemOpen
  \bibfield  {author} {\bibinfo {author} {\bibfnamefont {R.}~\bibnamefont {Stockill}}, \bibinfo {author} {\bibfnamefont {M.}~\bibnamefont {Forsch}}, \bibinfo {author} {\bibfnamefont {F.}~\bibnamefont {Hijazi}}, \bibinfo {author} {\bibfnamefont {G.}~\bibnamefont {Beaudoin}}, \bibinfo {author} {\bibfnamefont {K.}~\bibnamefont {Pantzas}}, \bibinfo {author} {\bibfnamefont {I.}~\bibnamefont {Sagnes}}, \bibinfo {author} {\bibfnamefont {R.}~\bibnamefont {Braive}},\ and\ \bibinfo {author} {\bibfnamefont {S.}~\bibnamefont {Gröblacher}},\ }\bibfield  {title} {\bibinfo {title} {Ultra-low-noise microwave to optics conversion in gallium phosphide},\ }\href {https://doi.org/10.1038/s41467-022-34338-x} {\bibfield  {journal} {\bibinfo  {journal} {Nature Communications}\ }\textbf {\bibinfo {volume} {13}},\ \bibinfo {pages} {6583} (\bibinfo {year} {2022})}\BibitemShut {NoStop}%
\bibitem [{\citenamefont {Arnold}\ \emph {et~al.}(2025)\citenamefont {Arnold}, \citenamefont {Werner}, \citenamefont {Sahu}, \citenamefont {Kapoor}, \citenamefont {Qiu},\ and\ \citenamefont {Fink}}]{Arnold2025}%
  \BibitemOpen
  \bibfield  {author} {\bibinfo {author} {\bibfnamefont {G.}~\bibnamefont {Arnold}}, \bibinfo {author} {\bibfnamefont {T.}~\bibnamefont {Werner}}, \bibinfo {author} {\bibfnamefont {R.}~\bibnamefont {Sahu}}, \bibinfo {author} {\bibfnamefont {L.~N.}\ \bibnamefont {Kapoor}}, \bibinfo {author} {\bibfnamefont {L.}~\bibnamefont {Qiu}},\ and\ \bibinfo {author} {\bibfnamefont {J.~M.}\ \bibnamefont {Fink}},\ }\bibfield  {title} {\bibinfo {title} {All-optical superconducting qubit readout},\ }\href {https://doi.org/10.1038/s41567-024-02741-4} {\bibfield  {journal} {\bibinfo  {journal} {Nature Physics}\ }\textbf {\bibinfo {volume} {21}},\ \bibinfo {pages} {393} (\bibinfo {year} {2025})}\BibitemShut {NoStop}%
\bibitem [{\citenamefont {Sahu}\ \emph {et~al.}(2022)\citenamefont {Sahu}, \citenamefont {Hease}, \citenamefont {Rueda}, \citenamefont {Arnold}, \citenamefont {Qiu},\ and\ \citenamefont {Fink}}]{sahu_quantum-enabled_2022}%
  \BibitemOpen
  \bibfield  {author} {\bibinfo {author} {\bibfnamefont {R.}~\bibnamefont {Sahu}}, \bibinfo {author} {\bibfnamefont {W.}~\bibnamefont {Hease}}, \bibinfo {author} {\bibfnamefont {A.}~\bibnamefont {Rueda}}, \bibinfo {author} {\bibfnamefont {G.}~\bibnamefont {Arnold}}, \bibinfo {author} {\bibfnamefont {L.}~\bibnamefont {Qiu}},\ and\ \bibinfo {author} {\bibfnamefont {J.~M.}\ \bibnamefont {Fink}},\ }\bibfield  {title} {\bibinfo {title} {Quantum-enabled operation of a microwave-optical interface},\ }\href {https://doi.org/10.1038/s41467-022-28924-2} {\bibfield  {journal} {\bibinfo  {journal} {Nature Communications}\ }\textbf {\bibinfo {volume} {13}},\ \bibinfo {pages} {1276} (\bibinfo {year} {2022})}\BibitemShut {NoStop}%
\bibitem [{\citenamefont {Shi}\ and\ \citenamefont {Zhuang}(2024)}]{shi_overcoming_2024}%
  \BibitemOpen
  \bibfield  {author} {\bibinfo {author} {\bibfnamefont {H.}~\bibnamefont {Shi}}\ and\ \bibinfo {author} {\bibfnamefont {Q.}~\bibnamefont {Zhuang}},\ }\bibfield  {title} {\bibinfo {title} {Overcoming the fundamental limit of quantum transduction via intraband entanglement},\ }\href {https://doi.org/10.1364/OPTICAQ.540881} {\bibfield  {journal} {\bibinfo  {journal} {Optica Quantum}\ }\textbf {\bibinfo {volume} {2}},\ \bibinfo {pages} {475} (\bibinfo {year} {2024})}\BibitemShut {NoStop}%
\bibitem [{\citenamefont {Caleffi}\ \emph {et~al.}(2026)\citenamefont {Caleffi}, \citenamefont {d’Avossa}, \citenamefont {Han},\ and\ \citenamefont {Sara~Cacciapuoti}}]{caleffi_quantum_2026}%
  \BibitemOpen
  \bibfield  {author} {\bibinfo {author} {\bibfnamefont {M.}~\bibnamefont {Caleffi}}, \bibinfo {author} {\bibfnamefont {L.}~\bibnamefont {d’Avossa}}, \bibinfo {author} {\bibfnamefont {X.}~\bibnamefont {Han}},\ and\ \bibinfo {author} {\bibfnamefont {A.}~\bibnamefont {Sara~Cacciapuoti}},\ }\bibfield  {title} {\bibinfo {title} {Quantum {Transduction}: {Enabling} {Quantum} {Networking}},\ }\href {https://doi.org/10.1109/COMST.2025.3631150} {\bibfield  {journal} {\bibinfo  {journal} {IEEE Communications Surveys \& Tutorials}\ }\textbf {\bibinfo {volume} {28}},\ \bibinfo {pages} {4195} (\bibinfo {year} {2026})}\BibitemShut {NoStop}%
\bibitem [{\citenamefont {Wang}\ \emph {et~al.}(2022)\citenamefont {Wang}, \citenamefont {Li},\ and\ \citenamefont {Jiang}}]{wang_quantum_2022}%
  \BibitemOpen
  \bibfield  {author} {\bibinfo {author} {\bibfnamefont {C.~H.}\ \bibnamefont {Wang}}, \bibinfo {author} {\bibfnamefont {F.}~\bibnamefont {Li}},\ and\ \bibinfo {author} {\bibfnamefont {L.}~\bibnamefont {Jiang}},\ }\bibfield  {title} {\bibinfo {title} {Quantum capacities of transducers},\ }\bibfield  {journal} {\bibinfo  {journal} {Nature Communications}\ }\textbf {\bibinfo {volume} {13}},\ \href {https://doi.org/10.1038/s41467-022-34373-8} {10.1038/s41467-022-34373-8} (\bibinfo {year} {2022})\BibitemShut {NoStop}%
\bibitem [{\citenamefont {Duan}\ \emph {et~al.}(2001)\citenamefont {Duan}, \citenamefont {Lukin}, \citenamefont {Cirac},\ and\ \citenamefont {Zoller}}]{Duan2001}%
  \BibitemOpen
  \bibfield  {author} {\bibinfo {author} {\bibfnamefont {L.~M.}\ \bibnamefont {Duan}}, \bibinfo {author} {\bibfnamefont {M.~D.}\ \bibnamefont {Lukin}}, \bibinfo {author} {\bibfnamefont {J.~I.}\ \bibnamefont {Cirac}},\ and\ \bibinfo {author} {\bibfnamefont {P.}~\bibnamefont {Zoller}},\ }\bibfield  {title} {\bibinfo {title} {Long-distance quantum communication with atomic ensembles and linear optics},\ }\bibfield  {journal} {\bibinfo  {journal} {Nature}\ }\textbf {\bibinfo {volume} {414}},\ \href {https://doi.org/10.1038/35106500} {10.1038/35106500} (\bibinfo {year} {2001})\BibitemShut {NoStop}%
\bibitem [{\citenamefont {Zhong}\ \emph {et~al.}(2020{\natexlab{b}})\citenamefont {Zhong}, \citenamefont {Han}, \citenamefont {Tang},\ and\ \citenamefont {Jiang}}]{zhong2020A}%
  \BibitemOpen
  \bibfield  {author} {\bibinfo {author} {\bibfnamefont {C.}~\bibnamefont {Zhong}}, \bibinfo {author} {\bibfnamefont {X.}~\bibnamefont {Han}}, \bibinfo {author} {\bibfnamefont {H.~X.}\ \bibnamefont {Tang}},\ and\ \bibinfo {author} {\bibfnamefont {L.}~\bibnamefont {Jiang}},\ }\bibfield  {title} {\bibinfo {title} {Entanglement of microwave-optical modes in a strongly coupled electro-optomechanical system},\ }\href {https://doi.org/10.1103/PhysRevA.101.032345} {\bibfield  {journal} {\bibinfo  {journal} {Phys. Rev. A}\ }\textbf {\bibinfo {volume} {101}},\ \bibinfo {pages} {032345} (\bibinfo {year} {2020}{\natexlab{b}})}\BibitemShut {NoStop}%
\bibitem [{\citenamefont {Duan}\ and\ \citenamefont {Kimble}(2004)}]{PhysRevLett.92.127902}%
  \BibitemOpen
  \bibfield  {author} {\bibinfo {author} {\bibfnamefont {L.-M.}\ \bibnamefont {Duan}}\ and\ \bibinfo {author} {\bibfnamefont {H.~J.}\ \bibnamefont {Kimble}},\ }\bibfield  {title} {\bibinfo {title} {Scalable photonic quantum computation through cavity-assisted interactions},\ }\href {https://doi.org/10.1103/PhysRevLett.92.127902} {\bibfield  {journal} {\bibinfo  {journal} {Phys. Rev. Lett.}\ }\textbf {\bibinfo {volume} {92}},\ \bibinfo {pages} {127902} (\bibinfo {year} {2004})}\BibitemShut {NoStop}%
\bibitem [{\citenamefont {Nemoto}\ \emph {et~al.}(2014)\citenamefont {Nemoto}, \citenamefont {Trupke}, \citenamefont {Devitt}, \citenamefont {Stephens}, \citenamefont {Scharfenberger}, \citenamefont {Buczak}, \citenamefont {N\"obauer}, \citenamefont {Everitt}, \citenamefont {Schmiedmayer},\ and\ \citenamefont {Munro}}]{PhysRevX.4.031022}%
  \BibitemOpen
  \bibfield  {author} {\bibinfo {author} {\bibfnamefont {K.}~\bibnamefont {Nemoto}}, \bibinfo {author} {\bibfnamefont {M.}~\bibnamefont {Trupke}}, \bibinfo {author} {\bibfnamefont {S.~J.}\ \bibnamefont {Devitt}}, \bibinfo {author} {\bibfnamefont {A.~M.}\ \bibnamefont {Stephens}}, \bibinfo {author} {\bibfnamefont {B.}~\bibnamefont {Scharfenberger}}, \bibinfo {author} {\bibfnamefont {K.}~\bibnamefont {Buczak}}, \bibinfo {author} {\bibfnamefont {T.}~\bibnamefont {N\"obauer}}, \bibinfo {author} {\bibfnamefont {M.~S.}\ \bibnamefont {Everitt}}, \bibinfo {author} {\bibfnamefont {J.}~\bibnamefont {Schmiedmayer}},\ and\ \bibinfo {author} {\bibfnamefont {W.~J.}\ \bibnamefont {Munro}},\ }\bibfield  {title} {\bibinfo {title} {Photonic architecture for scalable quantum information processing in diamond},\ }\href {https://doi.org/10.1103/PhysRevX.4.031022} {\bibfield  {journal} {\bibinfo  {journal} {Phys. Rev. X}\ }\textbf {\bibinfo {volume} {4}},\ \bibinfo {pages} {031022} (\bibinfo {year} {2014})}\BibitemShut {NoStop}%
\bibitem [{\citenamefont {Nguyen}\ \emph {et~al.}(2019{\natexlab{a}})\citenamefont {Nguyen}, \citenamefont {Sukachev}, \citenamefont {Bhaskar}, \citenamefont {Machielse}, \citenamefont {Levonian}, \citenamefont {Knall}, \citenamefont {Stroganov}, \citenamefont {Chia}, \citenamefont {Burek}, \citenamefont {Riedinger}, \citenamefont {Park}, \citenamefont {Lon\ifmmode~\check{c}\else \v{c}\fi{}ar},\ and\ \citenamefont {Lukin}}]{Nguyen2019PRB}%
  \BibitemOpen
  \bibfield  {author} {\bibinfo {author} {\bibfnamefont {C.~T.}\ \bibnamefont {Nguyen}}, \bibinfo {author} {\bibfnamefont {D.~D.}\ \bibnamefont {Sukachev}}, \bibinfo {author} {\bibfnamefont {M.~K.}\ \bibnamefont {Bhaskar}}, \bibinfo {author} {\bibfnamefont {B.}~\bibnamefont {Machielse}}, \bibinfo {author} {\bibfnamefont {D.~S.}\ \bibnamefont {Levonian}}, \bibinfo {author} {\bibfnamefont {E.~N.}\ \bibnamefont {Knall}}, \bibinfo {author} {\bibfnamefont {P.}~\bibnamefont {Stroganov}}, \bibinfo {author} {\bibfnamefont {C.}~\bibnamefont {Chia}}, \bibinfo {author} {\bibfnamefont {M.~J.}\ \bibnamefont {Burek}}, \bibinfo {author} {\bibfnamefont {R.}~\bibnamefont {Riedinger}}, \bibinfo {author} {\bibfnamefont {H.}~\bibnamefont {Park}}, \bibinfo {author} {\bibfnamefont {M.}~\bibnamefont {Lon\ifmmode~\check{c}\else \v{c}\fi{}ar}},\ and\ \bibinfo {author} {\bibfnamefont {M.~D.}\ \bibnamefont {Lukin}},\ }\bibfield  {title} {\bibinfo {title} {An integrated nanophotonic quantum register based on silicon-vacancy spins in
  diamond},\ }\href {https://doi.org/10.1103/PhysRevB.100.165428} {\bibfield  {journal} {\bibinfo  {journal} {Phys. Rev. B}\ }\textbf {\bibinfo {volume} {100}},\ \bibinfo {pages} {165428} (\bibinfo {year} {2019}{\natexlab{a}})}\BibitemShut {NoStop}%
\bibitem [{\citenamefont {Raymer}\ \emph {et~al.}(2024)\citenamefont {Raymer}, \citenamefont {Embleton},\ and\ \citenamefont {Shapiro}}]{PhysRevApplied.22.044013}%
  \BibitemOpen
  \bibfield  {author} {\bibinfo {author} {\bibfnamefont {M.~G.}\ \bibnamefont {Raymer}}, \bibinfo {author} {\bibfnamefont {C.}~\bibnamefont {Embleton}},\ and\ \bibinfo {author} {\bibfnamefont {J.~H.}\ \bibnamefont {Shapiro}},\ }\bibfield  {title} {\bibinfo {title} {The duan-kimble cavity-atom quantum memory loading scheme revisited},\ }\href {https://doi.org/10.1103/PhysRevApplied.22.044013} {\bibfield  {journal} {\bibinfo  {journal} {Phys. Rev. Appl.}\ }\textbf {\bibinfo {volume} {22}},\ \bibinfo {pages} {044013} (\bibinfo {year} {2024})}\BibitemShut {NoStop}%
\bibitem [{\citenamefont {Sipahigil}\ \emph {et~al.}(2016)\citenamefont {Sipahigil}, \citenamefont {Evans}, \citenamefont {Sukachev}, \citenamefont {Burek}, \citenamefont {Borregaard}, \citenamefont {Bhaskar}, \citenamefont {Nguyen}, \citenamefont {Pacheco}, \citenamefont {Atikian}, \citenamefont {Meuwly}, \citenamefont {Camacho}, \citenamefont {Jelezko}, \citenamefont {Bielejec}, \citenamefont {Park}, \citenamefont {Lončar},\ and\ \citenamefont {Lukin}}]{Sipahigil2016integrated}%
  \BibitemOpen
  \bibfield  {author} {\bibinfo {author} {\bibfnamefont {A.}~\bibnamefont {Sipahigil}}, \bibinfo {author} {\bibfnamefont {R.~E.}\ \bibnamefont {Evans}}, \bibinfo {author} {\bibfnamefont {D.~D.}\ \bibnamefont {Sukachev}}, \bibinfo {author} {\bibfnamefont {M.~J.}\ \bibnamefont {Burek}}, \bibinfo {author} {\bibfnamefont {J.}~\bibnamefont {Borregaard}}, \bibinfo {author} {\bibfnamefont {M.~K.}\ \bibnamefont {Bhaskar}}, \bibinfo {author} {\bibfnamefont {C.~T.}\ \bibnamefont {Nguyen}}, \bibinfo {author} {\bibfnamefont {J.~L.}\ \bibnamefont {Pacheco}}, \bibinfo {author} {\bibfnamefont {H.~A.}\ \bibnamefont {Atikian}}, \bibinfo {author} {\bibfnamefont {C.}~\bibnamefont {Meuwly}}, \bibinfo {author} {\bibfnamefont {R.~M.}\ \bibnamefont {Camacho}}, \bibinfo {author} {\bibfnamefont {F.}~\bibnamefont {Jelezko}}, \bibinfo {author} {\bibfnamefont {E.}~\bibnamefont {Bielejec}}, \bibinfo {author} {\bibfnamefont {H.}~\bibnamefont {Park}}, \bibinfo {author} {\bibfnamefont {M.}~\bibnamefont {Lončar}},\ and\ \bibinfo {author}
  {\bibfnamefont {M.~D.}\ \bibnamefont {Lukin}},\ }\bibfield  {title} {\bibinfo {title} {An integrated diamond nanophotonics platform for quantum-optical networks},\ }\href {https://doi.org/10.1126/science.aah6875} {\bibfield  {journal} {\bibinfo  {journal} {Science}\ }\textbf {\bibinfo {volume} {354}},\ \bibinfo {pages} {847} (\bibinfo {year} {2016})}\BibitemShut {NoStop}%
\bibitem [{\citenamefont {Nguyen}\ \emph {et~al.}(2019{\natexlab{b}})\citenamefont {Nguyen}, \citenamefont {Sukachev}, \citenamefont {Bhaskar}, \citenamefont {Machielse}, \citenamefont {Levonian}, \citenamefont {Knall}, \citenamefont {Stroganov}, \citenamefont {Riedinger}, \citenamefont {Park}, \citenamefont {Lon\ifmmode~\check{c}\else \v{c}\fi{}ar},\ and\ \citenamefont {Lukin}}]{Nguyen2019PRL}%
  \BibitemOpen
  \bibfield  {author} {\bibinfo {author} {\bibfnamefont {C.~T.}\ \bibnamefont {Nguyen}}, \bibinfo {author} {\bibfnamefont {D.~D.}\ \bibnamefont {Sukachev}}, \bibinfo {author} {\bibfnamefont {M.~K.}\ \bibnamefont {Bhaskar}}, \bibinfo {author} {\bibfnamefont {B.}~\bibnamefont {Machielse}}, \bibinfo {author} {\bibfnamefont {D.~S.}\ \bibnamefont {Levonian}}, \bibinfo {author} {\bibfnamefont {E.~N.}\ \bibnamefont {Knall}}, \bibinfo {author} {\bibfnamefont {P.}~\bibnamefont {Stroganov}}, \bibinfo {author} {\bibfnamefont {R.}~\bibnamefont {Riedinger}}, \bibinfo {author} {\bibfnamefont {H.}~\bibnamefont {Park}}, \bibinfo {author} {\bibfnamefont {M.}~\bibnamefont {Lon\ifmmode~\check{c}\else \v{c}\fi{}ar}},\ and\ \bibinfo {author} {\bibfnamefont {M.~D.}\ \bibnamefont {Lukin}},\ }\bibfield  {title} {\bibinfo {title} {Quantum network nodes based on diamond qubits with an efficient nanophotonic interface},\ }\href {https://doi.org/10.1103/PhysRevLett.123.183602} {\bibfield  {journal} {\bibinfo  {journal} {Phys. Rev.
  Lett.}\ }\textbf {\bibinfo {volume} {123}},\ \bibinfo {pages} {183602} (\bibinfo {year} {2019}{\natexlab{b}})}\BibitemShut {NoStop}%
\bibitem [{\citenamefont {Tchebotareva}\ \emph {et~al.}(2019)\citenamefont {Tchebotareva}, \citenamefont {Hermans}, \citenamefont {Humphreys}, \citenamefont {Voigt}, \citenamefont {Harmsma}, \citenamefont {Cheng}, \citenamefont {Verlaan}, \citenamefont {Dijkhuizen}, \citenamefont {de~Jong}, \citenamefont {Dr\'eau},\ and\ \citenamefont {Hanson}}]{hanson2019}%
  \BibitemOpen
  \bibfield  {author} {\bibinfo {author} {\bibfnamefont {A.}~\bibnamefont {Tchebotareva}}, \bibinfo {author} {\bibfnamefont {S.~L.~N.}\ \bibnamefont {Hermans}}, \bibinfo {author} {\bibfnamefont {P.~C.}\ \bibnamefont {Humphreys}}, \bibinfo {author} {\bibfnamefont {D.}~\bibnamefont {Voigt}}, \bibinfo {author} {\bibfnamefont {P.~J.}\ \bibnamefont {Harmsma}}, \bibinfo {author} {\bibfnamefont {L.~K.}\ \bibnamefont {Cheng}}, \bibinfo {author} {\bibfnamefont {A.~L.}\ \bibnamefont {Verlaan}}, \bibinfo {author} {\bibfnamefont {N.}~\bibnamefont {Dijkhuizen}}, \bibinfo {author} {\bibfnamefont {W.}~\bibnamefont {de~Jong}}, \bibinfo {author} {\bibfnamefont {A.}~\bibnamefont {Dr\'eau}},\ and\ \bibinfo {author} {\bibfnamefont {R.}~\bibnamefont {Hanson}},\ }\bibfield  {title} {\bibinfo {title} {Entanglement between a diamond spin qubit and a photonic time-bin qubit at telecom wavelength},\ }\href {https://doi.org/10.1103/PhysRevLett.123.063601} {\bibfield  {journal} {\bibinfo  {journal} {Phys. Rev. Lett.}\ }\textbf {\bibinfo
  {volume} {123}},\ \bibinfo {pages} {063601} (\bibinfo {year} {2019})}\BibitemShut {NoStop}%
\bibitem [{\citenamefont {Rosenthal}\ \emph {et~al.}(2023)\citenamefont {Rosenthal}, \citenamefont {Anderson}, \citenamefont {Kleidermacher}, \citenamefont {Stein}, \citenamefont {Lee}, \citenamefont {Grzesik}, \citenamefont {Scuri}, \citenamefont {Rugar}, \citenamefont {Riedel}, \citenamefont {Aghaeimeibodi}, \citenamefont {Ahn}, \citenamefont {Van~Gasse},\ and\ \citenamefont {Vu\ifmmode \check{c}\else \v{c}\fi{}kovi\ifmmode~\acute{c}\else \'{c}\fi{}}}]{Rosenthal2023}%
  \BibitemOpen
  \bibfield  {author} {\bibinfo {author} {\bibfnamefont {E.~I.}\ \bibnamefont {Rosenthal}}, \bibinfo {author} {\bibfnamefont {C.~P.}\ \bibnamefont {Anderson}}, \bibinfo {author} {\bibfnamefont {H.~C.}\ \bibnamefont {Kleidermacher}}, \bibinfo {author} {\bibfnamefont {A.~J.}\ \bibnamefont {Stein}}, \bibinfo {author} {\bibfnamefont {H.}~\bibnamefont {Lee}}, \bibinfo {author} {\bibfnamefont {J.}~\bibnamefont {Grzesik}}, \bibinfo {author} {\bibfnamefont {G.}~\bibnamefont {Scuri}}, \bibinfo {author} {\bibfnamefont {A.~E.}\ \bibnamefont {Rugar}}, \bibinfo {author} {\bibfnamefont {D.}~\bibnamefont {Riedel}}, \bibinfo {author} {\bibfnamefont {S.}~\bibnamefont {Aghaeimeibodi}}, \bibinfo {author} {\bibfnamefont {G.~H.}\ \bibnamefont {Ahn}}, \bibinfo {author} {\bibfnamefont {K.}~\bibnamefont {Van~Gasse}},\ and\ \bibinfo {author} {\bibfnamefont {J.}~\bibnamefont {Vu\ifmmode \check{c}\else \v{c}\fi{}kovi\ifmmode~\acute{c}\else \'{c}\fi{}}},\ }\bibfield  {title} {\bibinfo {title} {Microwave spin control of a tin-vacancy
  qubit in diamond},\ }\href {https://doi.org/10.1103/PhysRevX.13.031022} {\bibfield  {journal} {\bibinfo  {journal} {Phys. Rev. X}\ }\textbf {\bibinfo {volume} {13}},\ \bibinfo {pages} {031022} (\bibinfo {year} {2023})}\BibitemShut {NoStop}%
\bibitem [{\citenamefont {Guo}\ \emph {et~al.}(2023)\citenamefont {Guo}, \citenamefont {Stramma}, \citenamefont {Li}, \citenamefont {Roth}, \citenamefont {Huang}, \citenamefont {Jin}, \citenamefont {Parker}, \citenamefont {Martínez}, \citenamefont {Shofer}, \citenamefont {Michaels}, \citenamefont {Purser}, \citenamefont {Appel}, \citenamefont {Alexeev}, \citenamefont {Liu}, \citenamefont {Ferrari}, \citenamefont {Awschalom}, \citenamefont {Delegan}, \citenamefont {Pingault}, \citenamefont {Galli}, \citenamefont {Heremans}, \citenamefont {Atatüre},\ and\ \citenamefont {High}}]{Guo2023}%
  \BibitemOpen
  \bibfield  {author} {\bibinfo {author} {\bibfnamefont {X.}~\bibnamefont {Guo}}, \bibinfo {author} {\bibfnamefont {A.~M.}\ \bibnamefont {Stramma}}, \bibinfo {author} {\bibfnamefont {Z.}~\bibnamefont {Li}}, \bibinfo {author} {\bibfnamefont {W.~G.}\ \bibnamefont {Roth}}, \bibinfo {author} {\bibfnamefont {B.}~\bibnamefont {Huang}}, \bibinfo {author} {\bibfnamefont {Y.}~\bibnamefont {Jin}}, \bibinfo {author} {\bibfnamefont {R.~A.}\ \bibnamefont {Parker}}, \bibinfo {author} {\bibfnamefont {J.~A.}\ \bibnamefont {Martínez}}, \bibinfo {author} {\bibfnamefont {N.}~\bibnamefont {Shofer}}, \bibinfo {author} {\bibfnamefont {C.~P.}\ \bibnamefont {Michaels}}, \bibinfo {author} {\bibfnamefont {C.~P.}\ \bibnamefont {Purser}}, \bibinfo {author} {\bibfnamefont {M.~H.}\ \bibnamefont {Appel}}, \bibinfo {author} {\bibfnamefont {E.~M.}\ \bibnamefont {Alexeev}}, \bibinfo {author} {\bibfnamefont {T.}~\bibnamefont {Liu}}, \bibinfo {author} {\bibfnamefont {A.~C.}\ \bibnamefont {Ferrari}}, \bibinfo {author} {\bibfnamefont {D.~D.}\
  \bibnamefont {Awschalom}}, \bibinfo {author} {\bibfnamefont {N.}~\bibnamefont {Delegan}}, \bibinfo {author} {\bibfnamefont {B.}~\bibnamefont {Pingault}}, \bibinfo {author} {\bibfnamefont {G.}~\bibnamefont {Galli}}, \bibinfo {author} {\bibfnamefont {F.~J.}\ \bibnamefont {Heremans}}, \bibinfo {author} {\bibfnamefont {M.}~\bibnamefont {Atatüre}},\ and\ \bibinfo {author} {\bibfnamefont {A.~A.}\ \bibnamefont {High}},\ }\bibfield  {title} {\bibinfo {title} {Microwave-based quantum control and coherence protection of tin-vacancy spin qubits in a strain-tuned diamond-membrane heterostructure},\ }\bibfield  {journal} {\bibinfo  {journal} {Physical Review X}\ }\textbf {\bibinfo {volume} {13}},\ \href {https://doi.org/10.1103/PhysRevX.13.041037} {10.1103/PhysRevX.13.041037} (\bibinfo {year} {2023})\BibitemShut {NoStop}%
\bibitem [{\citenamefont {Harris}\ \emph {et~al.}(2025)\citenamefont {Harris}, \citenamefont {Christen}, \citenamefont {Patomäki}, \citenamefont {Raniwala}, \citenamefont {Sirotin}, \citenamefont {Colangelo}, \citenamefont {Chen}, \citenamefont {Errando-Herranz}, \citenamefont {Starling}, \citenamefont {Murphy}, \citenamefont {Shtyrkova}, \citenamefont {Medeiros}, \citenamefont {Trusheim}, \citenamefont {Berggren}, \citenamefont {Dixon},\ and\ \citenamefont {Englund}}]{harris2025high}%
  \BibitemOpen
  \bibfield  {author} {\bibinfo {author} {\bibfnamefont {I.~B.~W.}\ \bibnamefont {Harris}}, \bibinfo {author} {\bibfnamefont {I.}~\bibnamefont {Christen}}, \bibinfo {author} {\bibfnamefont {S.~M.}\ \bibnamefont {Patomäki}}, \bibinfo {author} {\bibfnamefont {H.}~\bibnamefont {Raniwala}}, \bibinfo {author} {\bibfnamefont {M.}~\bibnamefont {Sirotin}}, \bibinfo {author} {\bibfnamefont {M.}~\bibnamefont {Colangelo}}, \bibinfo {author} {\bibfnamefont {K.~C.}\ \bibnamefont {Chen}}, \bibinfo {author} {\bibfnamefont {C.}~\bibnamefont {Errando-Herranz}}, \bibinfo {author} {\bibfnamefont {D.~J.}\ \bibnamefont {Starling}}, \bibinfo {author} {\bibfnamefont {R.}~\bibnamefont {Murphy}}, \bibinfo {author} {\bibfnamefont {K.}~\bibnamefont {Shtyrkova}}, \bibinfo {author} {\bibfnamefont {O.}~\bibnamefont {Medeiros}}, \bibinfo {author} {\bibfnamefont {M.~E.}\ \bibnamefont {Trusheim}}, \bibinfo {author} {\bibfnamefont {K.~K.}\ \bibnamefont {Berggren}}, \bibinfo {author} {\bibfnamefont {P.~B.}\ \bibnamefont {Dixon}},\ and\
  \bibinfo {author} {\bibfnamefont {D.}~\bibnamefont {Englund}},\ }\href {https://arxiv.org/abs/2505.09267} {\bibinfo {title} {High-fidelity control of a strongly coupled electro-nuclear spin-photon interface}} (\bibinfo {year} {2025}),\ \Eprint {https://arxiv.org/abs/2505.09267} {arXiv:2505.09267 [quant-ph]} \BibitemShut {NoStop}%
\bibitem [{\citenamefont {Heo}\ \emph {et~al.}(2025)\citenamefont {Heo}, \citenamefont {Li}, \citenamefont {Wang}, \citenamefont {Higginbotham}, \citenamefont {High},\ and\ \citenamefont {Jiang}}]{PFMO_device_design_2025}%
  \BibitemOpen
  \bibfield  {author} {\bibinfo {author} {\bibfnamefont {J.}~\bibnamefont {Heo}}, \bibinfo {author} {\bibfnamefont {F.}~\bibnamefont {Li}}, \bibinfo {author} {\bibfnamefont {Z.}~\bibnamefont {Wang}}, \bibinfo {author} {\bibfnamefont {A.~P.}\ \bibnamefont {Higginbotham}}, \bibinfo {author} {\bibfnamefont {A.~A.}\ \bibnamefont {High}},\ and\ \bibinfo {author} {\bibfnamefont {L.}~\bibnamefont {Jiang}},\ }\href {https://arxiv.org/abs/2512.05096} {\bibinfo {title} {Pump free microwave-optical quantum transduction}} (\bibinfo {year} {2025}),\ \Eprint {https://arxiv.org/abs/2512.05096} {arXiv:2512.05096 [quant-ph]} \BibitemShut {NoStop}%
\bibitem [{\citenamefont {Verd\'u}\ \emph {et~al.}(2009)\citenamefont {Verd\'u}, \citenamefont {Zoubi}, \citenamefont {Koller}, \citenamefont {Majer}, \citenamefont {Ritsch},\ and\ \citenamefont {Schmiedmayer}}]{Verd2009}%
  \BibitemOpen
  \bibfield  {author} {\bibinfo {author} {\bibfnamefont {J.}~\bibnamefont {Verd\'u}}, \bibinfo {author} {\bibfnamefont {H.}~\bibnamefont {Zoubi}}, \bibinfo {author} {\bibfnamefont {C.}~\bibnamefont {Koller}}, \bibinfo {author} {\bibfnamefont {J.}~\bibnamefont {Majer}}, \bibinfo {author} {\bibfnamefont {H.}~\bibnamefont {Ritsch}},\ and\ \bibinfo {author} {\bibfnamefont {J.}~\bibnamefont {Schmiedmayer}},\ }\bibfield  {title} {\bibinfo {title} {Strong magnetic coupling of an ultracold gas to a superconducting waveguide cavity},\ }\href {https://doi.org/10.1103/PhysRevLett.103.043603} {\bibfield  {journal} {\bibinfo  {journal} {Phys. Rev. Lett.}\ }\textbf {\bibinfo {volume} {103}},\ \bibinfo {pages} {043603} (\bibinfo {year} {2009})}\BibitemShut {NoStop}%
\bibitem [{\citenamefont {Bernon}\ \emph {et~al.}(2013)\citenamefont {Bernon}, \citenamefont {Hattermann}, \citenamefont {Bothner}, \citenamefont {Knufinke}, \citenamefont {Weiss}, \citenamefont {Jessen}, \citenamefont {Cano}, \citenamefont {Kemmler}, \citenamefont {Kleiner}, \citenamefont {Koelle},\ and\ \citenamefont {Fort{\'a}gh}}]{Bernon2013}%
  \BibitemOpen
  \bibfield  {author} {\bibinfo {author} {\bibfnamefont {S.}~\bibnamefont {Bernon}}, \bibinfo {author} {\bibfnamefont {H.}~\bibnamefont {Hattermann}}, \bibinfo {author} {\bibfnamefont {D.}~\bibnamefont {Bothner}}, \bibinfo {author} {\bibfnamefont {M.}~\bibnamefont {Knufinke}}, \bibinfo {author} {\bibfnamefont {P.}~\bibnamefont {Weiss}}, \bibinfo {author} {\bibfnamefont {F.}~\bibnamefont {Jessen}}, \bibinfo {author} {\bibfnamefont {D.}~\bibnamefont {Cano}}, \bibinfo {author} {\bibfnamefont {M.}~\bibnamefont {Kemmler}}, \bibinfo {author} {\bibfnamefont {R.}~\bibnamefont {Kleiner}}, \bibinfo {author} {\bibfnamefont {D.}~\bibnamefont {Koelle}},\ and\ \bibinfo {author} {\bibfnamefont {J.}~\bibnamefont {Fort{\'a}gh}},\ }\bibfield  {title} {\bibinfo {title} {Manipulation and coherence of ultra-cold atoms on a superconducting atom chip},\ }\href {https://doi.org/10.1038/ncomms3380} {\bibfield  {journal} {\bibinfo  {journal} {Nature Communications}\ }\textbf {\bibinfo {volume} {4}},\ \bibinfo {pages} {2380} (\bibinfo
  {year} {2013})}\BibitemShut {NoStop}%
\bibitem [{\citenamefont {Hattermann}\ \emph {et~al.}(2017)\citenamefont {Hattermann}, \citenamefont {Bothner}, \citenamefont {Ley}, \citenamefont {Ferdinand}, \citenamefont {Wiedmaier}, \citenamefont {S{\'a}rk{\'a}ny}, \citenamefont {Kleiner}, \citenamefont {Koelle},\ and\ \citenamefont {Fort{\'a}gh}}]{Hattermann2017}%
  \BibitemOpen
  \bibfield  {author} {\bibinfo {author} {\bibfnamefont {H.}~\bibnamefont {Hattermann}}, \bibinfo {author} {\bibfnamefont {D.}~\bibnamefont {Bothner}}, \bibinfo {author} {\bibfnamefont {L.~Y.}\ \bibnamefont {Ley}}, \bibinfo {author} {\bibfnamefont {B.}~\bibnamefont {Ferdinand}}, \bibinfo {author} {\bibfnamefont {D.}~\bibnamefont {Wiedmaier}}, \bibinfo {author} {\bibfnamefont {L.}~\bibnamefont {S{\'a}rk{\'a}ny}}, \bibinfo {author} {\bibfnamefont {R.}~\bibnamefont {Kleiner}}, \bibinfo {author} {\bibfnamefont {D.}~\bibnamefont {Koelle}},\ and\ \bibinfo {author} {\bibfnamefont {J.}~\bibnamefont {Fort{\'a}gh}},\ }\bibfield  {title} {\bibinfo {title} {Coupling ultracold atoms to a superconducting coplanar waveguide resonator},\ }\href {https://doi.org/10.1038/s41467-017-02439-7} {\bibfield  {journal} {\bibinfo  {journal} {Nature Communications}\ }\textbf {\bibinfo {volume} {8}},\ \bibinfo {pages} {2254} (\bibinfo {year} {2017})}\BibitemShut {NoStop}%
\bibitem [{\citenamefont {Chen}\ \emph {et~al.}(2022)\citenamefont {Chen}, \citenamefont {Szurek}, \citenamefont {Hu}, \citenamefont {de~Hond}, \citenamefont {Braverman},\ and\ \citenamefont {Vuletic}}]{Chen2022}%
  \BibitemOpen
  \bibfield  {author} {\bibinfo {author} {\bibfnamefont {Y.-T.}\ \bibnamefont {Chen}}, \bibinfo {author} {\bibfnamefont {M.}~\bibnamefont {Szurek}}, \bibinfo {author} {\bibfnamefont {B.}~\bibnamefont {Hu}}, \bibinfo {author} {\bibfnamefont {J.}~\bibnamefont {de~Hond}}, \bibinfo {author} {\bibfnamefont {B.}~\bibnamefont {Braverman}},\ and\ \bibinfo {author} {\bibfnamefont {V.}~\bibnamefont {Vuletic}},\ }\bibfield  {title} {\bibinfo {title} {High finesse bow-tie cavity for strong atom-photon coupling in rydberg arrays},\ }\href {https://doi.org/10.1364/OE.469644} {\bibfield  {journal} {\bibinfo  {journal} {Opt. Express}\ }\textbf {\bibinfo {volume} {30}},\ \bibinfo {pages} {37426} (\bibinfo {year} {2022})}\BibitemShut {NoStop}%
\bibitem [{\citenamefont {Wilde}\ \emph {et~al.}(2025)\citenamefont {Wilde}, \citenamefont {Kaiser}, \citenamefont {Reinschmidt}, \citenamefont {G\"unther}, \citenamefont {Koelle}, \citenamefont {Fort\'agh}, \citenamefont {Kleiner},\ and\ \citenamefont {Bothner}}]{Wilde2025}%
  \BibitemOpen
  \bibfield  {author} {\bibinfo {author} {\bibfnamefont {B.}~\bibnamefont {Wilde}}, \bibinfo {author} {\bibfnamefont {M.}~\bibnamefont {Kaiser}}, \bibinfo {author} {\bibfnamefont {M.}~\bibnamefont {Reinschmidt}}, \bibinfo {author} {\bibfnamefont {A.}~\bibnamefont {G\"unther}}, \bibinfo {author} {\bibfnamefont {D.}~\bibnamefont {Koelle}}, \bibinfo {author} {\bibfnamefont {J.}~\bibnamefont {Fort\'agh}}, \bibinfo {author} {\bibfnamefont {R.}~\bibnamefont {Kleiner}},\ and\ \bibinfo {author} {\bibfnamefont {D.}~\bibnamefont {Bothner}},\ }\bibfield  {title} {\bibinfo {title} {Superconducting on-chip microwave cavity for tunable hybrid systems with optically trapped rydberg atoms},\ }\href {https://doi.org/10.1103/PhysRevApplied.23.064016} {\bibfield  {journal} {\bibinfo  {journal} {Phys. Rev. Appl.}\ }\textbf {\bibinfo {volume} {23}},\ \bibinfo {pages} {064016} (\bibinfo {year} {2025})}\BibitemShut {NoStop}%
\bibitem [{\citenamefont {Meesala}\ \emph {et~al.}(2018)\citenamefont {Meesala}, \citenamefont {Sohn}, \citenamefont {Pingault}, \citenamefont {Shao}, \citenamefont {Atikian}, \citenamefont {Holzgrafe}, \citenamefont {G{\"{u}}ndoğan}, \citenamefont {Stavrakas}, \citenamefont {Sipahigil}, \citenamefont {Chia}, \citenamefont {Evans}, \citenamefont {Burek}, \citenamefont {Zhang}, \citenamefont {Wu}, \citenamefont {Pacheco}, \citenamefont {Abraham}, \citenamefont {Bielejec}, \citenamefont {Lukin}, \citenamefont {Atat{\"{u}}re},\ and\ \citenamefont {Lon{\v{c}}ar}}]{Meesala2018}%
  \BibitemOpen
  \bibfield  {author} {\bibinfo {author} {\bibfnamefont {S.}~\bibnamefont {Meesala}}, \bibinfo {author} {\bibfnamefont {Y.-I.}\ \bibnamefont {Sohn}}, \bibinfo {author} {\bibfnamefont {B.}~\bibnamefont {Pingault}}, \bibinfo {author} {\bibfnamefont {L.}~\bibnamefont {Shao}}, \bibinfo {author} {\bibfnamefont {H.~A.}\ \bibnamefont {Atikian}}, \bibinfo {author} {\bibfnamefont {J.}~\bibnamefont {Holzgrafe}}, \bibinfo {author} {\bibfnamefont {M.}~\bibnamefont {G{\"{u}}ndoğan}}, \bibinfo {author} {\bibfnamefont {C.}~\bibnamefont {Stavrakas}}, \bibinfo {author} {\bibfnamefont {A.}~\bibnamefont {Sipahigil}}, \bibinfo {author} {\bibfnamefont {C.}~\bibnamefont {Chia}}, \bibinfo {author} {\bibfnamefont {R.}~\bibnamefont {Evans}}, \bibinfo {author} {\bibfnamefont {M.~J.}\ \bibnamefont {Burek}}, \bibinfo {author} {\bibfnamefont {M.}~\bibnamefont {Zhang}}, \bibinfo {author} {\bibfnamefont {L.}~\bibnamefont {Wu}}, \bibinfo {author} {\bibfnamefont {J.~L.}\ \bibnamefont {Pacheco}}, \bibinfo {author} {\bibfnamefont
  {J.}~\bibnamefont {Abraham}}, \bibinfo {author} {\bibfnamefont {E.}~\bibnamefont {Bielejec}}, \bibinfo {author} {\bibfnamefont {M.~D.}\ \bibnamefont {Lukin}}, \bibinfo {author} {\bibfnamefont {M.}~\bibnamefont {Atat{\"{u}}re}},\ and\ \bibinfo {author} {\bibfnamefont {M.}~\bibnamefont {Lon{\v{c}}ar}},\ }\bibfield  {title} {\bibinfo {title} {{Strain engineering of the silicon-vacancy center in diamond}},\ }\href {https://doi.org/10.1103/PhysRevB.97.205444} {\bibfield  {journal} {\bibinfo  {journal} {Physical Review B}\ }\textbf {\bibinfo {volume} {97}},\ \bibinfo {pages} {205444} (\bibinfo {year} {2018})}\BibitemShut {NoStop}%
\bibitem [{\citenamefont {Joe}\ \emph {et~al.}(2026)\citenamefont {Joe}, \citenamefont {Haas}, \citenamefont {Kuruma}, \citenamefont {Jin}, \citenamefont {Kang}, \citenamefont {Ding}, \citenamefont {Chia}, \citenamefont {Warner}, \citenamefont {Pingault}, \citenamefont {Machielse}, \citenamefont {Meesala},\ and\ \citenamefont {Lončar}}]{joe_purcell-enhanced_2026}%
  \BibitemOpen
  \bibfield  {author} {\bibinfo {author} {\bibfnamefont {G.}~\bibnamefont {Joe}}, \bibinfo {author} {\bibfnamefont {M.}~\bibnamefont {Haas}}, \bibinfo {author} {\bibfnamefont {K.}~\bibnamefont {Kuruma}}, \bibinfo {author} {\bibfnamefont {C.}~\bibnamefont {Jin}}, \bibinfo {author} {\bibfnamefont {D.~D.}\ \bibnamefont {Kang}}, \bibinfo {author} {\bibfnamefont {S.~W.}\ \bibnamefont {Ding}}, \bibinfo {author} {\bibfnamefont {C.}~\bibnamefont {Chia}}, \bibinfo {author} {\bibfnamefont {H.}~\bibnamefont {Warner}}, \bibinfo {author} {\bibfnamefont {B.}~\bibnamefont {Pingault}}, \bibinfo {author} {\bibfnamefont {B.}~\bibnamefont {Machielse}}, \bibinfo {author} {\bibfnamefont {S.}~\bibnamefont {Meesala}},\ and\ \bibinfo {author} {\bibfnamefont {M.}~\bibnamefont {Lončar}},\ }\bibfield  {title} {\bibinfo {title} {Purcell-enhanced spin–phonon coupling with a single colour centre},\ }\href {https://doi.org/10.1038/s41586-026-10495-7} {\bibfield  {journal} {\bibinfo  {journal} {Nature}\ }\textbf {\bibinfo {volume}
  {653}},\ \bibinfo {pages} {378} (\bibinfo {year} {2026})}\BibitemShut {NoStop}%
\bibitem [{\citenamefont {Arrangoiz-Arriola}\ \emph {et~al.}(2018)\citenamefont {Arrangoiz-Arriola}, \citenamefont {Wollack}, \citenamefont {Pechal}, \citenamefont {Witmer}, \citenamefont {Hill},\ and\ \citenamefont {Safavi-Naeini}}]{arrangoiz-arriola_coupling_2018}%
  \BibitemOpen
  \bibfield  {author} {\bibinfo {author} {\bibfnamefont {P.}~\bibnamefont {Arrangoiz-Arriola}}, \bibinfo {author} {\bibfnamefont {E.~A.}\ \bibnamefont {Wollack}}, \bibinfo {author} {\bibfnamefont {M.}~\bibnamefont {Pechal}}, \bibinfo {author} {\bibfnamefont {J.~D.}\ \bibnamefont {Witmer}}, \bibinfo {author} {\bibfnamefont {J.~T.}\ \bibnamefont {Hill}},\ and\ \bibinfo {author} {\bibfnamefont {A.~H.}\ \bibnamefont {Safavi-Naeini}},\ }\bibfield  {title} {\bibinfo {title} {Coupling a superconducting quantum circuit to a phononic crystal defect cavity},\ }\href {https://doi.org/10.1103/PhysRevX.8.031007} {\bibfield  {journal} {\bibinfo  {journal} {Phys. Rev. X}\ }\textbf {\bibinfo {volume} {8}},\ \bibinfo {pages} {031007} (\bibinfo {year} {2018})}\BibitemShut {NoStop}%
\bibitem [{\citenamefont {Stas}\ \emph {et~al.}(2022)\citenamefont {Stas}, \citenamefont {Huan}, \citenamefont {Machielse}, \citenamefont {Knall}, \citenamefont {Suleymanzade}, \citenamefont {Pingault}, \citenamefont {Sutula}, \citenamefont {Ding}, \citenamefont {Knaut}, \citenamefont {Assumpcao}, \citenamefont {Wei}, \citenamefont {Bhaskar}, \citenamefont {Riedinger}, \citenamefont {Sukachev}, \citenamefont {Park}, \citenamefont {Lončar}, \citenamefont {Levonian},\ and\ \citenamefont {Lukin}}]{Stas2022}%
  \BibitemOpen
  \bibfield  {author} {\bibinfo {author} {\bibfnamefont {P.-J.}\ \bibnamefont {Stas}}, \bibinfo {author} {\bibfnamefont {Y.~Q.}\ \bibnamefont {Huan}}, \bibinfo {author} {\bibfnamefont {B.}~\bibnamefont {Machielse}}, \bibinfo {author} {\bibfnamefont {E.~N.}\ \bibnamefont {Knall}}, \bibinfo {author} {\bibfnamefont {A.}~\bibnamefont {Suleymanzade}}, \bibinfo {author} {\bibfnamefont {B.}~\bibnamefont {Pingault}}, \bibinfo {author} {\bibfnamefont {M.}~\bibnamefont {Sutula}}, \bibinfo {author} {\bibfnamefont {S.~W.}\ \bibnamefont {Ding}}, \bibinfo {author} {\bibfnamefont {C.~M.}\ \bibnamefont {Knaut}}, \bibinfo {author} {\bibfnamefont {D.~R.}\ \bibnamefont {Assumpcao}}, \bibinfo {author} {\bibfnamefont {Y.-C.}\ \bibnamefont {Wei}}, \bibinfo {author} {\bibfnamefont {M.~K.}\ \bibnamefont {Bhaskar}}, \bibinfo {author} {\bibfnamefont {R.}~\bibnamefont {Riedinger}}, \bibinfo {author} {\bibfnamefont {D.~D.}\ \bibnamefont {Sukachev}}, \bibinfo {author} {\bibfnamefont {H.}~\bibnamefont {Park}}, \bibinfo {author}
  {\bibfnamefont {M.}~\bibnamefont {Lončar}}, \bibinfo {author} {\bibfnamefont {D.~S.}\ \bibnamefont {Levonian}},\ and\ \bibinfo {author} {\bibfnamefont {M.~D.}\ \bibnamefont {Lukin}},\ }\bibfield  {title} {\bibinfo {title} {Robust multi-qubit quantum network node with integrated error detection},\ }\href {https://doi.org/10.1126/science.add9771} {\bibfield  {journal} {\bibinfo  {journal} {Science}\ }\textbf {\bibinfo {volume} {378}},\ \bibinfo {pages} {557} (\bibinfo {year} {2022})}\BibitemShut {NoStop}%
\bibitem [{\citenamefont {Knaut}\ \emph {et~al.}(2024)\citenamefont {Knaut}, \citenamefont {Suleymanzade}, \citenamefont {Wei}, \citenamefont {Assumpcao}, \citenamefont {Stas}, \citenamefont {Huan}, \citenamefont {Machielse}, \citenamefont {Knall}, \citenamefont {Sutula}, \citenamefont {Baranes}, \citenamefont {Sinclair}, \citenamefont {De-Eknamkul}, \citenamefont {Levonian}, \citenamefont {Bhaskar}, \citenamefont {Park}, \citenamefont {Lon{\v c}ar},\ and\ \citenamefont {Lukin}}]{Knaut2024}%
  \BibitemOpen
  \bibfield  {author} {\bibinfo {author} {\bibfnamefont {C.~M.}\ \bibnamefont {Knaut}}, \bibinfo {author} {\bibfnamefont {A.}~\bibnamefont {Suleymanzade}}, \bibinfo {author} {\bibfnamefont {Y.~C.}\ \bibnamefont {Wei}}, \bibinfo {author} {\bibfnamefont {D.~R.}\ \bibnamefont {Assumpcao}}, \bibinfo {author} {\bibfnamefont {P.~J.}\ \bibnamefont {Stas}}, \bibinfo {author} {\bibfnamefont {Y.~Q.}\ \bibnamefont {Huan}}, \bibinfo {author} {\bibfnamefont {B.}~\bibnamefont {Machielse}}, \bibinfo {author} {\bibfnamefont {E.~N.}\ \bibnamefont {Knall}}, \bibinfo {author} {\bibfnamefont {M.}~\bibnamefont {Sutula}}, \bibinfo {author} {\bibfnamefont {G.}~\bibnamefont {Baranes}}, \bibinfo {author} {\bibfnamefont {N.}~\bibnamefont {Sinclair}}, \bibinfo {author} {\bibfnamefont {C.}~\bibnamefont {De-Eknamkul}}, \bibinfo {author} {\bibfnamefont {D.~S.}\ \bibnamefont {Levonian}}, \bibinfo {author} {\bibfnamefont {M.~K.}\ \bibnamefont {Bhaskar}}, \bibinfo {author} {\bibfnamefont {H.}~\bibnamefont {Park}}, \bibinfo {author}
  {\bibfnamefont {M.}~\bibnamefont {Lon{\v c}ar}},\ and\ \bibinfo {author} {\bibfnamefont {M.~D.}\ \bibnamefont {Lukin}},\ }\bibfield  {title} {\bibinfo {title} {Entanglement of nanophotonic quantum memory nodes in a telecom network},\ }\href {https://doi.org/10.1038/s41586-024-07252-z} {\bibfield  {journal} {\bibinfo  {journal} {Nature}\ }\textbf {\bibinfo {volume} {629}},\ \bibinfo {pages} {573} (\bibinfo {year} {2024})}\BibitemShut {NoStop}%
\bibitem [{\citenamefont {Wei}\ \emph {et~al.}(2025)\citenamefont {Wei}, \citenamefont {Stas}, \citenamefont {Suleymanzade}, \citenamefont {Baranes}, \citenamefont {Machado}, \citenamefont {Huan}, \citenamefont {Knaut}, \citenamefont {Ding}, \citenamefont {Merz}, \citenamefont {Knall}, \citenamefont {Yazlar}, \citenamefont {Sirotin}, \citenamefont {Wang}, \citenamefont {Machielse}, \citenamefont {Yelin}, \citenamefont {Borregaard}, \citenamefont {Park}, \citenamefont {Lončar},\ and\ \citenamefont {Lukin}}]{Wei2025}%
  \BibitemOpen
  \bibfield  {author} {\bibinfo {author} {\bibfnamefont {Y.-C.}\ \bibnamefont {Wei}}, \bibinfo {author} {\bibfnamefont {P.-J.}\ \bibnamefont {Stas}}, \bibinfo {author} {\bibfnamefont {A.}~\bibnamefont {Suleymanzade}}, \bibinfo {author} {\bibfnamefont {G.}~\bibnamefont {Baranes}}, \bibinfo {author} {\bibfnamefont {F.}~\bibnamefont {Machado}}, \bibinfo {author} {\bibfnamefont {Y.~Q.}\ \bibnamefont {Huan}}, \bibinfo {author} {\bibfnamefont {C.~M.}\ \bibnamefont {Knaut}}, \bibinfo {author} {\bibfnamefont {S.~W.}\ \bibnamefont {Ding}}, \bibinfo {author} {\bibfnamefont {M.}~\bibnamefont {Merz}}, \bibinfo {author} {\bibfnamefont {E.~N.}\ \bibnamefont {Knall}}, \bibinfo {author} {\bibfnamefont {U.}~\bibnamefont {Yazlar}}, \bibinfo {author} {\bibfnamefont {M.}~\bibnamefont {Sirotin}}, \bibinfo {author} {\bibfnamefont {I.~W.}\ \bibnamefont {Wang}}, \bibinfo {author} {\bibfnamefont {B.}~\bibnamefont {Machielse}}, \bibinfo {author} {\bibfnamefont {S.~F.}\ \bibnamefont {Yelin}}, \bibinfo {author} {\bibfnamefont
  {J.}~\bibnamefont {Borregaard}}, \bibinfo {author} {\bibfnamefont {H.}~\bibnamefont {Park}}, \bibinfo {author} {\bibfnamefont {M.}~\bibnamefont {Lončar}},\ and\ \bibinfo {author} {\bibfnamefont {M.~D.}\ \bibnamefont {Lukin}},\ }\bibfield  {title} {\bibinfo {title} {Universal distributed blind quantum computing with solid-state qubits},\ }\href {https://doi.org/10.1126/science.adu6894} {\bibfield  {journal} {\bibinfo  {journal} {Science}\ }\textbf {\bibinfo {volume} {388}},\ \bibinfo {pages} {509} (\bibinfo {year} {2025})}\BibitemShut {NoStop}%
\bibitem [{\citenamefont {Niemietz}\ \emph {et~al.}(2021)\citenamefont {Niemietz}, \citenamefont {Farrera}, \citenamefont {Langenfeld},\ and\ \citenamefont {Rempe}}]{Niemietz2021}%
  \BibitemOpen
  \bibfield  {author} {\bibinfo {author} {\bibfnamefont {D.}~\bibnamefont {Niemietz}}, \bibinfo {author} {\bibfnamefont {P.}~\bibnamefont {Farrera}}, \bibinfo {author} {\bibfnamefont {S.}~\bibnamefont {Langenfeld}},\ and\ \bibinfo {author} {\bibfnamefont {G.}~\bibnamefont {Rempe}},\ }\bibfield  {title} {\bibinfo {title} {Nondestructive detection of photonic qubits},\ }\href {https://doi.org/10.1038/s41586-021-03290-z} {\bibfield  {journal} {\bibinfo  {journal} {Nature}\ }\textbf {\bibinfo {volume} {591}},\ \bibinfo {pages} {570} (\bibinfo {year} {2021})}\BibitemShut {NoStop}%
\bibitem [{\citenamefont {Wenner}\ \emph {et~al.}(2014)\citenamefont {Wenner}, \citenamefont {Yin}, \citenamefont {Chen}, \citenamefont {Barends}, \citenamefont {Chiaro}, \citenamefont {Jeffrey}, \citenamefont {Kelly}, \citenamefont {Megrant}, \citenamefont {Mutus}, \citenamefont {Neill}, \citenamefont {O'Malley}, \citenamefont {Roushan}, \citenamefont {Sank}, \citenamefont {Vainsencher}, \citenamefont {White}, \citenamefont {Korotkov}, \citenamefont {Cleland},\ and\ \citenamefont {Martinis}}]{wenner_catching_2014}%
  \BibitemOpen
  \bibfield  {author} {\bibinfo {author} {\bibfnamefont {J.}~\bibnamefont {Wenner}}, \bibinfo {author} {\bibfnamefont {Y.}~\bibnamefont {Yin}}, \bibinfo {author} {\bibfnamefont {Y.}~\bibnamefont {Chen}}, \bibinfo {author} {\bibfnamefont {R.}~\bibnamefont {Barends}}, \bibinfo {author} {\bibfnamefont {B.}~\bibnamefont {Chiaro}}, \bibinfo {author} {\bibfnamefont {E.}~\bibnamefont {Jeffrey}}, \bibinfo {author} {\bibfnamefont {J.}~\bibnamefont {Kelly}}, \bibinfo {author} {\bibfnamefont {A.}~\bibnamefont {Megrant}}, \bibinfo {author} {\bibfnamefont {J.~Y.}\ \bibnamefont {Mutus}}, \bibinfo {author} {\bibfnamefont {C.}~\bibnamefont {Neill}}, \bibinfo {author} {\bibfnamefont {P.~J.~J.}\ \bibnamefont {O'Malley}}, \bibinfo {author} {\bibfnamefont {P.}~\bibnamefont {Roushan}}, \bibinfo {author} {\bibfnamefont {D.}~\bibnamefont {Sank}}, \bibinfo {author} {\bibfnamefont {A.}~\bibnamefont {Vainsencher}}, \bibinfo {author} {\bibfnamefont {T.~C.}\ \bibnamefont {White}}, \bibinfo {author} {\bibfnamefont {A.~N.}\ \bibnamefont
  {Korotkov}}, \bibinfo {author} {\bibfnamefont {A.~N.}\ \bibnamefont {Cleland}},\ and\ \bibinfo {author} {\bibfnamefont {J.~M.}\ \bibnamefont {Martinis}},\ }\bibfield  {title} {\bibinfo {title} {Catching time-reversed microwave coherent state photons with 99.4\% absorption efficiency},\ }\href {https://doi.org/10.1103/PhysRevLett.112.210501} {\bibfield  {journal} {\bibinfo  {journal} {Phys. Rev. Lett.}\ }\textbf {\bibinfo {volume} {112}},\ \bibinfo {pages} {210501} (\bibinfo {year} {2014})}\BibitemShut {NoStop}%
\bibitem [{\citenamefont {Axline}\ \emph {et~al.}(2018)\citenamefont {Axline}, \citenamefont {Burkhart}, \citenamefont {Pfaff}, \citenamefont {Zhang}, \citenamefont {Chou}, \citenamefont {Campagne-Ibarcq}, \citenamefont {Reinhold}, \citenamefont {Frunzio}, \citenamefont {Girvin}, \citenamefont {Jiang}, \citenamefont {Devoret},\ and\ \citenamefont {Schoelkopf}}]{axline_-demand_2018}%
  \BibitemOpen
  \bibfield  {author} {\bibinfo {author} {\bibfnamefont {C.}~\bibnamefont {Axline}}, \bibinfo {author} {\bibfnamefont {L.}~\bibnamefont {Burkhart}}, \bibinfo {author} {\bibfnamefont {W.}~\bibnamefont {Pfaff}}, \bibinfo {author} {\bibfnamefont {M.}~\bibnamefont {Zhang}}, \bibinfo {author} {\bibfnamefont {K.}~\bibnamefont {Chou}}, \bibinfo {author} {\bibfnamefont {P.}~\bibnamefont {Campagne-Ibarcq}}, \bibinfo {author} {\bibfnamefont {P.}~\bibnamefont {Reinhold}}, \bibinfo {author} {\bibfnamefont {L.}~\bibnamefont {Frunzio}}, \bibinfo {author} {\bibfnamefont {S.~M.}\ \bibnamefont {Girvin}}, \bibinfo {author} {\bibfnamefont {L.}~\bibnamefont {Jiang}}, \bibinfo {author} {\bibfnamefont {M.~H.}\ \bibnamefont {Devoret}},\ and\ \bibinfo {author} {\bibfnamefont {R.~J.}\ \bibnamefont {Schoelkopf}},\ }\bibfield  {title} {\bibinfo {title} {On-demand quantum state transfer and entanglement between remote microwave cavity memories},\ }\href {https://doi.org/10.1038/s41567-018-0115-y} {\bibfield  {journal} {\bibinfo
  {journal} {Nature Physics}\ }\textbf {\bibinfo {volume} {14}},\ \bibinfo {pages} {705} (\bibinfo {year} {2018})}\BibitemShut {NoStop}%
\bibitem [{\citenamefont {Campagne-Ibarcq}\ \emph {et~al.}(2018)\citenamefont {Campagne-Ibarcq}, \citenamefont {Zalys-Geller}, \citenamefont {Narla}, \citenamefont {Shankar}, \citenamefont {Reinhold}, \citenamefont {Burkhart}, \citenamefont {Axline}, \citenamefont {Pfaff}, \citenamefont {Frunzio}, \citenamefont {Schoelkopf},\ and\ \citenamefont {Devoret}}]{campagne-ibarcq_deterministic_2018}%
  \BibitemOpen
  \bibfield  {author} {\bibinfo {author} {\bibfnamefont {P.}~\bibnamefont {Campagne-Ibarcq}}, \bibinfo {author} {\bibfnamefont {E.}~\bibnamefont {Zalys-Geller}}, \bibinfo {author} {\bibfnamefont {A.}~\bibnamefont {Narla}}, \bibinfo {author} {\bibfnamefont {S.}~\bibnamefont {Shankar}}, \bibinfo {author} {\bibfnamefont {P.}~\bibnamefont {Reinhold}}, \bibinfo {author} {\bibfnamefont {L.~D.}\ \bibnamefont {Burkhart}}, \bibinfo {author} {\bibfnamefont {C.~J.}\ \bibnamefont {Axline}}, \bibinfo {author} {\bibfnamefont {W.}~\bibnamefont {Pfaff}}, \bibinfo {author} {\bibfnamefont {L.}~\bibnamefont {Frunzio}}, \bibinfo {author} {\bibfnamefont {R.~J.}\ \bibnamefont {Schoelkopf}},\ and\ \bibinfo {author} {\bibfnamefont {M.~H.}\ \bibnamefont {Devoret}},\ }\bibfield  {title} {\bibinfo {title} {Deterministic remote entanglement of superconducting circuits through microwave two-photon transitions},\ }\href {https://doi.org/10.1103/PhysRevLett.120.200501} {\bibfield  {journal} {\bibinfo  {journal} {Physical Review Letters}\
  }\textbf {\bibinfo {volume} {120}},\ \bibinfo {pages} {200501} (\bibinfo {year} {2018})}\BibitemShut {NoStop}%
\bibitem [{\citenamefont {Kurpiers}\ \emph {et~al.}(2018)\citenamefont {Kurpiers}, \citenamefont {Magnard}, \citenamefont {Walter}, \citenamefont {Royer}, \citenamefont {Pechal}, \citenamefont {Heinsoo}, \citenamefont {Salathé}, \citenamefont {Akin}, \citenamefont {Storz}, \citenamefont {Besse}, \citenamefont {Gasparinetti}, \citenamefont {Blais},\ and\ \citenamefont {Wallraff}}]{kurpiers_deterministic_2018}%
  \BibitemOpen
  \bibfield  {author} {\bibinfo {author} {\bibfnamefont {P.}~\bibnamefont {Kurpiers}}, \bibinfo {author} {\bibfnamefont {P.}~\bibnamefont {Magnard}}, \bibinfo {author} {\bibfnamefont {T.}~\bibnamefont {Walter}}, \bibinfo {author} {\bibfnamefont {B.}~\bibnamefont {Royer}}, \bibinfo {author} {\bibfnamefont {M.}~\bibnamefont {Pechal}}, \bibinfo {author} {\bibfnamefont {J.}~\bibnamefont {Heinsoo}}, \bibinfo {author} {\bibfnamefont {Y.}~\bibnamefont {Salathé}}, \bibinfo {author} {\bibfnamefont {A.}~\bibnamefont {Akin}}, \bibinfo {author} {\bibfnamefont {S.}~\bibnamefont {Storz}}, \bibinfo {author} {\bibfnamefont {J.-C.}\ \bibnamefont {Besse}}, \bibinfo {author} {\bibfnamefont {S.}~\bibnamefont {Gasparinetti}}, \bibinfo {author} {\bibfnamefont {A.}~\bibnamefont {Blais}},\ and\ \bibinfo {author} {\bibfnamefont {A.}~\bibnamefont {Wallraff}},\ }\bibfield  {title} {\bibinfo {title} {Deterministic {Quantum} {State} {Transfer} and {Generation} of {Remote} {Entanglement} using {Microwave} {Photons}},\ }\href
  {https://doi.org/10.1038/s41586-018-0195-y} {\bibfield  {journal} {\bibinfo  {journal} {Nature}\ }\textbf {\bibinfo {volume} {558}},\ \bibinfo {pages} {264} (\bibinfo {year} {2018})}\BibitemShut {NoStop}%
\bibitem [{\citenamefont {Kurpiers}\ \emph {et~al.}(2019)\citenamefont {Kurpiers}, \citenamefont {Pechal}, \citenamefont {Royer}, \citenamefont {Magnard}, \citenamefont {Walter}, \citenamefont {Heinsoo}, \citenamefont {Salath\'e}, \citenamefont {Akin}, \citenamefont {Storz}, \citenamefont {Besse}, \citenamefont {Gasparinetti}, \citenamefont {Blais},\ and\ \citenamefont {Wallraff}}]{PhysRevApplied.12.044067}%
  \BibitemOpen
  \bibfield  {author} {\bibinfo {author} {\bibfnamefont {P.}~\bibnamefont {Kurpiers}}, \bibinfo {author} {\bibfnamefont {M.}~\bibnamefont {Pechal}}, \bibinfo {author} {\bibfnamefont {B.}~\bibnamefont {Royer}}, \bibinfo {author} {\bibfnamefont {P.}~\bibnamefont {Magnard}}, \bibinfo {author} {\bibfnamefont {T.}~\bibnamefont {Walter}}, \bibinfo {author} {\bibfnamefont {J.}~\bibnamefont {Heinsoo}}, \bibinfo {author} {\bibfnamefont {Y.}~\bibnamefont {Salath\'e}}, \bibinfo {author} {\bibfnamefont {A.}~\bibnamefont {Akin}}, \bibinfo {author} {\bibfnamefont {S.}~\bibnamefont {Storz}}, \bibinfo {author} {\bibfnamefont {J.-C.}\ \bibnamefont {Besse}}, \bibinfo {author} {\bibfnamefont {S.}~\bibnamefont {Gasparinetti}}, \bibinfo {author} {\bibfnamefont {A.}~\bibnamefont {Blais}},\ and\ \bibinfo {author} {\bibfnamefont {A.}~\bibnamefont {Wallraff}},\ }\bibfield  {title} {\bibinfo {title} {Quantum communication with time-bin encoded microwave photons},\ }\href {https://doi.org/10.1103/PhysRevApplied.12.044067} {\bibfield
  {journal} {\bibinfo  {journal} {Phys. Rev. Appl.}\ }\textbf {\bibinfo {volume} {12}},\ \bibinfo {pages} {044067} (\bibinfo {year} {2019})}\BibitemShut {NoStop}%
\bibitem [{\citenamefont {Jerger}\ \emph {et~al.}(2016)\citenamefont {Jerger}, \citenamefont {Macha}, \citenamefont {Hamann}, \citenamefont {Reshitnyk}, \citenamefont {Juliusson},\ and\ \citenamefont {Fedorov}}]{jerger_realization_2016}%
  \BibitemOpen
  \bibfield  {author} {\bibinfo {author} {\bibfnamefont {M.}~\bibnamefont {Jerger}}, \bibinfo {author} {\bibfnamefont {P.}~\bibnamefont {Macha}}, \bibinfo {author} {\bibfnamefont {A.~R.}\ \bibnamefont {Hamann}}, \bibinfo {author} {\bibfnamefont {Y.}~\bibnamefont {Reshitnyk}}, \bibinfo {author} {\bibfnamefont {K.}~\bibnamefont {Juliusson}},\ and\ \bibinfo {author} {\bibfnamefont {A.}~\bibnamefont {Fedorov}},\ }\bibfield  {title} {\bibinfo {title} {Realization of a {Binary}-{Outcome} {Projection} {Measurement} of a {Three}-{Level} {Superconducting} {Quantum} {System}},\ }\href {https://doi.org/10.1103/PhysRevApplied.6.014014} {\bibfield  {journal} {\bibinfo  {journal} {Physical Review Applied}\ }\textbf {\bibinfo {volume} {6}},\ \bibinfo {pages} {014014} (\bibinfo {year} {2016})}\BibitemShut {NoStop}%
\bibitem [{\citenamefont {Rondin}\ \emph {et~al.}(2014)\citenamefont {Rondin}, \citenamefont {Tetienne}, \citenamefont {Hingant}, \citenamefont {Roch}, \citenamefont {Maletinsky},\ and\ \citenamefont {Jacques}}]{rondin_magnetometry_2014}%
  \BibitemOpen
  \bibfield  {author} {\bibinfo {author} {\bibfnamefont {L.}~\bibnamefont {Rondin}}, \bibinfo {author} {\bibfnamefont {J.-P.}\ \bibnamefont {Tetienne}}, \bibinfo {author} {\bibfnamefont {T.}~\bibnamefont {Hingant}}, \bibinfo {author} {\bibfnamefont {J.-F.}\ \bibnamefont {Roch}}, \bibinfo {author} {\bibfnamefont {P.}~\bibnamefont {Maletinsky}},\ and\ \bibinfo {author} {\bibfnamefont {V.}~\bibnamefont {Jacques}},\ }\bibfield  {title} {\bibinfo {title} {Magnetometry with nitrogen-vacancy defects in diamond},\ }\href {https://doi.org/10.1088/0034-4885/77/5/056503} {\bibfield  {journal} {\bibinfo  {journal} {Reports on Progress in Physics}\ }\textbf {\bibinfo {volume} {77}},\ \bibinfo {pages} {056503} (\bibinfo {year} {2014})},\ \bibinfo {note} {publisher: IOP Publishing}\BibitemShut {NoStop}%
\bibitem [{\citenamefont {B{\"o}hi}\ \emph {et~al.}(2009)\citenamefont {B{\"o}hi}, \citenamefont {Riedel}, \citenamefont {Hoffrogge}, \citenamefont {Reichel}, \citenamefont {H{\"a}nsch},\ and\ \citenamefont {Treutlein}}]{Bohi2009}%
  \BibitemOpen
  \bibfield  {author} {\bibinfo {author} {\bibfnamefont {P.}~\bibnamefont {B{\"o}hi}}, \bibinfo {author} {\bibfnamefont {M.~F.}\ \bibnamefont {Riedel}}, \bibinfo {author} {\bibfnamefont {J.}~\bibnamefont {Hoffrogge}}, \bibinfo {author} {\bibfnamefont {J.}~\bibnamefont {Reichel}}, \bibinfo {author} {\bibfnamefont {T.~W.}\ \bibnamefont {H{\"a}nsch}},\ and\ \bibinfo {author} {\bibfnamefont {P.}~\bibnamefont {Treutlein}},\ }\bibfield  {title} {\bibinfo {title} {Coherent manipulation of bose--einstein condensates with state-dependent microwave potentials on an atom chip},\ }\href {https://doi.org/10.1038/nphys1329} {\bibfield  {journal} {\bibinfo  {journal} {Nature Physics}\ }\textbf {\bibinfo {volume} {5}},\ \bibinfo {pages} {592} (\bibinfo {year} {2009})}\BibitemShut {NoStop}%
\bibitem [{\citenamefont {Sukachev}\ \emph {et~al.}(2017)\citenamefont {Sukachev}, \citenamefont {Sipahigil}, \citenamefont {Nguyen}, \citenamefont {Bhaskar}, \citenamefont {Evans}, \citenamefont {Jelezko},\ and\ \citenamefont {Lukin}}]{sukachev_silicon-vacancy_2017}%
  \BibitemOpen
  \bibfield  {author} {\bibinfo {author} {\bibfnamefont {D.~D.}\ \bibnamefont {Sukachev}}, \bibinfo {author} {\bibfnamefont {A.}~\bibnamefont {Sipahigil}}, \bibinfo {author} {\bibfnamefont {C.~T.}\ \bibnamefont {Nguyen}}, \bibinfo {author} {\bibfnamefont {M.~K.}\ \bibnamefont {Bhaskar}}, \bibinfo {author} {\bibfnamefont {R.~E.}\ \bibnamefont {Evans}}, \bibinfo {author} {\bibfnamefont {F.}~\bibnamefont {Jelezko}},\ and\ \bibinfo {author} {\bibfnamefont {M.~D.}\ \bibnamefont {Lukin}},\ }\bibfield  {title} {\bibinfo {title} {Silicon-vacancy spin qubit in diamond: A quantum memory exceeding 10 ms with single-shot state readout},\ }\href {https://doi.org/10.1103/PhysRevLett.119.223602} {\bibfield  {journal} {\bibinfo  {journal} {Phys. Rev. Lett.}\ }\textbf {\bibinfo {volume} {119}},\ \bibinfo {pages} {223602} (\bibinfo {year} {2017})}\BibitemShut {NoStop}%
\bibitem [{\citenamefont {Pingault}\ \emph {et~al.}(2017)\citenamefont {Pingault}, \citenamefont {Jarausch}, \citenamefont {Hepp}, \citenamefont {Klintberg}, \citenamefont {Becker}, \citenamefont {Markham}, \citenamefont {Becher},\ and\ \citenamefont {Atatüre}}]{pingault_coherent_2017}%
  \BibitemOpen
  \bibfield  {author} {\bibinfo {author} {\bibfnamefont {B.}~\bibnamefont {Pingault}}, \bibinfo {author} {\bibfnamefont {D.-D.}\ \bibnamefont {Jarausch}}, \bibinfo {author} {\bibfnamefont {C.}~\bibnamefont {Hepp}}, \bibinfo {author} {\bibfnamefont {L.}~\bibnamefont {Klintberg}}, \bibinfo {author} {\bibfnamefont {J.~N.}\ \bibnamefont {Becker}}, \bibinfo {author} {\bibfnamefont {M.}~\bibnamefont {Markham}}, \bibinfo {author} {\bibfnamefont {C.}~\bibnamefont {Becher}},\ and\ \bibinfo {author} {\bibfnamefont {M.}~\bibnamefont {Atatüre}},\ }\bibfield  {title} {\bibinfo {title} {Coherent control of the silicon-vacancy spin in diamond},\ }\href {https://doi.org/10.1038/ncomms15579} {\bibfield  {journal} {\bibinfo  {journal} {Nature Communications}\ }\textbf {\bibinfo {volume} {8}},\ \bibinfo {pages} {15579} (\bibinfo {year} {2017})}\BibitemShut {NoStop}%
\bibitem [{\citenamefont {Fuchs}\ \emph {et~al.}(2009)\citenamefont {Fuchs}, \citenamefont {Dobrovitski}, \citenamefont {Toyli}, \citenamefont {Heremans},\ and\ \citenamefont {Awschalom}}]{fuchs_gigahertz_2009}%
  \BibitemOpen
  \bibfield  {author} {\bibinfo {author} {\bibfnamefont {G.~D.}\ \bibnamefont {Fuchs}}, \bibinfo {author} {\bibfnamefont {V.~V.}\ \bibnamefont {Dobrovitski}}, \bibinfo {author} {\bibfnamefont {D.~M.}\ \bibnamefont {Toyli}}, \bibinfo {author} {\bibfnamefont {F.~J.}\ \bibnamefont {Heremans}},\ and\ \bibinfo {author} {\bibfnamefont {D.~D.}\ \bibnamefont {Awschalom}},\ }\bibfield  {title} {\bibinfo {title} {Gigahertz {Dynamics} of a {Strongly} {Driven} {Single} {Quantum} {Spin}},\ }\href {https://doi.org/10.1126/science.1181193} {\bibfield  {journal} {\bibinfo  {journal} {Science}\ }\textbf {\bibinfo {volume} {326}},\ \bibinfo {pages} {1520} (\bibinfo {year} {2009})}\BibitemShut {NoStop}%
\bibitem [{\citenamefont {Dolde}\ \emph {et~al.}(2014)\citenamefont {Dolde}, \citenamefont {Bergholm}, \citenamefont {Wang}, \citenamefont {Jakobi}, \citenamefont {Naydenov}, \citenamefont {Pezzagna}, \citenamefont {Meijer}, \citenamefont {Jelezko}, \citenamefont {Neumann}, \citenamefont {Schulte-Herbrüggen}, \citenamefont {Biamonte},\ and\ \citenamefont {Wrachtrup}}]{dolde_high-fidelity_2014}%
  \BibitemOpen
  \bibfield  {author} {\bibinfo {author} {\bibfnamefont {F.}~\bibnamefont {Dolde}}, \bibinfo {author} {\bibfnamefont {V.}~\bibnamefont {Bergholm}}, \bibinfo {author} {\bibfnamefont {Y.}~\bibnamefont {Wang}}, \bibinfo {author} {\bibfnamefont {I.}~\bibnamefont {Jakobi}}, \bibinfo {author} {\bibfnamefont {B.}~\bibnamefont {Naydenov}}, \bibinfo {author} {\bibfnamefont {S.}~\bibnamefont {Pezzagna}}, \bibinfo {author} {\bibfnamefont {J.}~\bibnamefont {Meijer}}, \bibinfo {author} {\bibfnamefont {F.}~\bibnamefont {Jelezko}}, \bibinfo {author} {\bibfnamefont {P.}~\bibnamefont {Neumann}}, \bibinfo {author} {\bibfnamefont {T.}~\bibnamefont {Schulte-Herbrüggen}}, \bibinfo {author} {\bibfnamefont {J.}~\bibnamefont {Biamonte}},\ and\ \bibinfo {author} {\bibfnamefont {J.}~\bibnamefont {Wrachtrup}},\ }\bibfield  {title} {\bibinfo {title} {High-fidelity spin entanglement using optimal control},\ }\href {https://doi.org/10.1038/ncomms4371} {\bibfield  {journal} {\bibinfo  {journal} {Nature Communications}\ }\textbf {\bibinfo
  {volume} {5}},\ \bibinfo {pages} {3371} (\bibinfo {year} {2014})}\BibitemShut {NoStop}%
\bibitem [{\citenamefont {Karapatzakis}\ \emph {et~al.}(2024)\citenamefont {Karapatzakis}, \citenamefont {Resch}, \citenamefont {Schrodin}, \citenamefont {Fuchs}, \citenamefont {Kieschnick}, \citenamefont {Heupel}, \citenamefont {Kussi}, \citenamefont {Sürgers}, \citenamefont {Popov}, \citenamefont {Meijer}, \citenamefont {Becher}, \citenamefont {Wernsdorfer},\ and\ \citenamefont {Hunger}}]{karapatzakis_microwave_2024}%
  \BibitemOpen
  \bibfield  {author} {\bibinfo {author} {\bibfnamefont {I.}~\bibnamefont {Karapatzakis}}, \bibinfo {author} {\bibfnamefont {J.}~\bibnamefont {Resch}}, \bibinfo {author} {\bibfnamefont {M.}~\bibnamefont {Schrodin}}, \bibinfo {author} {\bibfnamefont {P.}~\bibnamefont {Fuchs}}, \bibinfo {author} {\bibfnamefont {M.}~\bibnamefont {Kieschnick}}, \bibinfo {author} {\bibfnamefont {J.}~\bibnamefont {Heupel}}, \bibinfo {author} {\bibfnamefont {L.}~\bibnamefont {Kussi}}, \bibinfo {author} {\bibfnamefont {C.}~\bibnamefont {Sürgers}}, \bibinfo {author} {\bibfnamefont {C.}~\bibnamefont {Popov}}, \bibinfo {author} {\bibfnamefont {J.}~\bibnamefont {Meijer}}, \bibinfo {author} {\bibfnamefont {C.}~\bibnamefont {Becher}}, \bibinfo {author} {\bibfnamefont {W.}~\bibnamefont {Wernsdorfer}},\ and\ \bibinfo {author} {\bibfnamefont {D.}~\bibnamefont {Hunger}},\ }\bibfield  {title} {\bibinfo {title} {Microwave {Control} of the {Tin}-{Vacancy} {Spin} {Qubit} in {Diamond} with a {Superconducting} {Waveguide}},\ }\href
  {https://doi.org/10.1103/PhysRevX.14.031036} {\bibfield  {journal} {\bibinfo  {journal} {Physical Review X}\ }\textbf {\bibinfo {volume} {14}},\ \bibinfo {pages} {031036} (\bibinfo {year} {2024})}\BibitemShut {NoStop}%
\bibitem [{\citenamefont {Doherty}\ \emph {et~al.}(2013)\citenamefont {Doherty}, \citenamefont {Manson}, \citenamefont {Delaney}, \citenamefont {Jelezko}, \citenamefont {Wrachtrup},\ and\ \citenamefont {Hollenberg}}]{doherty_nitrogen-vacancy_2013}%
  \BibitemOpen
  \bibfield  {author} {\bibinfo {author} {\bibfnamefont {M.~W.}\ \bibnamefont {Doherty}}, \bibinfo {author} {\bibfnamefont {N.~B.}\ \bibnamefont {Manson}}, \bibinfo {author} {\bibfnamefont {P.}~\bibnamefont {Delaney}}, \bibinfo {author} {\bibfnamefont {F.}~\bibnamefont {Jelezko}}, \bibinfo {author} {\bibfnamefont {J.}~\bibnamefont {Wrachtrup}},\ and\ \bibinfo {author} {\bibfnamefont {L.~C.}\ \bibnamefont {Hollenberg}},\ }\bibfield  {title} {\bibinfo {title} {The nitrogen-vacancy colour centre in diamond},\ }\href {https://doi.org/https://doi.org/10.1016/j.physrep.2013.02.001} {\bibfield  {journal} {\bibinfo  {journal} {Physics Reports}\ }\textbf {\bibinfo {volume} {528}},\ \bibinfo {pages} {1} (\bibinfo {year} {2013})},\ \bibinfo {note} {the nitrogen-vacancy colour centre in diamond}\BibitemShut {NoStop}%
\bibitem [{\citenamefont {Magnard}\ \emph {et~al.}(2018)\citenamefont {Magnard}, \citenamefont {Kurpiers}, \citenamefont {Royer}, \citenamefont {Walter}, \citenamefont {Besse}, \citenamefont {Gasparinetti}, \citenamefont {Pechal}, \citenamefont {Heinsoo}, \citenamefont {Storz}, \citenamefont {Blais},\ and\ \citenamefont {Wallraff}}]{magnard_fast_2018}%
  \BibitemOpen
  \bibfield  {author} {\bibinfo {author} {\bibfnamefont {P.}~\bibnamefont {Magnard}}, \bibinfo {author} {\bibfnamefont {P.}~\bibnamefont {Kurpiers}}, \bibinfo {author} {\bibfnamefont {B.}~\bibnamefont {Royer}}, \bibinfo {author} {\bibfnamefont {T.}~\bibnamefont {Walter}}, \bibinfo {author} {\bibfnamefont {J.-C.}\ \bibnamefont {Besse}}, \bibinfo {author} {\bibfnamefont {S.}~\bibnamefont {Gasparinetti}}, \bibinfo {author} {\bibfnamefont {M.}~\bibnamefont {Pechal}}, \bibinfo {author} {\bibfnamefont {J.}~\bibnamefont {Heinsoo}}, \bibinfo {author} {\bibfnamefont {S.}~\bibnamefont {Storz}}, \bibinfo {author} {\bibfnamefont {A.}~\bibnamefont {Blais}},\ and\ \bibinfo {author} {\bibfnamefont {A.}~\bibnamefont {Wallraff}},\ }\bibfield  {title} {\bibinfo {title} {Fast and unconditional all-microwave reset of a superconducting qubit},\ }\href {https://doi.org/10.1103/PhysRevLett.121.060502} {\bibfield  {journal} {\bibinfo  {journal} {Phys. Rev. Lett.}\ }\textbf {\bibinfo {volume} {121}},\ \bibinfo {pages} {060502}
  (\bibinfo {year} {2018})}\BibitemShut {NoStop}%
\bibitem [{\citenamefont {Egger}\ \emph {et~al.}(2018)\citenamefont {Egger}, \citenamefont {Werninghaus}, \citenamefont {Ganzhorn}, \citenamefont {Salis}, \citenamefont {Fuhrer}, \citenamefont {M\"uller},\ and\ \citenamefont {Filipp}}]{egger_pulsed_2018}%
  \BibitemOpen
  \bibfield  {author} {\bibinfo {author} {\bibfnamefont {D.}~\bibnamefont {Egger}}, \bibinfo {author} {\bibfnamefont {M.}~\bibnamefont {Werninghaus}}, \bibinfo {author} {\bibfnamefont {M.}~\bibnamefont {Ganzhorn}}, \bibinfo {author} {\bibfnamefont {G.}~\bibnamefont {Salis}}, \bibinfo {author} {\bibfnamefont {A.}~\bibnamefont {Fuhrer}}, \bibinfo {author} {\bibfnamefont {P.}~\bibnamefont {M\"uller}},\ and\ \bibinfo {author} {\bibfnamefont {S.}~\bibnamefont {Filipp}},\ }\bibfield  {title} {\bibinfo {title} {Pulsed reset protocol for fixed-frequency superconducting qubits},\ }\href {https://doi.org/10.1103/PhysRevApplied.10.044030} {\bibfield  {journal} {\bibinfo  {journal} {Phys. Rev. Appl.}\ }\textbf {\bibinfo {volume} {10}},\ \bibinfo {pages} {044030} (\bibinfo {year} {2018})}\BibitemShut {NoStop}%
\bibitem [{\citenamefont {Zhou}\ \emph {et~al.}(2021)\citenamefont {Zhou}, \citenamefont {Zhang}, \citenamefont {Yin}, \citenamefont {Huai}, \citenamefont {Gu}, \citenamefont {Xu}, \citenamefont {Allcock}, \citenamefont {Liu}, \citenamefont {Xi}, \citenamefont {Yu}, \citenamefont {Zhang}, \citenamefont {Zhang}, \citenamefont {Li}, \citenamefont {Song}, \citenamefont {Wang}, \citenamefont {Zheng}, \citenamefont {An}, \citenamefont {Zheng},\ and\ \citenamefont {Zhang}}]{zhou_rapid_2021}%
  \BibitemOpen
  \bibfield  {author} {\bibinfo {author} {\bibfnamefont {Y.}~\bibnamefont {Zhou}}, \bibinfo {author} {\bibfnamefont {Z.}~\bibnamefont {Zhang}}, \bibinfo {author} {\bibfnamefont {Z.}~\bibnamefont {Yin}}, \bibinfo {author} {\bibfnamefont {S.}~\bibnamefont {Huai}}, \bibinfo {author} {\bibfnamefont {X.}~\bibnamefont {Gu}}, \bibinfo {author} {\bibfnamefont {X.}~\bibnamefont {Xu}}, \bibinfo {author} {\bibfnamefont {J.}~\bibnamefont {Allcock}}, \bibinfo {author} {\bibfnamefont {F.}~\bibnamefont {Liu}}, \bibinfo {author} {\bibfnamefont {G.}~\bibnamefont {Xi}}, \bibinfo {author} {\bibfnamefont {Q.}~\bibnamefont {Yu}}, \bibinfo {author} {\bibfnamefont {H.}~\bibnamefont {Zhang}}, \bibinfo {author} {\bibfnamefont {M.}~\bibnamefont {Zhang}}, \bibinfo {author} {\bibfnamefont {H.}~\bibnamefont {Li}}, \bibinfo {author} {\bibfnamefont {X.}~\bibnamefont {Song}}, \bibinfo {author} {\bibfnamefont {Z.}~\bibnamefont {Wang}}, \bibinfo {author} {\bibfnamefont {D.}~\bibnamefont {Zheng}}, \bibinfo {author} {\bibfnamefont
  {S.}~\bibnamefont {An}}, \bibinfo {author} {\bibfnamefont {Y.}~\bibnamefont {Zheng}},\ and\ \bibinfo {author} {\bibfnamefont {S.}~\bibnamefont {Zhang}},\ }\bibfield  {title} {\bibinfo {title} {Rapid and unconditional parametric reset protocol for tunable superconducting qubits},\ }\href {https://doi.org/10.1038/s41467-021-26205-y} {\bibfield  {journal} {\bibinfo  {journal} {Nature Communications}\ }\textbf {\bibinfo {volume} {12}},\ \bibinfo {pages} {5924} (\bibinfo {year} {2021})}\BibitemShut {NoStop}%
\bibitem [{\citenamefont {Sillanpää}\ \emph {et~al.}(2007)\citenamefont {Sillanpää}, \citenamefont {Park},\ and\ \citenamefont {Simmonds}}]{sillanpaa_coherent_2007}%
  \BibitemOpen
  \bibfield  {author} {\bibinfo {author} {\bibfnamefont {M.~A.}\ \bibnamefont {Sillanpää}}, \bibinfo {author} {\bibfnamefont {J.~I.}\ \bibnamefont {Park}},\ and\ \bibinfo {author} {\bibfnamefont {R.~W.}\ \bibnamefont {Simmonds}},\ }\bibfield  {title} {\bibinfo {title} {Coherent quantum state storage and transfer between two phase qubits via a resonant cavity},\ }\href {https://doi.org/10.1038/nature06124} {\bibfield  {journal} {\bibinfo  {journal} {Nature}\ }\textbf {\bibinfo {volume} {449}},\ \bibinfo {pages} {438} (\bibinfo {year} {2007})}\BibitemShut {NoStop}%
\bibitem [{\citenamefont {Hofheinz}\ \emph {et~al.}(2008)\citenamefont {Hofheinz}, \citenamefont {Weig}, \citenamefont {Ansmann}, \citenamefont {Bialczak}, \citenamefont {Lucero}, \citenamefont {Neeley}, \citenamefont {O’Connell}, \citenamefont {Wang}, \citenamefont {Martinis},\ and\ \citenamefont {Cleland}}]{hofheinz_generation_2008}%
  \BibitemOpen
  \bibfield  {author} {\bibinfo {author} {\bibfnamefont {M.}~\bibnamefont {Hofheinz}}, \bibinfo {author} {\bibfnamefont {E.~M.}\ \bibnamefont {Weig}}, \bibinfo {author} {\bibfnamefont {M.}~\bibnamefont {Ansmann}}, \bibinfo {author} {\bibfnamefont {R.~C.}\ \bibnamefont {Bialczak}}, \bibinfo {author} {\bibfnamefont {E.}~\bibnamefont {Lucero}}, \bibinfo {author} {\bibfnamefont {M.}~\bibnamefont {Neeley}}, \bibinfo {author} {\bibfnamefont {A.~D.}\ \bibnamefont {O’Connell}}, \bibinfo {author} {\bibfnamefont {H.}~\bibnamefont {Wang}}, \bibinfo {author} {\bibfnamefont {J.~M.}\ \bibnamefont {Martinis}},\ and\ \bibinfo {author} {\bibfnamefont {A.~N.}\ \bibnamefont {Cleland}},\ }\bibfield  {title} {\bibinfo {title} {Generation of {Fock} states in a superconducting quantum circuit},\ }\href {https://doi.org/10.1038/nature07136} {\bibfield  {journal} {\bibinfo  {journal} {Nature}\ }\textbf {\bibinfo {volume} {454}},\ \bibinfo {pages} {310} (\bibinfo {year} {2008})}\BibitemShut {NoStop}%
\bibitem [{\citenamefont {Zeytinoğlu}\ \emph {et~al.}(2015)\citenamefont {Zeytinoğlu}, \citenamefont {Pechal}, \citenamefont {Berger}, \citenamefont {Abdumalikov}, \citenamefont {Wallraff},\ and\ \citenamefont {Filipp}}]{zeytinoglu_microwave-induced_2015}%
  \BibitemOpen
  \bibfield  {author} {\bibinfo {author} {\bibfnamefont {S.}~\bibnamefont {Zeytinoğlu}}, \bibinfo {author} {\bibfnamefont {M.}~\bibnamefont {Pechal}}, \bibinfo {author} {\bibfnamefont {S.}~\bibnamefont {Berger}}, \bibinfo {author} {\bibfnamefont {A.~A.}\ \bibnamefont {Abdumalikov}}, \bibinfo {author} {\bibfnamefont {A.}~\bibnamefont {Wallraff}},\ and\ \bibinfo {author} {\bibfnamefont {S.}~\bibnamefont {Filipp}},\ }\bibfield  {title} {\bibinfo {title} {Microwave-induced amplitude- and phase-tunable qubit-resonator coupling in circuit quantum electrodynamics},\ }\href {https://doi.org/10.1103/PhysRevA.91.043846} {\bibfield  {journal} {\bibinfo  {journal} {Physical Review A}\ }\textbf {\bibinfo {volume} {91}},\ \bibinfo {pages} {043846} (\bibinfo {year} {2015})}\BibitemShut {NoStop}%
\bibitem [{\citenamefont {Tavis}\ and\ \citenamefont {Cummings}(1968)}]{Tavis1968}%
  \BibitemOpen
  \bibfield  {author} {\bibinfo {author} {\bibfnamefont {M.}~\bibnamefont {Tavis}}\ and\ \bibinfo {author} {\bibfnamefont {F.~W.}\ \bibnamefont {Cummings}},\ }\bibfield  {title} {\bibinfo {title} {Exact solution for an $n$-molecule---radiation-field hamiltonian},\ }\href {https://doi.org/10.1103/PhysRev.170.379} {\bibfield  {journal} {\bibinfo  {journal} {Phys. Rev.}\ }\textbf {\bibinfo {volume} {170}},\ \bibinfo {pages} {379} (\bibinfo {year} {1968})}\BibitemShut {NoStop}%
\bibitem [{\citenamefont {Plankensteiner}\ \emph {et~al.}(2019)\citenamefont {Plankensteiner}, \citenamefont {Sommer}, \citenamefont {Reitz}, \citenamefont {Ritsch},\ and\ \citenamefont {Genes}}]{Plankensteiner2019}%
  \BibitemOpen
  \bibfield  {author} {\bibinfo {author} {\bibfnamefont {D.}~\bibnamefont {Plankensteiner}}, \bibinfo {author} {\bibfnamefont {C.}~\bibnamefont {Sommer}}, \bibinfo {author} {\bibfnamefont {M.}~\bibnamefont {Reitz}}, \bibinfo {author} {\bibfnamefont {H.}~\bibnamefont {Ritsch}},\ and\ \bibinfo {author} {\bibfnamefont {C.}~\bibnamefont {Genes}},\ }\bibfield  {title} {\bibinfo {title} {Enhanced collective purcell effect of coupled quantum emitter systems},\ }\href {https://doi.org/10.1103/PhysRevA.99.043843} {\bibfield  {journal} {\bibinfo  {journal} {Phys. Rev. A}\ }\textbf {\bibinfo {volume} {99}},\ \bibinfo {pages} {043843} (\bibinfo {year} {2019})}\BibitemShut {NoStop}%
\bibitem [{Note1()}]{Note1}%
  \BibitemOpen
  \bibinfo {note} {Rydberg blockade can be used in a neutral atom ensemble for the preparation of the single collective excitation \cite {Mei2009}. Since Rydberg blockade relies on optical driving, the single-excitation preparation step should be temporally separated from the entangling step to avoid pump-induced heating. After this preparation step, no optical drive is required, and the entangling process remains pump-free.}\BibitemShut {Stop}%
\bibitem [{Note2()}]{Note2}%
  \BibitemOpen
  \bibinfo {note} {Preparation of the single collective excitation can be heralded by recapturing the emitted microwave photon with the transmon. We apply $\pi _{g_a0}$ before the initially prepared excitation decays and apply the same pulse again after the decay, leaving the ensemble in the state $\ket {W_0}$ while emitting the microwave photon used for heralding.}\BibitemShut {Stop}%
\bibitem [{\citenamefont {Wesenberg}\ \emph {et~al.}(2011)\citenamefont {Wesenberg}, \citenamefont {Kurucz},\ and\ \citenamefont {M\o{}lmer}}]{Wesenberg2011}%
  \BibitemOpen
  \bibfield  {author} {\bibinfo {author} {\bibfnamefont {J.~H.}\ \bibnamefont {Wesenberg}}, \bibinfo {author} {\bibfnamefont {Z.}~\bibnamefont {Kurucz}},\ and\ \bibinfo {author} {\bibfnamefont {K.}~\bibnamefont {M\o{}lmer}},\ }\bibfield  {title} {\bibinfo {title} {Dynamics of the collective modes of an inhomogeneous spin ensemble in a cavity},\ }\href {https://doi.org/10.1103/PhysRevA.83.023826} {\bibfield  {journal} {\bibinfo  {journal} {Phys. Rev. A}\ }\textbf {\bibinfo {volume} {83}},\ \bibinfo {pages} {023826} (\bibinfo {year} {2011})}\BibitemShut {NoStop}%
\bibitem [{\citenamefont {Kurucz}\ \emph {et~al.}(2011)\citenamefont {Kurucz}, \citenamefont {Wesenberg},\ and\ \citenamefont {M\o{}lmer}}]{Kurucz2011}%
  \BibitemOpen
  \bibfield  {author} {\bibinfo {author} {\bibfnamefont {Z.}~\bibnamefont {Kurucz}}, \bibinfo {author} {\bibfnamefont {J.~H.}\ \bibnamefont {Wesenberg}},\ and\ \bibinfo {author} {\bibfnamefont {K.}~\bibnamefont {M\o{}lmer}},\ }\bibfield  {title} {\bibinfo {title} {Spectroscopic properties of inhomogeneously broadened spin ensembles in a cavity},\ }\href {https://doi.org/10.1103/PhysRevA.83.053852} {\bibfield  {journal} {\bibinfo  {journal} {Phys. Rev. A}\ }\textbf {\bibinfo {volume} {83}},\ \bibinfo {pages} {053852} (\bibinfo {year} {2011})}\BibitemShut {NoStop}%
\bibitem [{\citenamefont {Putz}\ \emph {et~al.}(2014)\citenamefont {Putz}, \citenamefont {Krimer}, \citenamefont {Ams{\"u}ss}, \citenamefont {Valookaran}, \citenamefont {N{\"o}bauer}, \citenamefont {Schmiedmayer}, \citenamefont {Rotter},\ and\ \citenamefont {Majer}}]{Putz2014}%
  \BibitemOpen
  \bibfield  {author} {\bibinfo {author} {\bibfnamefont {S.}~\bibnamefont {Putz}}, \bibinfo {author} {\bibfnamefont {D.~O.}\ \bibnamefont {Krimer}}, \bibinfo {author} {\bibfnamefont {R.}~\bibnamefont {Ams{\"u}ss}}, \bibinfo {author} {\bibfnamefont {A.}~\bibnamefont {Valookaran}}, \bibinfo {author} {\bibfnamefont {T.}~\bibnamefont {N{\"o}bauer}}, \bibinfo {author} {\bibfnamefont {J.}~\bibnamefont {Schmiedmayer}}, \bibinfo {author} {\bibfnamefont {S.}~\bibnamefont {Rotter}},\ and\ \bibinfo {author} {\bibfnamefont {J.}~\bibnamefont {Majer}},\ }\bibfield  {title} {\bibinfo {title} {Protecting a spin ensemble against decoherence in the strong-coupling regime of cavity qed},\ }\href {https://doi.org/10.1038/nphys3050} {\bibfield  {journal} {\bibinfo  {journal} {Nature Physics}\ }\textbf {\bibinfo {volume} {10}},\ \bibinfo {pages} {720} (\bibinfo {year} {2014})}\BibitemShut {NoStop}%
\bibitem [{\citenamefont {Julsgaard}\ \emph {et~al.}(2013)\citenamefont {Julsgaard}, \citenamefont {Grezes}, \citenamefont {Bertet},\ and\ \citenamefont {M\o{}lmer}}]{Julsgaard2013}%
  \BibitemOpen
  \bibfield  {author} {\bibinfo {author} {\bibfnamefont {B.}~\bibnamefont {Julsgaard}}, \bibinfo {author} {\bibfnamefont {C.}~\bibnamefont {Grezes}}, \bibinfo {author} {\bibfnamefont {P.}~\bibnamefont {Bertet}},\ and\ \bibinfo {author} {\bibfnamefont {K.}~\bibnamefont {M\o{}lmer}},\ }\bibfield  {title} {\bibinfo {title} {Quantum memory for microwave photons in an inhomogeneously broadened spin ensemble},\ }\href {https://doi.org/10.1103/PhysRevLett.110.250503} {\bibfield  {journal} {\bibinfo  {journal} {Phys. Rev. Lett.}\ }\textbf {\bibinfo {volume} {110}},\ \bibinfo {pages} {250503} (\bibinfo {year} {2013})}\BibitemShut {NoStop}%
\bibitem [{\citenamefont {Kumar}\ \emph {et~al.}(2023)\citenamefont {Kumar}, \citenamefont {Suleymanzade}, \citenamefont {Stone}, \citenamefont {Taneja}, \citenamefont {Anferov}, \citenamefont {Schuster},\ and\ \citenamefont {Simon}}]{kumar_quantum-enabled_2023}%
  \BibitemOpen
  \bibfield  {author} {\bibinfo {author} {\bibfnamefont {A.}~\bibnamefont {Kumar}}, \bibinfo {author} {\bibfnamefont {A.}~\bibnamefont {Suleymanzade}}, \bibinfo {author} {\bibfnamefont {M.}~\bibnamefont {Stone}}, \bibinfo {author} {\bibfnamefont {L.}~\bibnamefont {Taneja}}, \bibinfo {author} {\bibfnamefont {A.}~\bibnamefont {Anferov}}, \bibinfo {author} {\bibfnamefont {D.~I.}\ \bibnamefont {Schuster}},\ and\ \bibinfo {author} {\bibfnamefont {J.}~\bibnamefont {Simon}},\ }\bibfield  {title} {\bibinfo {title} {Quantum-enabled millimetre wave to optical transduction using neutral atoms},\ }\href {https://doi.org/10.1038/s41586-023-05740-2} {\bibfield  {journal} {\bibinfo  {journal} {Nature}\ }\textbf {\bibinfo {volume} {615}},\ \bibinfo {pages} {614} (\bibinfo {year} {2023})}\BibitemShut {NoStop}%
\bibitem [{Note3()}]{Note3}%
  \BibitemOpen
  \bibinfo {note} {It is worth mentioning that the strong spin-phonon coupling ($g_{sp}/\kappa _m>1$) also makes the platform suitable for the coherent swap scheme. A simple estimation, assuming sequential swaps between the spin-phonon and the phonon-microwave, gives the swapping time $t_{\protect \text {swap}}=\pi /(2g_{sp})+\pi /(2g_{pe})$. Using the same set of parameter values as in the above paragraph, except for a reduced microwave loss rate of $0.1$ MHz, the swap time takes roughly $1~\mu $s, and the microwave retrieval is realized with fidelity $0.7$. (see Supplemental Material). Additional schemes that maps the spin state to a bosonic mode state has been considered, for example, in \cite {PhysRevLett.105.220501}.}\BibitemShut {Stop}%
\bibitem [{\citenamefont {Mei}\ \emph {et~al.}(2009)\citenamefont {Mei}, \citenamefont {Feng}, \citenamefont {Yu},\ and\ \citenamefont {Zhang}}]{Mei2009}%
  \BibitemOpen
  \bibfield  {author} {\bibinfo {author} {\bibfnamefont {F.}~\bibnamefont {Mei}}, \bibinfo {author} {\bibfnamefont {M.}~\bibnamefont {Feng}}, \bibinfo {author} {\bibfnamefont {Y.-F.}\ \bibnamefont {Yu}},\ and\ \bibinfo {author} {\bibfnamefont {Z.-M.}\ \bibnamefont {Zhang}},\ }\bibfield  {title} {\bibinfo {title} {Scalable quantum information processing with atomic ensembles and flying photons},\ }\href {https://doi.org/10.1103/PhysRevA.80.042319} {\bibfield  {journal} {\bibinfo  {journal} {Phys. Rev. A}\ }\textbf {\bibinfo {volume} {80}},\ \bibinfo {pages} {042319} (\bibinfo {year} {2009})}\BibitemShut {NoStop}%
\bibitem [{\citenamefont {Stannigel}\ \emph {et~al.}(2010)\citenamefont {Stannigel}, \citenamefont {Rabl}, \citenamefont {S\o{}rensen}, \citenamefont {Zoller},\ and\ \citenamefont {Lukin}}]{PhysRevLett.105.220501}%
  \BibitemOpen
  \bibfield  {author} {\bibinfo {author} {\bibfnamefont {K.}~\bibnamefont {Stannigel}}, \bibinfo {author} {\bibfnamefont {P.}~\bibnamefont {Rabl}}, \bibinfo {author} {\bibfnamefont {A.~S.}\ \bibnamefont {S\o{}rensen}}, \bibinfo {author} {\bibfnamefont {P.}~\bibnamefont {Zoller}},\ and\ \bibinfo {author} {\bibfnamefont {M.~D.}\ \bibnamefont {Lukin}},\ }\bibfield  {title} {\bibinfo {title} {Optomechanical transducers for long-distance quantum communication},\ }\href {https://doi.org/10.1103/PhysRevLett.105.220501} {\bibfield  {journal} {\bibinfo  {journal} {Phys. Rev. Lett.}\ }\textbf {\bibinfo {volume} {105}},\ \bibinfo {pages} {220501} (\bibinfo {year} {2010})}\BibitemShut {NoStop}%
\bibitem [{\citenamefont {Reiserer}\ and\ \citenamefont {Rempe}(2015)}]{cavityQED}%
  \BibitemOpen
  \bibfield  {author} {\bibinfo {author} {\bibfnamefont {A.}~\bibnamefont {Reiserer}}\ and\ \bibinfo {author} {\bibfnamefont {G.}~\bibnamefont {Rempe}},\ }\bibfield  {title} {\bibinfo {title} {Cavity-based quantum networks with single atoms and optical photons},\ }\href {https://doi.org/10.1103/RevModPhys.87.1379} {\bibfield  {journal} {\bibinfo  {journal} {Rev. Mod. Phys.}\ }\textbf {\bibinfo {volume} {87}},\ \bibinfo {pages} {1379} (\bibinfo {year} {2015})}\BibitemShut {NoStop}%
\bibitem [{\citenamefont {Alkauskas}\ \emph {et~al.}(2014)\citenamefont {Alkauskas}, \citenamefont {Buckley}, \citenamefont {Awschalom},\ and\ \citenamefont {Van~de Walle}}]{Alkauskas2014}%
  \BibitemOpen
  \bibfield  {author} {\bibinfo {author} {\bibfnamefont {A.}~\bibnamefont {Alkauskas}}, \bibinfo {author} {\bibfnamefont {B.~B.}\ \bibnamefont {Buckley}}, \bibinfo {author} {\bibfnamefont {D.~D.}\ \bibnamefont {Awschalom}},\ and\ \bibinfo {author} {\bibfnamefont {C.~G.}\ \bibnamefont {Van~de Walle}},\ }\bibfield  {title} {\bibinfo {title} {First-principles theory of the luminescence lineshape for the triplet transition in diamond nv centres},\ }\href {https://doi.org/10.1088/1367-2630/16/7/073026} {\bibfield  {journal} {\bibinfo  {journal} {New Journal of Physics}\ }\textbf {\bibinfo {volume} {16}},\ \bibinfo {pages} {073026} (\bibinfo {year} {2014})}\BibitemShut {NoStop}%
\bibitem [{\citenamefont {Haikka}\ \emph {et~al.}(2017)\citenamefont {Haikka}, \citenamefont {Kubo}, \citenamefont {Bienfait}, \citenamefont {Bertet},\ and\ \citenamefont {M\o{}lmer}}]{Haikka2017}%
  \BibitemOpen
  \bibfield  {author} {\bibinfo {author} {\bibfnamefont {P.}~\bibnamefont {Haikka}}, \bibinfo {author} {\bibfnamefont {Y.}~\bibnamefont {Kubo}}, \bibinfo {author} {\bibfnamefont {A.}~\bibnamefont {Bienfait}}, \bibinfo {author} {\bibfnamefont {P.}~\bibnamefont {Bertet}},\ and\ \bibinfo {author} {\bibfnamefont {K.}~\bibnamefont {M\o{}lmer}},\ }\bibfield  {title} {\bibinfo {title} {Proposal for detecting a single electron spin in a microwave resonator},\ }\href {https://doi.org/10.1103/PhysRevA.95.022306} {\bibfield  {journal} {\bibinfo  {journal} {Phys. Rev. A}\ }\textbf {\bibinfo {volume} {95}},\ \bibinfo {pages} {022306} (\bibinfo {year} {2017})}\BibitemShut {NoStop}%
\bibitem [{\citenamefont {Ding}\ \emph {et~al.}(2024)\citenamefont {Ding}, \citenamefont {Haas}, \citenamefont {Guo}, \citenamefont {Kuruma}, \citenamefont {Jin}, \citenamefont {Li}, \citenamefont {Awschalom}, \citenamefont {Delegan}, \citenamefont {Heremans}, \citenamefont {High},\ and\ \citenamefont {Loncar}}]{ding2024high}%
  \BibitemOpen
  \bibfield  {author} {\bibinfo {author} {\bibfnamefont {S.~W.}\ \bibnamefont {Ding}}, \bibinfo {author} {\bibfnamefont {M.}~\bibnamefont {Haas}}, \bibinfo {author} {\bibfnamefont {X.}~\bibnamefont {Guo}}, \bibinfo {author} {\bibfnamefont {K.}~\bibnamefont {Kuruma}}, \bibinfo {author} {\bibfnamefont {C.}~\bibnamefont {Jin}}, \bibinfo {author} {\bibfnamefont {Z.}~\bibnamefont {Li}}, \bibinfo {author} {\bibfnamefont {D.~D.}\ \bibnamefont {Awschalom}}, \bibinfo {author} {\bibfnamefont {N.}~\bibnamefont {Delegan}}, \bibinfo {author} {\bibfnamefont {F.~J.}\ \bibnamefont {Heremans}}, \bibinfo {author} {\bibfnamefont {A.~A.}\ \bibnamefont {High}},\ and\ \bibinfo {author} {\bibfnamefont {M.}~\bibnamefont {Loncar}},\ }\bibfield  {title} {\bibinfo {title} {High-q cavity interface for color centers in thin film diamond},\ }\href {https://doi.org/10.1038/s41467-024-50667-5} {\bibfield  {journal} {\bibinfo  {journal} {Nature Communications}\ }\textbf {\bibinfo {volume} {15}},\ \bibinfo {pages} {6358} (\bibinfo {year}
  {2024})}\BibitemShut {NoStop}%
\bibitem [{\citenamefont {Myers}(2016)}]{Myers2014}%
  \BibitemOpen
  \bibfield  {author} {\bibinfo {author} {\bibfnamefont {B.~A.}\ \bibnamefont {Myers}},\ }\emph {\bibinfo {title} {Quantum decoherence of near-surface nitrogen-vacancy centers in diamond and implications for nanoscale imaging}},\ \href@noop {} {Ph.D. thesis},\ \bibinfo  {school} {UC Santa Barbara} (\bibinfo {year} {2016})\BibitemShut {NoStop}%
\bibitem [{\citenamefont {Tiecke}\ \emph {et~al.}(2014)\citenamefont {Tiecke}, \citenamefont {Thompson}, \citenamefont {de~Leon}, \citenamefont {Liu}, \citenamefont {Vuleti{\'{c}}},\ and\ \citenamefont {Lukin}}]{Tiecke2014}%
  \BibitemOpen
  \bibfield  {author} {\bibinfo {author} {\bibfnamefont {T.~G.}\ \bibnamefont {Tiecke}}, \bibinfo {author} {\bibfnamefont {J.~D.}\ \bibnamefont {Thompson}}, \bibinfo {author} {\bibfnamefont {N.~P.}\ \bibnamefont {de~Leon}}, \bibinfo {author} {\bibfnamefont {L.~R.}\ \bibnamefont {Liu}}, \bibinfo {author} {\bibfnamefont {V.}~\bibnamefont {Vuleti{\'{c}}}},\ and\ \bibinfo {author} {\bibfnamefont {M.~D.}\ \bibnamefont {Lukin}},\ }\bibfield  {title} {\bibinfo {title} {Nanophotonic quantum phase switch with a single atom},\ }\href {https://doi.org/10.1038/nature13188} {\bibfield  {journal} {\bibinfo  {journal} {Nature}\ }\textbf {\bibinfo {volume} {508}},\ \bibinfo {pages} {241} (\bibinfo {year} {2014})}\BibitemShut {NoStop}%
\bibitem [{\citenamefont {Ðorđević}\ \emph {et~al.}(2021)\citenamefont {Ðorđević}, \citenamefont {Samutpraphoot}, \citenamefont {Ocola}, \citenamefont {Bernien}, \citenamefont {Grinkemeyer}, \citenamefont {Dimitrova}, \citenamefont {Vuletić},\ and\ \citenamefont {Lukin}}]{Dordevic2021}%
  \BibitemOpen
  \bibfield  {author} {\bibinfo {author} {\bibfnamefont {T.}~\bibnamefont {Ðorđević}}, \bibinfo {author} {\bibfnamefont {P.}~\bibnamefont {Samutpraphoot}}, \bibinfo {author} {\bibfnamefont {P.~L.}\ \bibnamefont {Ocola}}, \bibinfo {author} {\bibfnamefont {H.}~\bibnamefont {Bernien}}, \bibinfo {author} {\bibfnamefont {B.}~\bibnamefont {Grinkemeyer}}, \bibinfo {author} {\bibfnamefont {I.}~\bibnamefont {Dimitrova}}, \bibinfo {author} {\bibfnamefont {V.}~\bibnamefont {Vuletić}},\ and\ \bibinfo {author} {\bibfnamefont {M.~D.}\ \bibnamefont {Lukin}},\ }\bibfield  {title} {\bibinfo {title} {Entanglement transport and a nanophotonic interface for atoms in optical tweezers},\ }\href {https://doi.org/10.1126/science.abi9917} {\bibfield  {journal} {\bibinfo  {journal} {Science}\ }\textbf {\bibinfo {volume} {373}},\ \bibinfo {pages} {1511} (\bibinfo {year} {2021})}\BibitemShut {NoStop}%
\bibitem [{\citenamefont {Beck}\ \emph {et~al.}(2016)\citenamefont {Beck}, \citenamefont {Isaacs}, \citenamefont {Booth}, \citenamefont {Pritchard}, \citenamefont {Saffman},\ and\ \citenamefont {McDermott}}]{Beck2016}%
  \BibitemOpen
  \bibfield  {author} {\bibinfo {author} {\bibfnamefont {M.~A.}\ \bibnamefont {Beck}}, \bibinfo {author} {\bibfnamefont {J.~A.}\ \bibnamefont {Isaacs}}, \bibinfo {author} {\bibfnamefont {D.}~\bibnamefont {Booth}}, \bibinfo {author} {\bibfnamefont {J.~D.}\ \bibnamefont {Pritchard}}, \bibinfo {author} {\bibfnamefont {M.}~\bibnamefont {Saffman}},\ and\ \bibinfo {author} {\bibfnamefont {R.}~\bibnamefont {McDermott}},\ }\bibfield  {title} {\bibinfo {title} {Optimized coplanar waveguide resonators for a superconductor–atom interface},\ }\href {https://doi.org/10.1063/1.4962172} {\bibfield  {journal} {\bibinfo  {journal} {Applied Physics Letters}\ }\textbf {\bibinfo {volume} {109}},\ \bibinfo {pages} {092602} (\bibinfo {year} {2016})}\BibitemShut {NoStop}%
\bibitem [{\citenamefont {Tuchendler}\ \emph {et~al.}(2008)\citenamefont {Tuchendler}, \citenamefont {Lance}, \citenamefont {Browaeys}, \citenamefont {Sortais},\ and\ \citenamefont {Grangier}}]{Tuchendler2008}%
  \BibitemOpen
  \bibfield  {author} {\bibinfo {author} {\bibfnamefont {C.}~\bibnamefont {Tuchendler}}, \bibinfo {author} {\bibfnamefont {A.~M.}\ \bibnamefont {Lance}}, \bibinfo {author} {\bibfnamefont {A.}~\bibnamefont {Browaeys}}, \bibinfo {author} {\bibfnamefont {Y.~R.~P.}\ \bibnamefont {Sortais}},\ and\ \bibinfo {author} {\bibfnamefont {P.}~\bibnamefont {Grangier}},\ }\bibfield  {title} {\bibinfo {title} {Energy distribution and cooling of a single atom in an optical tweezer},\ }\href {https://doi.org/10.1103/PhysRevA.78.033425} {\bibfield  {journal} {\bibinfo  {journal} {Phys. Rev. A}\ }\textbf {\bibinfo {volume} {78}},\ \bibinfo {pages} {033425} (\bibinfo {year} {2008})}\BibitemShut {NoStop}%
\bibitem [{\citenamefont {Lee}\ \emph {et~al.}(2014)\citenamefont {Lee}, \citenamefont {Vrijsen}, \citenamefont {Teper}, \citenamefont {Hosten},\ and\ \citenamefont {Kasevich}}]{Lee2014}%
  \BibitemOpen
  \bibfield  {author} {\bibinfo {author} {\bibfnamefont {J.}~\bibnamefont {Lee}}, \bibinfo {author} {\bibfnamefont {G.}~\bibnamefont {Vrijsen}}, \bibinfo {author} {\bibfnamefont {I.}~\bibnamefont {Teper}}, \bibinfo {author} {\bibfnamefont {O.}~\bibnamefont {Hosten}},\ and\ \bibinfo {author} {\bibfnamefont {M.~A.}\ \bibnamefont {Kasevich}},\ }\bibfield  {title} {\bibinfo {title} {Many-atom–cavity qed system with homogeneous atom–cavity coupling},\ }\href {https://doi.org/10.1364/OL.39.004005} {\bibfield  {journal} {\bibinfo  {journal} {Opt. Lett.}\ }\textbf {\bibinfo {volume} {39}},\ \bibinfo {pages} {4005} (\bibinfo {year} {2014})}\BibitemShut {NoStop}%
\bibitem [{\citenamefont {Brennecke}\ \emph {et~al.}(2007)\citenamefont {Brennecke}, \citenamefont {Donner}, \citenamefont {Ritter}, \citenamefont {Bourdel}, \citenamefont {K{\"o}hl},\ and\ \citenamefont {Esslinger}}]{Brennecke2007}%
  \BibitemOpen
  \bibfield  {author} {\bibinfo {author} {\bibfnamefont {F.}~\bibnamefont {Brennecke}}, \bibinfo {author} {\bibfnamefont {T.}~\bibnamefont {Donner}}, \bibinfo {author} {\bibfnamefont {S.}~\bibnamefont {Ritter}}, \bibinfo {author} {\bibfnamefont {T.}~\bibnamefont {Bourdel}}, \bibinfo {author} {\bibfnamefont {M.}~\bibnamefont {K{\"o}hl}},\ and\ \bibinfo {author} {\bibfnamefont {T.}~\bibnamefont {Esslinger}},\ }\bibfield  {title} {\bibinfo {title} {Cavity qed with a bose--einstein condensate},\ }\href {https://doi.org/10.1038/nature06120} {\bibfield  {journal} {\bibinfo  {journal} {Nature}\ }\textbf {\bibinfo {volume} {450}},\ \bibinfo {pages} {268} (\bibinfo {year} {2007})}\BibitemShut {NoStop}%
\end{thebibliography}%

\clearpage

\onecolumngrid
\begin{appendix}

\startcontents[app]
\printcontents[app]{l}{1}{}

\renewcommand{\thesection}{\Alph{section}}
\renewcommand{\thesubsection}{\thesection.\arabic{subsection}}

\section{Spin-Photon Entanglement Schemes}\label{sec:schemes}

The spin-photon entanglement is created by spin-dependent optical reflection. In the setting of cavity quantum electrodynamics (cavity QED), an atom couples to a cavity mode through the Jaynes-Cummings Hamiltonian. Using the input-output relation, a photon impinging on the cavity gets reflected depending on its carrier frequency $\omega$ by \cite{cavityQED}
\begin{equation}\label{refspec}
\mathcal{R}(\omega)=|r(\omega)|^2=\left|1-\frac{2 \kappa_c}{i\left(\omega-\omega_c\right)+\kappa+g^2 /\left(i\left(\omega-\omega_a\right)+\gamma\right)}\right|^2,
\end{equation}
where $\omega_c$ and $\omega_a$ are the resonance frequencies of the cavity and the two-level atom, respectively. The $g$ denotes the atom-cavity coupling strength; $\gamma$ is the atom's decay rate. The cavity couples to the environment with a rate $\kappa_c$ and has an internal loss rate of $\kappa_i$. The total loss rate is $\kappa =\kappa_i+\kappa_c$. Throughout the optical cavity-QED sections, $\kappa$ and $\gamma$ denote field/coherence HWHM rates, so that $C=g^2/(\kappa\gamma)$. This is consistent with the convention used in the companion paper \cite{PFMO_device_design_2025}. 

There are two different reflection schemes: amplitude-based \cite{Nguyen2019PRB} and phase-shift-based \cite{PhysRevLett.92.127902,PhysRevApplied.22.044013}. Both schemes require strong spin-photon interaction with high cooperativity $C = \frac{g^2}{\kappa\gamma} \gg 1$. In the amplitude-based protocol, one spin state results in high reflectivity $\mathcal{R}_{\max}=1$, and the other spin state results in a non-reflective $\mathcal{R}_{\min}=0$, which carves out a Bell state from the initial spin-photon state. More specifically, the protocol starts with a spin-photon initial state of $1/2(\ket{E}+\ket{L})(\ket{0}+\ket{1})$, where $E$ and $L$ denote the early and late time bins of the photon state. $0$ and $1$ denote the spin qubit states. Suppose the $\ket{0}$ state is highly reflective, and $\ket{1}$ is non-reflective. The state after the reflection of the first time-bin is $1/\sqrt{3}(\ket{E}\ket{0}+\ket{L}(\ket{0}+\ket{1}))$. Inserting a $\pi$-pulse on the spin states between the two time bins, the state after the reflection of the late time bin is 
$1/\sqrt{2}(\ket{E}\ket{1}+\ket{L}\ket{0}).$ The fidelity of the final Bell state depends critically on the non-reflective state $\mathcal{R}_{\min}=0$. The heralding probability of the scheme depends on $\mathcal{R}_{\max}$. 

In the phase-shift-based scheme, the two spin states reflect with a phase difference of $\pi$. In the original protocol \cite{PhysRevLett.92.127902}, the initial state is prepared in $1/2(\ket{H}+\ket{V})(\ket{0}+\ket{1})$, where the $H$ and $V$ denote horizontally polarized and vertically polarized states of a photon. Here, $0$ and $1$ denote the spin qubit states. The vertically polarized photon does not enter the cavity, so the cavity only modifies the reflection of the horizontally polarized photon. The $\ket{0}$ state creates a $\pi$ phase shift on the reflected state, whereas the $\ket{1}$ state does not. So the final state after reflection is $1/2(-\ket{H}\ket{0}+\ket{H}\ket{1}+\ket{V}\ket{0}+\ket{V}\ket{1}).$ In the more recent paper \cite{PhysRevApplied.22.044013}, the authors refer to the original scheme as the “on-off''scheme and propose a more symmetric phase-shift scheme: the “push-pull" scheme. The proposed scheme starts with the same initial state. The $\ket{0}$ and $\ket{1}$ spin states imprint $\pm\pi/2$ phase-shift on the reflected state, resulting in the final entangled state $1/2(-i\ket{H}\ket{0}+i\ket{H}\ket{1}+\ket{V}\ket{0}+\ket{V}\ket{1}).$ The successful creation of phase contrast relies on a high-Q cavity and the designed detuning between the two spin states and the cavity. 

In the main text, we focus on the amplitude-based scheme, but we can also apply the phase-shift based scheme for a higher overall entangling rate. The performance of the schemes depends on the cooperativity of the system. We summarize and compare the schemes in Table~\ref{schemecompare}. In short, the phase-shift scheme, compared with the amplitude-based scheme, creates Bell pairs with a higher success probability and lower fidelity. In addition, the phase-shift scheme receives infidelity contribution from the photon wavepacket distortion, which affects the photon interference.

\subsection{Amplitude-Based Scheme}

The scheme works in the nearly critically coupled ($\kappa_c \sim \kappa_i$), nearly resonant regime. Denoting the atom-cavity detuning to be $\Delta_a = \omega_a-\omega_c$, the hierarchy of parameters follows $\gamma<g<\Delta_a<\kappa\ll \omega_c$, where the atom locally modifies the cavity reflection spectrum \cite{Nguyen2019PRB}. 

In our protocol, we ensure that $\mathcal{R}_{\min}=0$ to guarantee the generation of unit-fidelity spin-photon Bell state. The deviation from this condition is roughly captured by imbalance between the loss rate as $\mathcal{R}_{\min}=(\kappa
_c-\kappa_i)^2/\kappa^2 + O(\gamma/\kappa)$. Satisfying the non-reflective condition at zeroth order (which means critically coupled $\kappa_c=\kappa_i$), the maximal reflection point can be shown to be $\mathcal{R}_{\max}=C^2/(C+1)^2 + O(\Delta_a^2/\kappa^2)$, where the cooperativity has the convention $C=g^2/\kappa\gamma$. The formula is also presented in \cite{cavityQED,PhysRevX.4.031022}. 

The amplitude-based scheme has the advantage that it can guarantee unit Bell fidelity, however, due to the non-reflective state, the probability of successfully creating a spin-photon Bell state is bounded by 0.5, resulting in the final success probability to be $P = 0.5 \cdot C^2/(C+1)^2$. To avoid this intrinsic reduction in success probability, we may consider the phase-shift protocol. 

\begin{table*}[t]
\caption{Comparison between three different spin-photon entanglement schemes. In the phase-shift schemes, the performance metrics are expanded around small intrinsic loss rate $\kappa_i/\kappa_c$. Following the convention in \cite{PhysRevApplied.22.044013}, the cooperativity $\mathcal{C}=g^2/\kappa_c\gamma$ is used in the context of phase-shift scheme. $\langle\omega^2\rangle$ is the second moment of the input photon spectral density.\label{schemecompare}} 
\begin{center}
\begin{tabular}{|c|c|c|c|}
\hline
 &  Amplitude-based & Phase-shift (on-off) & Phase-shift (push-pull) \\
\hline
Success Probability &  $0.5 \frac{C^2}{(C+1)^2} - O(\frac{\Delta^2}{\kappa^2})$ & $\frac{3}{4} + \frac{(\mathcal{C}-1)^2}{4(\mathcal{C}+1)^2} - O(\frac{\kappa_i}{\kappa_c}) $ & $\frac{\mathcal{C}}{1+\mathcal{C}}-O(\frac{ \kappa_i}{ \kappa_c})$  \\
\hline
State Fidelity & $1 - O(\frac{\gamma}{\kappa})$   & $\frac{(1+2\mathcal{C})^2}{4(1+\mathcal{C}+\mathcal{C}^2)} - O(\frac{\kappa_i}{\kappa_c}) $   & $\frac{\mathcal{C}+\sqrt{-1+\mathcal{C}^2}}{2 \mathcal{C}} - O(\frac{\kappa_i}{\kappa_c})$ \\
\hline
Waveform Distortion & Irrelevant   & $\propto\left\vert r_{\text{ON/OFF}}(0) +1/2\,  r''_{\text{ON/OFF}}(0) \langle\omega^2\rangle\right \vert^2$   & $\propto\left\vert r_{\pm}(0) +1/2\,  r''_{\pm}(0) \langle\omega^2\rangle\right \vert^2$  \\
\hline
\end{tabular}
\end{center}
\end{table*}

\subsection{Phase-Shift Scheme}
Since the phase-shift scheme creates entanglement by phase difference, instead of erasing part of the state, the scheme achieves unit success probability in the high cooperativity ($C \gg1$), over-coupled ($\kappa_c\gg\kappa_i$) regime \cite{PhysRevLett.92.127902,PhysRevApplied.22.044013}. In the original spin-photon entanglement protocol, one requires two different spin transitions. One transition is on resonance with the cavity frequency, creating atom-cavity dressed state; the other transition is far off-resonant, effectively creating an ``empty cavity". The incoming photon at the cavity frequency will enter the ``empty cavity" and bounce off the dressed state, creating an effective $\pi$-phase difference between these two cases. If we denote the reflection coefficient of the on-resonance condition to be $r_{\text{ON}}(\omega)$ and the off-resonance condition to be $r_{\text{OFF}}(\omega)$, the ``on-off" scheme ideally realizes $|\arg(r_{\text{ON}}(\omega))-\arg(r_{\text{OFF}}(\omega))|=\pi$. In real experiments, the reflection is constrained by the cavity cooperativity and intrinsic loss. Following the convention in \cite{PhysRevApplied.22.044013}, $$r_{\text{ON}}=\frac{\left(1-\kappa_i / \kappa_c \right)-\mathcal{C}}{\left(1+\kappa_i / \kappa_c \right)+\mathcal{C}}, \quad \quad r_{\text{OFF}}=\frac{1-\kappa_i / \kappa_c }{1+\kappa_i / \kappa_c },$$
where $\mathcal{C}=g^2/\kappa_c\gamma$ is the cooperativity defined only using the coupling loss. The probability of generating a Bell pair using this scheme is estimated to be $P =1/4\left((r_{\text{ON}})^2 + (r_{\text{OFF}})^2 +2\right)$. The fidelity is given by $F = \left(-r_{\text{ON}}+r_{\text{OFF}}+2\right)^2/16P$. We can expand the expression for the probability and fidelity in the small intrinsic loss regime, and the result is summarized in Table~\ref{schemecompare}. In the lossless limit, the success probability is shown to be greater than $3/4$, exceeding the highest achievable probability of $1/2$ of the amplitude-based scheme.

Despite a higher probability of success, one drawback of the scheme is that $|r_{\text{ON}}(\omega)| = |r_{\text{OFF}}(\omega)|$ is usually not guaranteed, thus contributing to infidelity in the final Bell state. To overcome this issue, a more symmetric scheme uses two transitions $\pm \Delta$ detuned from the cavity resonance. This design contributes to equal and opposite phase shifts of $\pm \pi/2$ on the incoming photon at zero detuning $\omega = \omega_c$ as long as the detuning satisfies $\Delta^2/\gamma^2 = ((\mathcal{C}+\kappa_i^2/\kappa_c^2)-1)/(1-\kappa_i^2/\kappa_c^2)$ \cite{PhysRevApplied.22.044013}. The $\pm\Delta$ detuned states result in reflection coefficients $r_-(\omega)$ and $r_+(\omega)$ respectively, and it is guaranteed that $|r_-(\omega_c)| = |r_+(\omega_c)|$, as shown in \cite{PhysRevApplied.22.044013}: 
\begin{equation}
r_{\pm}=\pm i \Delta \frac{\sqrt{\gamma^2 \kappa_c^2+\Delta^2\left(\kappa_c^2-\kappa_i^2\right)}-\gamma \kappa_c}{\Delta^2\left(\kappa_c+\kappa_i\right)}. 
\end{equation}
The probability and fidelity are similarly given by $P =1/4\left(|r_+|^2 + |r_-|^2 +2\right)$ and $F = \left(-ir_{+}+ir_{-}+2\right)^2/16P$. We can again expand to see the leading order behavior of the scheme in Table~\ref{schemecompare}. The comparison shows that in the $C\gg1$ regime, the fidelity of the push-pull scheme upper bounds that of the on-off scheme.

Compared with the amplitude-based scheme, one additional concern for the phase-shift scheme is waveform distortion. The phase-shift scheme interferes a photon reflected from the cavity and a photon reflected from the mirror. It is important for photons to be indistinguishable. This is not a problem for the amplitude-based scheme, since all the outgoing photons are reflected in the same way by the cavity-atom system. Such waveform distortion can also be factored into the decrease in Bell state fidelity. In the above fidelity estimation, this distortion contribution is neglected because we fix the incoming photon frequency. To see the waveform distortion, we need the response of the atom-cavity system to a photon wave-packet. Suppose the incoming photon pulse has a smooth temporal shape given by $f(t)$ (e.g. a Gaussian pulse). Its spectrum is given by the Fourier transform $\mathcal{F}[f(t)] = f(\omega)$. After the reflection, the pulse's waveform in frequency domain is altered by $r(\omega)f(\omega)$. To quantify waveform distortion, we calculate the overlap between the reflected signal waveform and the original waveform: 
\begin{equation}
   F_{\text{signal}}= \frac{1}{\mathcal{N}}\left\vert\int_{-\infty}^\infty r(\omega)|f(\omega)|^2 d \omega   \right\vert^2, 
\end{equation}
where we assume $f(\omega)$ is square-normalized, and the normalization factor $\mathcal{N}= \int|r(\omega)|^2|f(\omega)|^2d\omega$ is the reflection probability of the photon. Given $r(\omega)$ of the system, the distortion is minimized when the input signal is monochromatic (i.e. a Dirac delta function in the frequency domain). For an arbitrary input signal $f(t)$, we can estimate $F_{\text{signal}}$ in the following way. We expand $r(\omega)$ around the carrier frequency $r(\omega) = r(0) + r'(0)\omega + \frac{1}{2}r''(0)\omega^2 +...$. Integrating against the normalized spectral density:
\begin{equation}
  \int_{-\infty}^\infty r(\omega)|f(\omega)|^2 d \omega = \int_{-\infty}^\infty \left(r(0) + r'(0)\omega + \frac{1}{2}r''(0)\omega^2\right) |f(\omega)|^2 = r(0) + r'(0)\langle\omega\rangle +\frac{1}{2}r''(0) \langle \omega^2\rangle + ...
\end{equation}
In the above expression, the moment of the input pulse spectral density is defined as $\langle\omega^n\rangle = \int |f(\omega)|^2\omega^n d\omega$. For a symmetric spectrum the first moment vanishes. This gives us the final signal fidelity: 
\begin{equation}
   F_{\text{signal}} \approx \frac{1}{\mathcal{N}}\left\vert r(0) + \frac{1}{2} r''(0) \langle\omega^2\rangle\right \vert^2.
\end{equation}
When $r(\omega)$ peaks at zero, the second derivative is negative, and therefore as shown by the formula, the infidelity is given by the second moment of the signal in frequency space. Intuitively, the photon bandwidth should be narrow compared with the characteristic spectral scale of the reflection coefficient so that the signal can be minimally distorted. 

We summarize the three different schemes in Table~\ref{schemecompare}. Opting for a high fidelity protocol and a simple demonstration, we focus on the amplitude-based scheme in the main text. Nevertheless, to pursue a higher generation rate, future experiments may also adopt the phase-shift protocol, which requires a high quality, over-coupled spin-photon interface, careful design of the cavity detuning, and pulse shaping to minimize the waveform distortion.

\section{Figures of Merit Analysis}\label{figmerit}

\subsection{Microwave Photon Emission Probability}
\label{mw_prob_calculation}

\begin{figure}[t!]
    \centering
    \includegraphics[width=0.95\columnwidth]{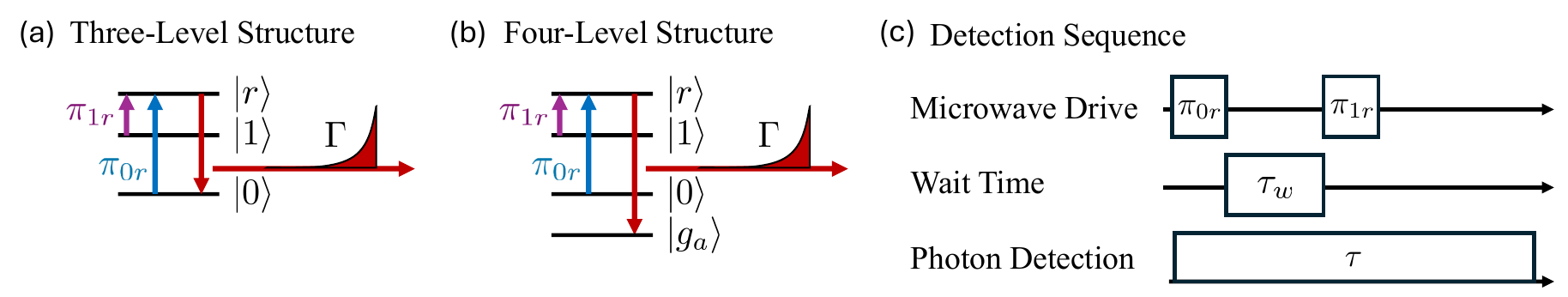}
    \caption{Microwave retrieval using (a) a three-level structure and (b) a four-level structure containing the ground state $\ket{g_a}$. (c) Pulse and detection sequence.}
    \label{MW_detection_prob_cal}
\end{figure}

We consider the microwave retrieval sequence shown in Fig.~\ref{MW_detection_prob_cal}. Fig.~\ref{MW_detection_prob_cal}(a) shows the three-level microwave manifold spanned by $\{\ket{0},\ket{1},\ket{r}\}$, which corresponds to the single-spin protocol, while Fig.~\ref{MW_detection_prob_cal}(b) shows the four-level manifold spanned by $\{\ket{g_a},\ket{0},\ket{1},\ket{r}\}$, which is relevant to the ensemble protocol. Within the collective single-excitation manifold, the dynamics of the ensemble under the applied drives are analogous to those of the single-spin system.

The initial density matrix is $\rho(0)=\ket{\psi(0)}\bra{\psi(0)}$, where $\ket{\psi(0)}=(\ket{0}+\ket{1})/\sqrt{2}$. The retrieval sequence consists of a resonant $\pi_{0r}$ pulse, a waiting time $\tau_w$, and a resonant $\pi_{1r}$ pulse. The two pulses are applied over $0<t\leq\tau_\pi$ and $\tau_\pi+\tau_w<t\leq2\tau_\pi+\tau_w$, respectively. Here, $\tau_\pi=\pi/\Omega$, where $\Omega$ is the Rabi frequency. The pulse Hamiltonians are
\begin{equation}
    H_{jr}=\frac{\hbar\Omega}{2}\left(\ket{r}\bra{j}+\ket{j}\bra{r}\right),\qquad j\in\{0,1\}.
    \label{mw_drive_hamiltonian}
\end{equation}
Accordingly, $H(t)=H_{0r}$ during the first pulse, $H(t)=H_{1r}$ during the second pulse, and $H(t)=0$ otherwise.

The Purcell-enhanced microwave emission channels in the three- and four-level manifolds are described by the jump operators
\begin{equation}
    J_3=\sqrt{\Gamma}\ket{0}\bra{r},\qquad J_4=\sqrt{\Gamma}\ket{g_a}\bra{r},
    \label{mw_jump_operators}
\end{equation}
respectively, where $\Gamma$ is the Purcell-enhanced emission rate. The density matrix evolves according to
\begin{equation}
    \dot{\rho}=-\frac{i}{\hbar}[H(t),\rho]+\mathcal{D}(J_\nu)\rho,\qquad
    \mathcal{D}(J)\rho=J\rho J^\dagger-\frac{1}{2}\left(J^\dagger J\rho+\rho J^\dagger J\right),
    \label{mw_retrieval_master_equation}
\end{equation}
where $\nu=3$ or $4$ denotes the corresponding manifold. The microwave emission accumulated over a total detection time $\tau$ is
\begin{equation}
    p_{\mathrm{mw},\nu}(\tau)=\int_0^\tau\mathrm{Tr}\left(J_\nu^\dagger J_\nu\rho(t)\right)dt
    =\int_0^\tau\Gamma\rho_{rr}(t)\,dt,
    \label{mw_integrated_emission}
\end{equation}
where $\rho_{xy}=\bra{x}\rho\ket{y}$.

For compact notation, we define
\begin{align}
    \mathcal{I}(A,B,C;\alpha,\omega;t)
    &\equiv\int_0^t e^{-\alpha s}\left(A+B\cos(\omega s)+C\sin(\omega s)\right)ds \nonumber\\
    &=A\frac{1-e^{-\alpha t}}{\alpha}
    +B\frac{e^{-\alpha t}\left(-\alpha\cos(\omega t)+\omega\sin(\omega t)\right)+\alpha}{\alpha^2+\omega^2}+C\frac{e^{-\alpha t}\left(-\alpha\sin(\omega t)-\omega\cos(\omega t)\right)+\omega}{\alpha^2+\omega^2}.
    \label{mw_integral_function}
\end{align}

\emph{Three-Level Microwave Manifold}\textemdash
In the three-level manifold, the first pulse drives the $\ket{0}\leftrightarrow\ket{r}$ transition, while $\ket{r}$ decays back to $\ket{0}$. An emission during the first pulse can therefore be followed by re-excitation and an additional emission. Thus, for finite-duration pulses, $p_{\mathrm{mw},3}$ represents the mean number of emitted microwave photons. In the fast-drive limit used for the single-spin protocol, emission during the control pulses and re-excitation during the first pulse are negligible, so $p_{\mathrm{mw},3}$ reduces to the microwave-photon detection probability.

During the first pulse, Eq.~\eqref{mw_retrieval_master_equation} with $H(t)=H_{0r}$ and $J_\nu=J_3$ gives
\begin{equation}
    \dot{\rho}_{rr}=\Omega\,\mathrm{Im}\rho_{0r}-\Gamma\rho_{rr},\qquad
    \mathrm{Im}\dot{\rho}_{0r}=-\frac{\Omega}{2}\left(2\rho_{rr}-\frac{1}{2}\right)-\frac{\Gamma}{2}\mathrm{Im}\rho_{0r}.
    \label{mw_first_pulse_eoms}
\end{equation}
Here, $\rho_{00}+\rho_{rr}=1/2$. We consider the underdamped regime $\Omega>\Gamma/2$ and define
\begin{equation}
    p_{\mathrm{ss}}=\frac{\Omega^2}{2\left(2\Omega^2+\Gamma^2\right)},\qquad
    \omega_1=\sqrt{\Omega^2-\frac{\Gamma^2}{16}},
    \label{mw_first_pulse_parameters}
\end{equation}
where $p_{\mathrm{ss}}$ is the steady-state $\ket{r}$ population and $\omega_1$ is the damped oscillation frequency during the first pulse. Solving Eq.~\eqref{mw_first_pulse_eoms} with $\rho_{rr}(0)=0$ and $\mathrm{Im}\rho_{0r}(0)=0$ gives
\begin{equation}
    \rho_{rr}^{(1)}(t)=p_{\mathrm{ss}}\left(1-e^{-3\Gamma t/4}
    \left(\cos(\omega_1t)+\frac{3\Gamma}{4\omega_1}\sin(\omega_1t)\right)\right),
    \qquad 0<t\leq\tau_\pi,
    \label{mw_emit_first_pulse_solution}
\end{equation}
and the microwave emission accumulated during the first pulse is
\begin{equation}
    p_{\mathrm{mw},3}^{(1)}(t)
    =\Gamma p_{\mathrm{ss}}\left(t-\mathcal{I}\left(0,1,\frac{3\Gamma}{4\omega_1};\frac{3\Gamma}{4},\omega_1;t\right)\right).
    \label{mw_first_pulse_emission}
\end{equation}

The first pulse also generates the $\ket{1}$--$\ket{r}$ coherence required for the second pulse. The relevant equations are
\begin{equation}
    \dot{\rho}_{10}=i\frac{\Omega}{2}\rho_{1r},\qquad
    \dot{\rho}_{1r}=i\frac{\Omega}{2}\rho_{10}-\frac{\Gamma}{2}\rho_{1r}.
    \label{mw_residual_coherence_eoms}
\end{equation}
With $\rho_{10}(0)=1/2$ and $\rho_{1r}(0)=0$, and using $\dot{\rho}_{rr}=-\Gamma\rho_{rr}$ and $\dot{\rho}_{1r}=-(\Gamma/2)\rho_{1r}$ during the waiting interval, the population and coherence at the beginning of the second pulse are
\begin{equation}
    X\equiv\rho_{rr}(\tau_\pi+\tau_w)=\rho_{rr}^{(1)}(\tau_\pi)e^{-\Gamma\tau_w},\qquad
    Y\equiv\mathrm{Im}\rho_{1r}(\tau_\pi+\tau_w)
    =\frac{\Omega}{2\omega_2}e^{-\Gamma(\tau_\pi/4+\tau_w/2)}
    \sin\left(\frac{\omega_2\tau_\pi}{2}\right),
    \label{mw_second_pulse_initial_conditions}
\end{equation}
where $\omega_2=\sqrt{\Omega^2-\Gamma^2/4}$. The microwave emission accumulated over a waiting interval $t$ is $p_{\mathrm{mw},3}^{(w)}(t)=\rho_{rr}^{(1)}(\tau_\pi)(1-e^{-\Gamma t})$.

Let $s=t-(\tau_\pi+\tau_w)$ denote the time from the beginning of the second pulse. Equation~\eqref{mw_retrieval_master_equation} with $H(t)=H_{1r}$ and $J_\nu=J_3$ gives
\begin{equation}
    \dot{\rho}_{rr}=\Omega\,\mathrm{Im}\rho_{1r}-\Gamma\rho_{rr},\qquad
    \mathrm{Im}\dot{\rho}_{1r}=\frac{\Omega}{2}\left(\rho_{11}-\rho_{rr}\right)-\frac{\Gamma}{2}\mathrm{Im}\rho_{1r},\qquad
    \dot{\rho}_{11}=-\Omega\,\mathrm{Im}\rho_{1r}.
    \label{mw_second_pulse_eoms}
\end{equation}
With $\rho_{rr}(s=0)=X$, $\rho_{11}(s=0)=1/2$, and $\mathrm{Im}\rho_{1r}(s=0)=Y$, the solution is
\begin{equation}
    \rho_{rr}^{(2)}(s)=e^{-\Gamma s/2}\left(A+B\cos(\omega_2s)+C\sin(\omega_2s)\right),
    \qquad 0<s\leq\tau_\pi,
    \label{mw_emit_second_pulse_solution}
\end{equation}
where
\begin{equation}
    B=\frac{\left(2\Omega^2-\Gamma^2\right)X-\Omega^2+2\Omega\Gamma Y}{4\omega_2^2},\qquad
    C=\frac{\Omega Y-\Gamma X/2}{\omega_2},\qquad A=X-B.
    \label{mw_second_pulse_coefficients}
\end{equation}
The microwave emission accumulated during the second pulse is
\begin{equation}
    p_{\mathrm{mw},3}^{(2)}(s;\tau_w)
    =\Gamma\mathcal{I}\left(A,B,C;\frac{\Gamma}{2},\omega_2;s\right).
    \label{mw_second_pulse_emission}
\end{equation}

Combining the four stages gives the finite-drive result
\begin{equation}
    p_{\mathrm{mw},3}(\tau;\tau_w)=
    \begin{cases}
        p_{\mathrm{mw},3}^{(1)}(\tau),
        & 0<\tau\leq\tau_\pi,\\
        p_{\mathrm{mw},3}^{(1)}(\tau_\pi)
        +\rho_{rr}^{(1)}(\tau_\pi)\left(1-e^{-\Gamma(\tau-\tau_\pi)}\right),
        & \tau_\pi<\tau\leq\tau_\pi+\tau_w,\\
        p_{\mathrm{mw},3}^{(1)}(\tau_\pi)
        +\rho_{rr}^{(1)}(\tau_\pi)\left(1-e^{-\Gamma\tau_w}\right)
        +p_{\mathrm{mw},3}^{(2)}(\tau-\tau_\pi-\tau_w;\tau_w),
        & \tau_\pi+\tau_w<\tau\leq2\tau_\pi+\tau_w,\\
        p_{\mathrm{mw},3}^{(1)}(\tau_\pi)
        +\rho_{rr}^{(1)}(\tau_\pi)\left(1-e^{-\Gamma\tau_w}\right)
        +p_{\mathrm{mw},3}^{(2)}(\tau_\pi;\tau_w)
        +\rho_{rr}^{(2)}(\tau_\pi)\left(1-e^{-\Gamma(\tau-2\tau_\pi-\tau_w)}\right),
        & \tau>2\tau_\pi+\tau_w.
    \end{cases}
    \label{mw_total_three_level}
\end{equation}

For the single-spin protocol, $\tau_\pi\ll1/\Gamma$. In this limit, emission during the control pulses is negligible, and $\rho_{rr}^{(1)}(\tau_\pi)\simeq1/2$, $p_{\mathrm{mw},3}^{(1)}(\tau_\pi)\simeq\pi\Gamma/(4\Omega)\ll1$. The waiting-time contribution becomes $p_{\mathrm{mw},3}^{(w)}(\tau_w)\simeq\frac{1}{2}(1-e^{-\Gamma\tau_w})$. The second-pulse emission is also of order $\Gamma/\Omega$, and $\rho_{rr}^{(2)}(\tau_\pi)\simeq1/2$. Neglecting the pulse-emission terms of order $\Gamma/\Omega$, Eq.~\eqref{mw_total_three_level} reduces to
\begin{equation}
    p_{\mathrm{mw},3}(\tau;\tau_w)\simeq
    \begin{cases}
        0,
        & 0<\tau\leq\tau_\pi,\\
        \frac{1}{2}\left(1-e^{-\Gamma(\tau-\tau_\pi)}\right),
        & \tau_\pi<\tau\leq\tau_\pi+\tau_w,\\
        \frac{1}{2}\left(1-e^{-\Gamma\tau_w}\right),
        & \tau_\pi+\tau_w<\tau\leq2\tau_\pi+\tau_w,\\
        \frac{1}{2}\left(1-e^{-\Gamma\tau_w}\right)
        +\frac{1}{2}\left(1-e^{-\Gamma(\tau-2\tau_\pi-\tau_w)}\right),
        & \tau>2\tau_\pi+\tau_w.
    \end{cases}
    \label{mw_prob_three_level}
\end{equation}
For fixed $\tau$, the final expression is maximized at $\tau_w=(\tau-2\tau_\pi)/2$, giving $p_{\mathrm{mw},3}^{\mathrm{opt}}(\tau)\simeq1-e^{-\Gamma(\tau-2\tau_\pi)/2}$.

In the instantaneous-pulse limit $\tau_\pi\rightarrow0$, for $\tau>\tau_w$,
\begin{equation}
    p_{\mathrm{mw},3}(\tau;\tau_w)\simeq1-\frac{1}{2}\left(e^{-\Gamma\tau_w}+e^{-\Gamma(\tau-\tau_w)}\right),\qquad
    p_{\mathrm{mw},3}^{\mathrm{opt}}(\tau)\simeq1-e^{-\Gamma\tau/2}.
    \label{mw_instantaneous_drive_probability}
\end{equation}
The optimum occurs at $\tau_w=\tau/2$. This is the expression used for the single-spin protocol in the main text.

\emph{Four-Level Microwave Manifold}\textemdash
For the four-level manifold, the retrieval sequence and pulse Hamiltonians are the same as in the three-level case, but the Purcell-enhanced decay channel is $\ket{r}\rightarrow\ket{g_a}$. Since $\ket{g_a}$ is not driven by either pulse, a decay event removes the system from the driven subspace. Therefore, the integrated microwave emission can be directly interpreted as the detection probability. Primed quantities denote the corresponding four-level results.

During the first pulse, Eq.~\eqref{mw_retrieval_master_equation} with $H(t)=H_{0r}$ and $J_\nu=J_4$ gives
\begin{equation}
    \dot{\rho}_{rr}=\Omega\,\mathrm{Im}\rho_{0r}-\Gamma\rho_{rr},\qquad
    \dot{\rho}_{00}=-\Omega\,\mathrm{Im}\rho_{0r},\qquad
    \mathrm{Im}\dot{\rho}_{0r}=\frac{\Omega}{2}\left(\rho_{00}-\rho_{rr}\right)-\frac{\Gamma}{2}\mathrm{Im}\rho_{0r}.
    \label{mw_first_pulse_eoms_4level}
\end{equation}
With $\rho_{rr}(0)=0$, $\rho_{00}(0)=1/2$, and $\mathrm{Im}\rho_{0r}(0)=0$, the solution and accumulated emission probability are
\begin{equation}
    \rho_{rr}^{\prime(1)}(t)=\frac{\Omega^2}{4\omega_2^2}e^{-\Gamma t/2}\left(1-\cos(\omega_2t)\right),\qquad
    p_{\mathrm{mw},4}^{(1)}(t)=\Gamma\frac{\Omega^2}{4\omega_2^2}\mathcal{I}\left(1,-1,0;\frac{\Gamma}{2},\omega_2;t\right),
    \qquad 0<t\leq\tau_\pi.
    \label{mw_first_pulse_results_4level}
\end{equation}

At the beginning of the second pulse, $X'=\rho_{rr}^{\prime(1)}(\tau_\pi)e^{-\Gamma\tau_w}$, while the residual coherence remains $Y$ in Eq.~\eqref{mw_second_pulse_initial_conditions}. The second-pulse equations are therefore identical to Eq.~\eqref{mw_second_pulse_eoms}. With $\rho_{rr}(s=0)=X'$, $\rho_{11}(s=0)=1/2$, and $\mathrm{Im}\rho_{1r}(s=0)=Y$, the results are
\begin{equation}
    \rho_{rr}^{\prime(2)}(s)=e^{-\Gamma s/2}\left(A'+B'\cos(\omega_2s)+C'\sin(\omega_2s)\right),\qquad
    p_{\mathrm{mw},4}^{(2)}(s;\tau_w)=\Gamma\mathcal{I}\left(A',B',C';\frac{\Gamma}{2},\omega_2;s\right),
    \label{mw_second_pulse_results_4level}
\end{equation}
where $(A',B',C')$ are obtained from Eq.~\eqref{mw_second_pulse_coefficients} by replacing $X$ with $X'$.

The total finite-drive emission probability is
\begin{equation}
    p_{\mathrm{mw},4}(\tau;\tau_w)=
    \begin{cases}
        p_{\mathrm{mw},4}^{(1)}(\tau),
        & 0<\tau\leq\tau_\pi,\\
        p_{\mathrm{mw},4}^{(1)}(\tau_\pi)
        +\rho_{rr}^{\prime(1)}(\tau_\pi)\left(1-e^{-\Gamma(\tau-\tau_\pi)}\right),
        & \tau_\pi<\tau\leq\tau_\pi+\tau_w,\\
        p_{\mathrm{mw},4}^{(1)}(\tau_\pi)
        +\rho_{rr}^{\prime(1)}(\tau_\pi)\left(1-e^{-\Gamma\tau_w}\right)
        +p_{\mathrm{mw},4}^{(2)}(\tau-\tau_\pi-\tau_w;\tau_w),
        & \tau_\pi+\tau_w<\tau\leq2\tau_\pi+\tau_w,\\
        p_{\mathrm{mw},4}^{(1)}(\tau_\pi)
        +\rho_{rr}^{\prime(1)}(\tau_\pi)\left(1-e^{-\Gamma\tau_w}\right)
        +p_{\mathrm{mw},4}^{(2)}(\tau_\pi;\tau_w)
        +\rho_{rr}^{\prime(2)}(\tau_\pi)\left(1-e^{-\Gamma(\tau-2\tau_\pi-\tau_w)}\right),
        & \tau>2\tau_\pi+\tau_w.
    \end{cases}
    \label{mw_prob_four_level}
\end{equation}

In the fast-drive limit $\Gamma\tau_\pi\ll1$, each pulse transfers one half of the initial population to $\ket{r}$ before appreciable emission occurs. Equation~\eqref{mw_prob_four_level} therefore reduces to the same stage-resolved probability as Eq.~\eqref{mw_prob_three_level}, with the instantaneous-pulse optimum $p_{\mathrm{mw},4}^{\mathrm{opt}}(\tau)\simeq1-e^{-\Gamma\tau/2}$.

\subsection{Heralding Probability Estimation}

The microwave photon capture is heralded by measuring the transmon in the excited subspace $\{\ket{e},\ket{f}\}$. Leakage out of the subspace due to the transmon relaxation to $\ket{g}$ can lower the heralding probability. In the main text, we ignore this effect assuming $\tau \ll T_{1,t}$. This can be made more rigorous by taking into account the transmon relaxation, and expand the correction to the heralding probability in terms of $\tau/T_{1,t}$. 

To find the population in the excited subspace, we can solve the rate equation, which follows the cascade $\ket{f}\rightarrow\ket{e}\rightarrow\ket{g}$. Denoting the population in a state as $P_i$, $i=e,g,f$, we have the following equations:
\begin{align}
  \dot P_f &= -\Gamma_f\, P_f, \label{eq:rate_f}\\
  \dot P_e &= \Gamma_f\, P_f - \Gamma_1\, P_e, \label{eq:rate_e}\\
  \dot P_g &= \Gamma_1\, P_e,\label{eq:rate_g}
\end{align}
where $\Gamma_f$ and $\Gamma_1$ are the relaxation rate of $\ket{f}$ and $\ket{e}$ respectively. 

After capturing the first time-bin, the initial condition is
$(P_f,P_e,P_g)=(1,0,0)$ at $t=0$.
Solving Eqs.~\eqref{eq:rate_f}--\eqref{eq:rate_g}:
\begin{align}
  P_f(t) &= e^{-\Gamma_f t}, \label{eq:Pf}\\
  P_e(t) &= \frac{\Gamma_f}{\Gamma_f-\Gamma_1}
             \bigl(e^{-\Gamma_1 t}-e^{-\Gamma_f t}\bigr).
             \label{eq:Pe_from_f}
\end{align}
Denoting the time interval between the first time-bin and the second time-bin as $\Delta t$, and the interval between the second time-bin and the measurement as $t_m$, the survival probability in the herald subspace at the measurement time
$T\equiv\Delta t+t_m$ is
\begin{equation}
  S_{\mathrm{early}}(T)
    = P_f(T)+P_e(T)
    = e^{-\Gamma_f T}
      + \frac{\Gamma_f}{\Gamma_f-\Gamma_1}
        \bigl(e^{-\Gamma_1 T}-e^{-\Gamma_f T}\bigr).
  \label{eq:S_early_general}
\end{equation}
For a weakly anharmonic transmon, $\Gamma_f\approx 2\,\Gamma_1 = 2/T_{1,t}$, which
simplifies Eq.~\eqref{eq:S_early_general} to
\begin{equation}
    S_{\mathrm{early}}(T)
      = 2\,e^{-T/T_{1,t}} - e^{-2T/T_{1,t}}.
  \label{eq:S_early}
\end{equation}

The transmon is excited to $\ket{e}$ at $t=\Delta t$ and must survive
until $t=\Delta t+t_m$.  Simple exponential decay gives
\begin{equation}
    S_{\mathrm{late}}(t_m) = e^{-t_m/T_{1,t}}.
  \label{eq:S_late}
\end{equation}
Combining both capture paths (in equal superposition) with emission probability~$p_{\text{mw}}$ and capture
efficiency~$\eta$, the total excitation state population is
\begin{equation}
    P_{ef}
      = \frac{1}{2}p_{\text{mw}}\,\eta\Bigl[
          S_{\mathrm{early}}(\Delta t+t_m)
        + S_{\mathrm{late}}(t_m)
        \Bigr].
  \label{eq:P_success}
\end{equation}
Substituting Eqs.~\eqref{eq:S_early} and~\eqref{eq:S_late}
(with $\Gamma_f=2\,\Gamma_1$) yields
\begin{equation}
  P_{ef}
    = \frac{1}{2}p_{\text{mw}}\,\eta\left[
        \left(
          2\,e^{-(\Delta t+t_m)/T_{1,t}}
          - e^{-2(\Delta t+t_m)/T_{1,t}}\right)
      + e^{-t_m/T_{1,t}}
      \right].
  \label{eq:P_success_explicit}
\end{equation}
Assuming the measurement is taken immediately after the second time-bin capture, $t_m\sim0$, and the intervals before the two time-bins are roughly $\Delta t \sim \tau/2$, the expression simplifies to 
\begin{equation}
  P_{ef}
    = \frac{1}{2}p_{\text{mw}}\,\eta
        \left(
          2\,e^{-\tau/2T_{1,t}}
          - e^{-\tau/T_{1,t}}+1\right).
           \label{eq:limit_balanced}
\end{equation}
To quantify the leading relaxation error we expand
Eq.~\eqref{eq:limit_balanced} in powers of
$\tau/2T_{1,t}\ll 1$. The first-order term vanishes identically, and we are left with
\begin{equation}
    P_{ef}
      = p_{\text{mw}}\,\eta\left[
          1 - \frac{1}{8}\left(\frac{\tau}{T_{1,t}}\right)^{2}
          + \mathcal{O}\left(\frac{\tau}{T_{1,t}}\right)^{3}
        \right].
  \label{eq:P_short_time}
\end{equation}
Physically, the population that leaks out of $\ket{f}$ at
rate~$\Gamma_f$ is temporarily caught in~$\ket{e}$, still within the
herald subspace, so the net leakage out of
$\{\ket{e},\ket{f}\}$ begins only at second order. The shelving step provides first-order protection against energy relaxation during the inter-bin waiting time. Since the effect of transmon relaxation is only sub-leading when $\tau\ll T_{1,t}$, we can take $P_{ef} = p_{\text{mw}}$ when the capturing efficiency is high $\eta \rightarrow 1$. 

\subsection{Entanglement Fidelity of Microwave Retrieval}\label{sec:fidelity}

We characterize the microwave retrieval process by calculating its entanglement fidelity, which quantifies how well the microwave retrieval preserves the entanglement of the original spin–photon Bell state. For the initial Bell state, $\ket{\Phi}=(\ket{E}_o \ket{1}_m + \ket{L}_o \ket{0}_m)/\sqrt{2}$, where the subscripts $o$ and $m$ denote the optical photon and the microwave spin level, respectively, the dominant decoherence mechanisms are spin dephasing and transmon relaxation. With heralding or post-selection, we can restrict to the qubit subspace of the spin ($\ket{0}$  $\ket{1}$), or the transmon ($\ket{e}$  $\ket{f}$). In this subspace, the error processes can be modeled as a dephasing channel $\mathcal{D_m}$ and an amplitude-damping channel $\mathcal{L_m}$ from $\ket{f}$ to $\ket{e}$, parameterized by the spin dephasing time $T_{2,\text{s}}$ and the transmon relaxation time $T_{1,\text{t}}$, respectively. Defining the combined channel as $\mathcal{E_m} = \mathcal{L_m} \circ \mathcal{D_m}$, the entanglement fidelity is
\begin{equation}
    \mathcal{F}_{\text{mw}} = \bra{\Phi} (\mathbb{1}_o \otimes \mathcal{E}_m)(\ket{\Phi}\bra{\Phi}) \ket{\Phi}. 
\end{equation}
The density matrix of $\ket{\Phi}$ is
\begin{equation}
\rho =\ket{\Phi}\bra{\Phi}=\frac{1}{2}\left(\ket{E}\bra{E}_o\, \ket{1}\bra{1}_m + \ket{E}\bra{L}_o\, \ket{1}\bra{0}_m +\ket{L}\bra{E}_o\, \ket{0}\bra{1}_m +\ket{L}\bra{L}_o\, \ket{0}\bra{0}_m \right).
\end{equation}
Thus, the fidelity reduces to
\begin{equation}
     \mathcal{F}_{\text{mw}} = \frac{1}{4}\left(\bra{1}\mathcal{E}(\ket{1}\bra{1})\ket{1} + \bra{1}\mathcal{E}(\ket{1}\bra{0})\ket{0} + \bra{0}\mathcal{E}(\ket{0}\bra{1})\ket{1} + \bra{0}\mathcal{E}(\ket{0}\bra{0})\ket{0}\right),
\end{equation}
where the subscripts are dropped since only the microwave subsystem is affected by the channel.

For the qubit dephasing channel, the Kraus operators are $M_{d,0}=\sqrt{1-p_d} I$, $M_{d,1}= \left(\begin{array}{cc} \sqrt{p_d} & 0 \\ 0 & 0\end{array}\right)$, and $M_{d,2}=\left(\begin{array}{cc} 0 & 0 \\ 0 & \sqrt{p_d}\end{array}\right)$, where $p_d$ is the dephasing probability. For the amplitude damping channel on the same basis, the Kraus operators are  $M_{a,0}=\left(\begin{array}{cc}1 & 0 \\ 0 & \sqrt{1-p_a}\end{array}\right)$ and $ M_{a,1}=\left(\begin{array}{cc}0 & \sqrt{p_a} \\ 0 & 0\end{array}\right)$, where $p_a$ is the excited state decay probability. The action of the channel gives
\begin{align*}
    & \mathcal{E}(\ket{0}\bra{0}) = \ket{0}\bra{0}, \\
    & \mathcal{E}(\ket{0}\bra{1}) = \sqrt{1-p_a}(1-p_d)\ket{0}\bra{1}, \\
    & \mathcal{E}(\ket{1}\bra{0}) = \sqrt{1-p_a}(1-p_d)\ket{1}\bra{0}, \\
    & \mathcal{E}(\ket{1}\bra{1}) = p_a\ket{0}\bra{0}+(1-p_a)\ket{1}\bra{1}. \\    
\end{align*}
Substituting these into the fidelity expression yields
\begin{equation}
     \mathcal{F}_{\text{mw}} = \frac{1}{4}\left(1 + (1-p_a)+2\sqrt{1-p_a}(1-p_d)\right). 
\end{equation}

The damping parameter $p_a$ here is given by decay within the logical subspace $\ket{f}\rightarrow \ket{e}$, which occurs with a probability $1-e^{-2t/T_{1,\text{t}}}$ over a time duration $t$. Since the transmon is only excited to $\ket{f}$ after capturing the first microwave pulse, the damping time is in fact $t =\tau/2$, which gives $p_a = 1-e^{-\tau/T_{1,\text{t}}}$. Spin dephasing during the entire retrieval window gives $p_d = 1-e^{-\tau/T_{2,\text{s}}}$. Notice that the $T_{2,\text{s}}$ is the $T_2$ time of the spin without the presence of the resonator, where the decoherence process is dominated by the dephasing $T_{1,\text{s}} \gg T_{2,\text{s}}$. Substituting $p_a$ and $p_d$ yields the entanglement fidelity:
\begin{equation}
     \mathcal{F}_{\text{mw}}=\frac{1}{4}(1+e^{-\tau/T_{1,\text{t}}}+2e^{-\tau/2T_{1,\text{t}}}e^{-\tau/T_{2,\text{s}}}).
\end{equation}

\subsection{Performance Analysis with Transmon Measurement Error}

A successful microwave photon capture is heralded by measuring the transmon in the excited state. We use a probabilistic model to account for transmon measurement errors by defining variable $M$ as the measurement result and $E$ as the result of photon capture. $M=1$ (or $0$) corresponds to measuring the transmon in excited (or ground) state. $E=1$ (or $0$) corresponds to a photon capture (or loss). The probability of photon capture is $P(E=1) = p_\text{mw}=1-e^{-\Gamma\tau/2}$, and the measurement error is modeled as a binary symmetric channel described by the conditional probability $P(M=0 | E=1)= P(M=1 | E=0)=\epsilon$ and $P(M=0 | E=0)=P(M=1 | E=1)=1-\epsilon$. The probability of measuring the transmon in the excited state is $P(M=1)= p_\text{mw}(1-\epsilon)+\epsilon(1-p_\text{mw})= P_{\text{mw}}$. However, among these heralding events, some of them come from the measurement error instead of a real photon capture. The question is how much we can trust the heralding event. This defines the fidelity of measurement $F_M$, and can be calculated by Bayes' theorem: 
\begin{equation}
    F_M= P(E=1|M=1) = \frac{P(M=1|E=1)P(E=1)}{P(M=1)} = \frac{(1-\epsilon)p_\text{mw}}{(1-\epsilon)p_\text{mw} + \epsilon(1-p_\text{mw})}.
\end{equation}
The above equation implies that when there is no photon $p=0$, all the heralding clicks come from the measurement error, and we cannot trust the measurement result $F_M=0$. As the photon probability approaches unity $p_\text{mw} \rightarrow1$, the heralding also becomes more trustworthy $F_M \rightarrow1$.

\section{Single-Spin-Based MO-NODE}\label{sec:color}

\subsection{Scheme-Compatible Color Centers}

\begin{table*}
\begin{tabular}{|c|c|c|}
\hline  & NV$^-$\cite{Alkauskas2014} & Hyperfine ${}^{117}\text{SnV}^-$ \cite{harris2025high} \\
\hline Debye-Waller Factor ($\xi$) & 0.03 & 0.6 \\
\hline ZPL Wavelength (nm) & 637 & 620 \\
\hline DC Magnetic Field $\vec{B}$ & 0.1--1 mT & 0 \\
\hline Strain-induced splitting $\alpha$ & 0 & 900 GHz \\
\hline MW qubit frequency (GHz) & 2.9 & 0.6 \\
\hline Magnetic transition matrix element & $\gamma_e/\sqrt{2}$ & $\alpha\gamma_e/(2\sqrt{2(\alpha^2+\lambda^2)})$ \cite{PFMO_device_design_2025} \\
\hline
\end{tabular}
\caption{Comparison between different candidate color centers. $\gamma_e$ is the gyromagnetic ratio of the electron. The magnetic dipole of the tin-vacancy center depends on the strain $\alpha$ in the system, and the spin-orbit coupling $\lambda$.
}\label{colortab}
\end{table*}

To demonstrate the feasibility of our single-spin MO-NODE, we consider the NV$^-$ and the ${}^{117}\text{SnV}^-$ centers in diamond with corresponding device design in Ref.~\cite{PFMO_device_design_2025}. The properties of the two color centers are compared in Table~\ref{colortab}. A key operational difference lies in the magnetic field requirement. The NV$^-$ has a spin-1 ground state $m=\pm1,0$. Two of the levels ($\pm1$) are degenerate in the zero-field limit. To establish a long-lived qubit space, and strongly coupled readout level, the three levels need to be sufficiently split by a DC magnetic field. To create a megahertz level splitting between the $\pm1$ states, which makes the state decay negligible ($<10^{-3}$) during the readout, a small magnetic field ($0.1-1$mT) is needed. The advantage of the spin-1 ground state is that it enables strong spin-microwave coupling, with magnetic transition matrix element given by $\gamma_e\bra{0}\vec{S}\ket{1}=\gamma_e/\sqrt{2}$ \cite{Haikka2017}. The ${}^{117}\text{SnV}^-$ ground state can be split by strain engineering and strong hyperfine interaction, as demonstrated in Ref.~\cite{harris2025high}. Therefore, it allows the operation of the scheme in zero-field. However, due to the hybridization of electron-nuclear spin in the ground state, the ${}^{117}\text{SnV}^-$'s dipole moment is less than half of that of the NV$^-$, as described in our companion paper \cite{PFMO_device_design_2025}. The small magnetic transition matrix element results in a low microwave retrieval efficiency. Moreover, due to the low transition frequency $\sim0.6$ GHz, the ${}^{117}\text{SnV}^-$ faces higher thermal noise and potential frequency mismatch when coupled to a transmon qubit, which may be resolved by operating the scheme at a lower cryogenic temperature and employing a flux tunable superconducting qubit.

In terms of optical properties, the NV$^-$ has a very small ($\sim0.03$) Debye-Waller factor (D-W factor), compared with a D-W factor of 0.6 for ${}^{117}\text{SnV}^-$. This results in NV$^-$'s smaller optical cooperativity \cite{PFMO_device_design_2025}. Nevertheless, spin-photon entanglement protocol for NV$^-$ is presented and analyzed in \cite{PhysRevX.4.031022}. Given the high finesse of the optical cavity, and the large magnetic transition matrix element, the NV$^-$ still exceeds ${}^{117}\text{SnV}^-$ in both fidelity and overall efficiency. If the small magnetic field is compatible with the experimental design, NV$^-$ is advantageous due to its high efficiency and fidelity in Bell pair generation. The ${}^{117}\text{SnV}^-$ is favorable due to its zero-field protocol.

Our scheme can be generalized to other single defect centers provided that they satisfy the following requirements. For generating spin-photon entanglement, the system must demonstrate a robust spin initialization, well-resolved spin-dependent optical transitions, and microwave control over the spin levels. To realize our microwave readout scheme, the spin states must contain a qubit subspace and a gigahertz-range readout level that couples strongly to microwave. Moreover, the qubit subspace and the readout level must be connected with allowed transitions to enable microwave control.

\subsection{Figures of Merit at Different Spin Locations}
\begin{figure}[t!]
\centering
\includegraphics[width=0.95\columnwidth]{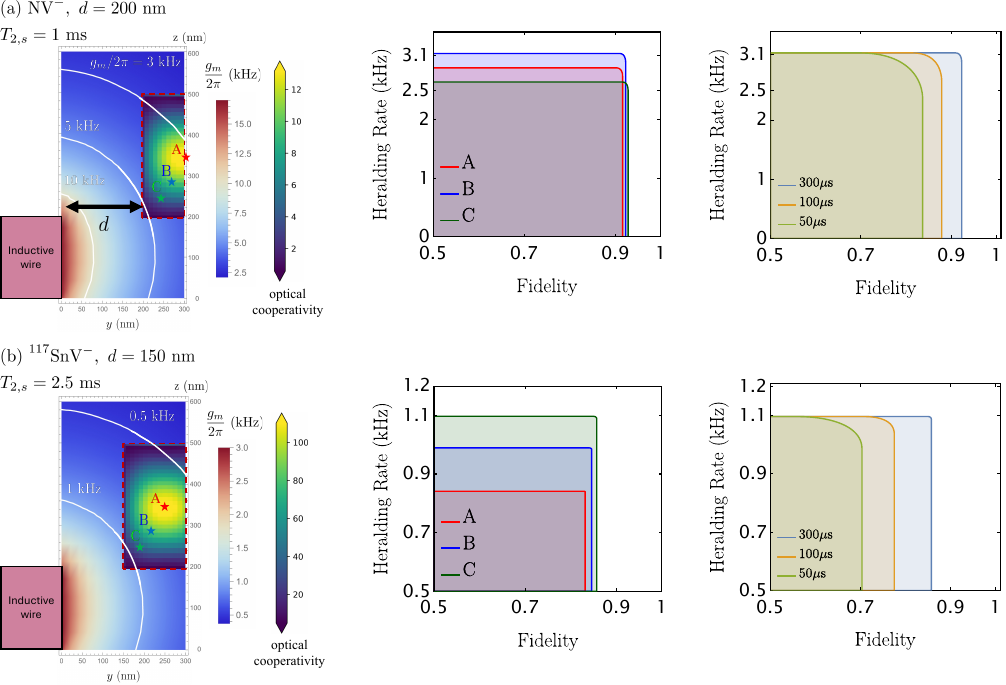}
\caption{Figures of merit for selected spin positions, indicated by stars, in the optical cavity for different color centers. The inductive wire generates zero-point magnetic-field fluctuations that couple to the spin magnetic dipole with coupling rate $g_m$ \cite{PFMO_device_design_2025}. The middle panels show the heralding rate and fidelity for the selected spin positions at a transmon lifetime of $T_{1,t}=300~\mu\mathrm{s}$. For each color center, the best-performing spin position is selected from the middle panels: position B for the NV$^-$ center and position C for the $^{117}\mathrm{SnV}^-$ center. The figures of merit are then compared as a function of the transmon lifetime $T_{1,t}$ in the right panels. All plots assume a transmon measurement error $\epsilon=0.01$.} \label{fom-spin}
\end{figure}

\begin{table*}
\begin{tabular}{|c|c|c|c|c|c|}
\hline Color Center & Location & $C$ & $g_m/2\pi$ (kHz) & Heralding Rate (kHz) & Fidelity\\
\hline
\multirow{3}{*}{NV$^-$}               & A & 13.6 & 3.18 & 2.7 & 0.92\\ \cline{2-6}
                        & B & 9.65 & 3.78 & 3.0 & 0.92 \\ \cline{2-6}
                        & C & 3.73 & 4.35 & 2.5 & 0.93\\ \hline
\multirow{3}{*}{${}^{117}\text{SnV}^-$} & A & 114 & 0.630 & 0.84 & 0.83 \\ \cline{2-6}
                        & B & 80.9 & 0.767 & 0.98 & 0.85 \\ \cline{2-6}
                        & C & 31.3 & 0.904 & 1.1 & 0.86 \\ \hline
\end{tabular}
\caption{Heralding rate at maximum fidelity for spins with different optical cooperativity $C$ and microwave coupling rate $g_m$. The transmon lifetime is taken to be $T_{1,\mathrm{t}} = 300~\mathrm{\mu s}$ and transmon measurement error $\epsilon=0.01$.}\label{fom comparison}
\end{table*}

To understand the performance of our scheme, we simulate the device architecture designed for the corresponding color centers described in Ref.~\cite{PFMO_device_design_2025}. The resulting simulated coupling rates $g_m$ for the NV$^-$ and $^{117}$SnV$^-$ centers are shown in Fig.~\ref{fom-spin}. The insets display the corresponding optical cooperativity where the cavity design follows Ref.~\cite{ding2024high}.

As the optical cavity approaches the microwave resonator, metal-induced losses increase, leading to a rapid degradation of the optical cooperativity. In contrast, the microwave coupling rate $g_m$ decreases as the spin is positioned farther from the resonator. Therefore, optimizing both the cavity placement and the spin location is crucial for achieving optimal device performance for pump-free quantum transduction. The middle panels of Fig.~\ref{fom-spin} show the calculated heralding rate and fidelity for selected spin positions, marked by stars, within the optical cavity. These positions are chosen to be at least $30~\mathrm{nm}$ away from the nearest diamond surface, outside the near-surface regime where charge noise and spin decoherence are most severe~\cite{Myers2014}. These calculations assume a transmon lifetime of $T_{1,\mathrm{t}} = 300~\mu\mathrm{s}$ and a transmon measurement error of $\epsilon=0.01$. The corresponding numerical values are listed in Table~\ref{fom comparison}. Selecting the best-performing spin location (position B for the NV$^-$ center and position C for the $^{117}\mathrm{SnV}^-$ center), the right panels of Fig.~\ref{fom-spin} show the effect of different transmon lifetimes $T_{1,\mathrm{t}}$. These are the plots shown in the main text. They can reach heralding rates beyond $1~\mathrm{kHz}$ with around $0.9$ fidelity.

Due to its larger magnetic transition matrix element, the NV$^-$ center achieves a microwave coupling rate $g_m$ that is 5.64 times larger than that of the $^{117}\mathrm{SnV}^-$ center. However, its small Debye-Waller factor suppresses the optical cooperativity, making it smaller by a factor of 18 than that of the $^{117}\mathrm{SnV}^-$ center for the same optical-cavity-inductive-wire separation. These trade-offs are directly reflected in the figures of merit for the selected spin positions.

\begin{figure*}[t!]
\centering
\includegraphics[width=\textwidth]{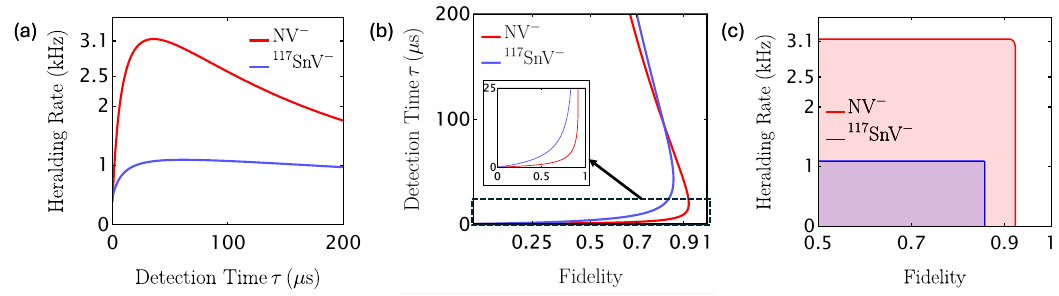}
\caption{(a) Intrinsic heralding rate and (b) fidelity as functions of the detection time $\tau$ for different diamond color centers, using the simulated parameters listed in Table~\ref{colortab}. The detection time considered here is limited by the transmon relaxation time $T_{1,t}=300$ µs. (c) Heralding rate as a function of fidelity, showing a sharp cutoff in the upper-right region corresponding to the maximum achievable heralding rate and fidelity, which occur at different detection times. The spin coherence times are taken to be $T_{2,s}=1~\mathrm{ms}$ for the NV$^-$ center~\cite{rondin_magnetometry_2014} and $2.5~\mathrm{ms}$ for $^{117}\mathrm{SnV}^-$~\cite{harris2025high}. All plots assume $\epsilon=0.01$.} \label{Fig_HR_F_single}
\end{figure*}

We conclude this section by discussing the dependence of the heralding rate and fidelity on the detection time. The heralding rate has a nonmonotonic dependence on $\tau$, as shown in Fig.~\ref{Fig_HR_F_single}(a). In the short-time limit, the heralding rate increases linearly with $\tau$, following $R_{\mathrm{herald}}\propto \epsilon / T_{\text{exe}} + \Gamma\tau / (2T_{\text{exe}})$ for $\tau<T_{\text{exe}}$. In the long-time limit, the microwave retrieval probability $p_\text{mw}$ saturates to 1, so the heralding rate decreases as $1/\tau$. The optimal detection time depends on $T_{\text{exe}}$ and $\Gamma$, and the optimal heralding rate is $\sim \frac{1}{2}\frac{C^2}{(1+C)^2}\frac{1}{2/\Gamma+T_{\text{exe}}}$.

The fidelity also shows a nonmonotonic dependence on $\tau$, as shown in Fig.~\ref{Fig_HR_F_single}(b). When $T_{1,t}$ and $T_{2,s}$ are much longer than $\tau$, the short-time behavior is governed by the measurement fidelity $F_M$, since the microwave time-bin photon must be emitted to obtain a nonzero measured fidelity. In this regime, $F_M$ increases linearly from zero as $\frac{\Gamma}{2\epsilon}\tau$ for $\tau<2\epsilon/\Gamma$. In the long-time limit, $F_M$ saturates to 1, while the full protocol fidelity is decreased by the decoherence in the system. The optimal detection time for the fidelity depends on $\Gamma$, $T_{1,t}$, and $T_{2,s}$.

As illustrated in Fig.~\ref{Fig_HR_F_single}(c), different combinations of heralding rate and fidelity can be achieved. These operating points can be extended to an achievable region, since both the heralding rate and the fidelity can always be reduced through discarding heralded trials and post-processing or by intentionally adding noise. These properties create a cutoff at the upper boundary of the rate-fidelity plot. Despite the difference between the optimal detection times for the heralding rate and fidelity in the parameter regime considered here, we can still achieve a kilohertz-scale heralding rate while maximizing the fidelity.

\section{Ensemble-Based MO-NODE}\label{ensemble_supplementary}

\subsection{Bright/Dark M-O Modes}\label{ensemble_bright_dark}

In MO-NODE, an entangled M-O pair is generated by the interaction between platform and cavity. For an ensemble, a critical factor governing the ensemble-cavity interaction is the collective mode in which the excitation is prepared. These modes are commonly classified as `bright' and `dark' \cite{Kurucz2011, Julsgaard2013}. In a bright mode, the individual emitter-cavity coupling amplitudes add constructively, allowing the ensemble to couple collectively to the cavity and potentially exhibit an enhanced interaction strength. In contrast, in a dark mode, these amplitudes interfere destructively, suppressing the net coupling to the cavity. Identifying the bright and dark modes is therefore essential for analyzing ensemble-cavity interactions.

\begin{figure}[t!]
\centering
\includegraphics[width=0.4\columnwidth]{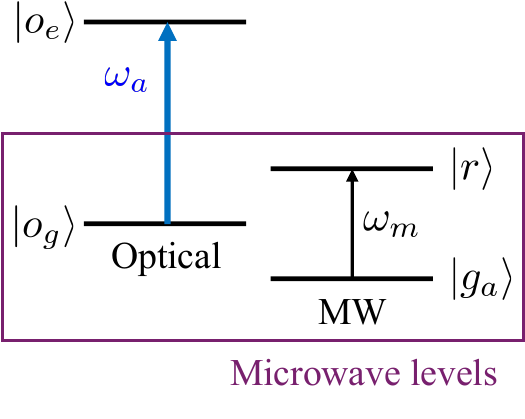}
\caption{Energy level diagram for ensemble-based MO-NODE where only $\ket{o_g}\leftrightarrow\ket{o_e}$ corresponds to optical transition.}\label{ensemble-level-model}
\end{figure}

Figure~\ref{ensemble-level-model} shows the simplified level structure for a single atom in ensemble. We assume that only the $\ket{o_g}\leftrightarrow\ket{o_e}$ transition is optically addressable and therefore coupled to the optical cavity. The microwave resonator couples the $\ket{g_a}\leftrightarrow\ket{r}$ transition. Here, $\ket{g_a}$, $\ket{r}$, and $\ket{o_g}$ belong to the microwave ground-state manifold, whereas $\ket{o_e}$ is the only optically excited state. Throughout this section, the subscripts $m$ and $o$ denote the microwave and optical modes, respectively.

We first consider the bright and dark collective modes associated with the microwave resonator. The microwave ensemble-cavity interaction Hamiltonian is $H_{m,\text{int}}=\sum_i \left(g_{m,i}^* a_m \sigma_{m,+,i} + g_{m,i} a_m^\dagger \sigma_{m,-,i}\right)$, where the subscript $i$ labels the $i$-th atom, $\sigma_{m,\pm,i}$ are the raising and lowering operators associated with the $\ket{g_a}\leftrightarrow\ket{r}$ transition, and $a_m$ is the annihilation operator of the microwave resonator mode. We define the cavity-coupled collective lowering operator as $S_{m,-} \equiv \sum_i g_{m,i} \sigma_{m,-,i}/G_m$ where $G_m = \sqrt{\sum_i |g_{m,i}|^2}$, collective coupling rate of microwave mode. Consider an arbitrary collective state in the single-excitation manifold, $\ket{M_\perp}=\sum_i d_{m,i}\ket{g_a g_a \cdots r_i \cdots g_a}$. Applying the collective lowering operator gives
\begin{equation}
    S_{m,-}\ket{M_\perp}=\frac{1}{G_m}\sum_i g_{m,i}d_{m,i} \ket{g_a g_a \cdots g_a}.
\end{equation}
This state becomes dark with respect to the microwave cavity when $\sum_i g_{m,i}d_{m,i}=0$. The corresponding bright state is obtained by applying the collective raising operator to the ensemble ground state:
\begin{equation}
    \ket{M_r} = \frac{1}{G_m}\sum_i g_{m,i}^* \ket{g_a g_a \cdots r_i \cdots g_a}.
\end{equation}
The bright state then satisfies $S_{m,-}\ket{M_r}=\ket{g_a g_a \cdots g_a}$, showing that it couples to the cavity with the collectively enhanced coupling strength $G_m$.

For the optical mode, however, there are no dark modes. The optical mode interaction Hamiltonian is $H_{o,\text{int}}=\sum_i \left(g_{o,i}^* a_o \sigma_{o,+,i} + g_{o,i} a_o^\dagger \sigma_{o,-,i}\right)$. The corresponding collective lowering operator is  $S_{o,-} \equiv \sum_i g_{o,i} \sigma_{o,-,i}/G_o$ where $G_o = \sqrt{\sum_i |g_{o,i}|^2}$. For an arbitrary collective state $\ket{O_\perp}=\sum_i c_{o,i}\ket{g_a g_a \cdots o_{e,i} \cdots g_a}$,
\begin{equation}
    S_{o,-}\ket{O_\perp}=\frac{1}{G_o} \sum_i g_{o,i}c_{o,i} \ket{g_a g_a \cdots o_{g,i} \cdots g_a}.
\end{equation}
This state becomes dark only when all $g_{o,i}c_{o,i}$ are 0. Provided $g_{o,i}\neq 0$ for every participating emitter, no optical dark state exists within the single-excitation manifold. Since each basis component is independent and orthogonal before and after the interaction, they do not interfere with one another, thus, any single collective excitation state prepared in optically addressable level can interact with optical cavity. However, this nature indicates that it cannot obtain collective enhancement in the coupling rate.

A key conclusion is that the optical and microwave coupling profiles need not be mode matched. The effect of optical-coupling nonuniformity is instead captured by the optical cavity reflectivity, as described in Sec.~\ref{ensemble_optical_reflection}. For collectively enhanced microwave-state retrieval, however, the spin-wave envelope must be matched to the microwave bright mode.

\subsection{Optical Reflection Considering Inhomogeneous Broadening and Coupling Rate}\label{ensemble_optical_reflection}

In this section, we analyze the optical reflectivity for an arbitrary single-collective-excitation input state in the presence of inhomogeneous broadening and nonuniform spin-cavity coupling across the ensemble.

The full Hamiltonian for the optical side is
\begin{equation}\label{tavis-cumming}
    H_o=\frac{1}{2}\sum_i \omega_{a,i}\sigma_{z,o,i} + \omega_c a_o^\dagger a_o + \sum_i \left(g_{o,i} a_o^\dagger \sigma_{-,o,i} + g_{o,i}^* a_o \sigma_{+,o,i}\right).
\end{equation}
In the frame rotating at the frequency $\omega$, the Heisenberg-Langevin equation gives,
\begin{align}
    &\dot a_o=-\left(i(\omega_c-\omega) + \kappa\right)a_o-i\sum_i g_{o,i}\sigma_{-,o,i}+\sqrt{2\kappa_c}a_{o,\rm in},\\
    &\dot\sigma_{-,o,i}=-\left(i(\omega_{a,i} - \omega) + \gamma_i \right)\sigma_{-,o,i} + i g_{o,i}^* a_o\sigma_{z,o,i}.
\end{align}
Since the cavity response depends on which optical branch is occupied, the cavity field can become correlated with the branch degree of freedom in the presence of inhomogeneous broadening or nonuniform coupling. Therefore, in general, $\langle a_o\sigma_{z,o,i}\rangle\neq\langle a_o\rangle\langle\sigma_{z,o,i}\rangle$. We define the branch-resolved cavity operator $a_{o,i}\equiv a_oQ_i$, where
\begin{equation}
    Q_i\equiv\ket{g_a\cdots o_{g,i}\cdots g_a}\bra{g_a\cdots o_{g,i}\cdots g_a}+\ket{g_a\cdots o_{e,i}\cdots g_a}\bra{g_a\cdots o_{e,i}\cdots g_a},
\end{equation}
such that $a_o=\sum_i a_{o,i}$. Here, $\sigma_{z,o,i}=2\ket{g_a\cdots o_{e,i}\cdots g_a}\bra{g_a\cdots o_{e,i}\cdots g_a}-Q_i$, and therefore, in the weak-excitation limit, $\sigma_{z,o,i}\simeq-Q_i$.

The operator $Q_i$ is conserved by the optical interaction, i.e., $\dot Q_i=0$. We then obtain
\begin{equation}
    \dot{a}_{o,i}=-\left(i(\omega_c-\omega)+\kappa\right)a_{o,i}-i g_{o,i}\sigma_{-,o,i}+\sqrt{2\kappa_c}a_{o,\rm in}Q_i.
\end{equation}
For an arbitrary initial state $\ket{\psi}=\sum_i c_i \ket{g_a g_a \cdots o_{g,i} \cdots g_a}$, $\langle Q_i\rangle=|c_i|^2$. In the steady-state limit,
\begin{equation}
    \langle a_{o,i}\rangle=|c_i|^2\frac{\sqrt{2\kappa_c}\langle a_{o,\rm in}\rangle}{i(\omega_c-\omega)+\kappa+\frac{|g_{o,i}|^2}{i(\omega_{a,i}-\omega)+\gamma_i}}.
\end{equation}
Thus, the $i$th branch exhibits the single-emitter cavity reflection amplitude
\begin{equation}
    r_i(\omega)=1-\frac{2\kappa_c}{i(\omega_c-\omega)+\kappa+\frac{|g_{o,i}|^2}{i(\omega_{a,i}-\omega)+\gamma_i}}.
\end{equation}
Using the input-output relation, the reflected field can be written as $a_{o,\rm out}=\sum_i r_i(\omega)a_{o,\rm in}Q_i$. Since $Q_iQ_j=\delta_{ij}Q_i$, we obtain $a_{o,\rm out}^\dagger a_{o,\rm out}=\sum_i |r_i(\omega)|^2a_{o,\rm in}^\dagger a_{o,\rm in}Q_i$. Therefore, the total reflectivity is
\begin{align}\label{ensemble_R}
    \mathcal{R}(\omega)=\sum_i |c_i|^2\left|1-\frac{2\kappa_c}{i(\omega_c-\omega)+\kappa+\frac{|g_{o,i}|^2}{i(\omega_{a,i}-\omega)+\gamma_i}}\right|^2.
\end{align}
For the critically coupled case at cavity resonance, the reflectivity becomes
\begin{equation}
    \sum_i |c_i|^2\left|\frac{C_i}{C_i+1}\right|^2,
\end{equation}
where $C_i\equiv \frac{|g_{o,i}|^2}{\kappa(\gamma_i+i(\omega_{a,i}-\omega_c))}$. This explicitly shows that there is no collective enhancement on the optical side.

\subsection{Quadratic dark-state leakage from inhomogeneous optical scattering}
\label{app:dark_leakage}

We show here that inhomogeneous optical transition frequencies and nonuniform optical coupling strengths modify the collective-state amplitude at first order, while the resulting leakage probability into states orthogonal to the initial bright state begins at second order.

Consider a normalized collective single-excitation state
$\ket{B}=\sum_{i=1}^{N}c_i\ket{i}$ with $\sum_i |c_i|^2=1$,
where we abbreviate $\ket{i} =\ket{g_a\cdots o_{g,i}\cdots g_a}$. The full state of the photon-emitter system after the photon reflection is related to the initial state through the scattering operator. The reflection calculation in the above section shows that each branch evolves independently during reflection, so $\dot{Q}_i=0$. Branch conservation then implies that the full photon-emitter scattering operator is block diagonal on the asymptotic emitter manifold,
\begin{equation}
    S=\sum_i P_i\otimes S_i,
    \qquad
    P_i=\ket{i}\!\bra{i},
\end{equation}
where \(S_i\) is the optical scattering operator conditional on the stored excitation occupying branch \(i\), and for each branch we assume that the emitter state returns to the initial state in the asymptotic limit (i.e. $P_i =\ket{i}\!\bra{i} $). Projecting the outgoing field onto the reflected mode therefore defines the emitter Kraus operator
\begin{align}
    M_R
    &=\bra{1_R}S\ket{1_{\mathrm{in}}}=\sum_i\bra{1_R}S_i\ket{1_{\mathrm{in}}}
    P_i
    =
    \sum_i r_i P_i,
    \label{eq:reflection-kraus-from-scattering}
\end{align}
where \(r_i=\bra{1_R}S_i\ket{1_{\mathrm{in}}}\) is the matrix element between the incoming photon state $\ket{1_{\mathrm{in}}}$ and the reflected photon state $\ket{1_R}$, which by definition is the the complex branch-resolved reflection amplitude. Hence, for \(\ket{B}=\sum_i c_i\ket{i}\), the unnormalized emitter state conditioned on reflection is
\begin{equation}
    \ket{\widetilde{\psi}_R}
    =
    M_R\ket{B}
    =
    \sum_i c_i r_i\ket{i}.
    \label{eq:conditional_state}
\end{equation}
This form assumes branch-preserving scattering and a common reflected optical mode $\ket{1_R}$; for a finite-bandwidth pulse, \(r_i\) is replaced by the corresponding overlap between the branch-dependent reflected wavepacket and the detected mode.

Let $p_i=|c_i|^2$ and $\overline r_p=\sum_i p_i r_i$, and define the projector onto the subspace orthogonal to the initial bright state, $\Pi_D=\mathbb I_{\mathrm{1exc}}-\ket{B}\!\bra{B}$, where $\mathbb I_{\mathrm{1exc}}$ is the identity operator on the entire single excitation subspace. The total reflection probability is given in Eqn.~(\ref{ensemble_R}) $\mathcal{R}(\omega)=\sum_i p_i|r_i|^2$. The unnormalized emitter state overlaps with the dark subspace: 
\begin{equation}
    \mathcal P_D^{(R)}
    =
    \bra{\widetilde{\psi}_R}
    \Pi_D
    \ket{\widetilde{\psi}_R}
    \nonumber
    =
    \sum_i p_i|r_i|^2
    -
    \left|\sum_i p_i r_i\right|^2
    \nonumber
    =
    \sum_i p_i|r_i-\overline r_p|^2.
    \label{eq:dark_variance}
\end{equation}
Consequently, the dark-state probability conditioned on reflection is
\begin{equation}
    \mathcal P_{D|R}
    =
    \frac{
    \sum_i p_i|r_i-\overline r_p|^2
    }{
    \sum_i p_i|r_i|^2
    }.
    \label{eq:conditional_dark_probability}
\end{equation}
Dark-state leakage is therefore determined by the weighted variance of the reflection amplitudes. A branch-independent change in \(r_i\) changes the overall reflection amplitude but does not generate a dark component.

To determine the leading dependence on inhomogeneity, we consider cavity and emitter resonance, $\omega=\omega_c=\omega_a$, and write
$\omega_{a,i}=\omega_a+\delta_i,$ and $g_{o,i}=g(1+\epsilon_i),$
with $|\epsilon_i|\ll1,$ and $|\delta_i|/\gamma\ll1$.
We further assume critical coupling and define $C=\frac{g^2}{\kappa\gamma}$.

Expanding the reflection amplitude formula to first order gives
\begin{equation}
    r_i
    =
    \frac{C}{1+C}
    +
    \frac{C}{(1+C)^2}
    \left(
    2\epsilon_i
    -
    i\frac{\delta_i}{\gamma}
    \right)
    +
    \mathcal O(\eta^2),
    \label{eq:r_expansion}
\end{equation}
where $\eta = \max_i\left\{|\epsilon_i|,|\delta_i|/\gamma\right\}$. Introducing centered fluctuations $\widetilde\epsilon_i
    =
    \epsilon_i-\overline\epsilon_p$, $
    \widetilde\delta_i
    =
    \delta_i-\overline\delta_p,$
with \(\overline\epsilon_p=\sum_i p_i\epsilon_i\) and
\(\overline\delta_p=\sum_i p_i\delta_i\), one finds
\begin{equation}
    r_i-\overline r_p
    =
    \frac{C}{(1+C)^2}
    \left(
    2\widetilde\epsilon_i
    -
    i\frac{\widetilde\delta_i}{\gamma}
    \right)
    +
    \mathcal O(\eta^2).
    \label{eq:centered_r}
\end{equation}
At exact resonance, coupling inhomogeneity changes the real part of \(r_i\) to first order, whereas frequency inhomogeneity changes its imaginary part. Hence,
\begin{equation}
\begin{split}
    |r_i-\overline r_p|^2
    &=
    \frac{C^2}{(1+C)^4}
    \left(
    4\widetilde\epsilon_i^2
    +
    \frac{\widetilde\delta_i^2}{\gamma^2}
    \right)
    +
    \mathcal O(\eta^3).
\end{split}
    \label{eq:squared_difference}
\end{equation}
The leading mixed term vanishes because the coupling-inhomogeneity correction is real whereas the frequency-inhomogeneity correction is purely imaginary at exact resonance. Away from exact resonance, a covariance term proportional to
\(\operatorname{Cov}(\epsilon,\delta)\) may appear, but it remains second order.

Substituting Eq.~\eqref{eq:squared_difference} into
Eq.~\eqref{eq:conditional_dark_probability}, and using $\mathcal{R}(\omega) =\frac{C^2}{(1+C)^2}+\mathcal O(\eta)$,
gives
\begin{equation}
    \mathcal P_{D|R}
    =
    \frac{1}{(1+C)^2}
    \left[
    4\,\operatorname{Var}_p(\epsilon)
    +
    \frac{\operatorname{Var}_p(\delta)}{\gamma^2}
    \right]
    +
    \mathcal O(\eta^3). \label{eq:final_dark_leakage}
\end{equation}
Equivalently, since \(\epsilon_i=(g_i-g)/g\), we can replace $\operatorname{Var}_p(\epsilon)$ with
$\operatorname{Var}_p(g)/{g^2}$.

The branch-dependent dark-state amplitude is therefore linear in either the frequency or coupling inhomogeneity, whereas the corresponding leakage probability is quadratic. Only deviations from the ensemble mean contribute; uniform shifts of all transition frequencies or coupling strengths do not generate dark-state leakage.

\subsection{Application to $^{87}\mathrm{Rb}$ Atom Ensemble and Challenges}\label{ensemble_Rb_application}

\begin{figure}[t!]
\centering
\includegraphics[width=0.8\columnwidth]{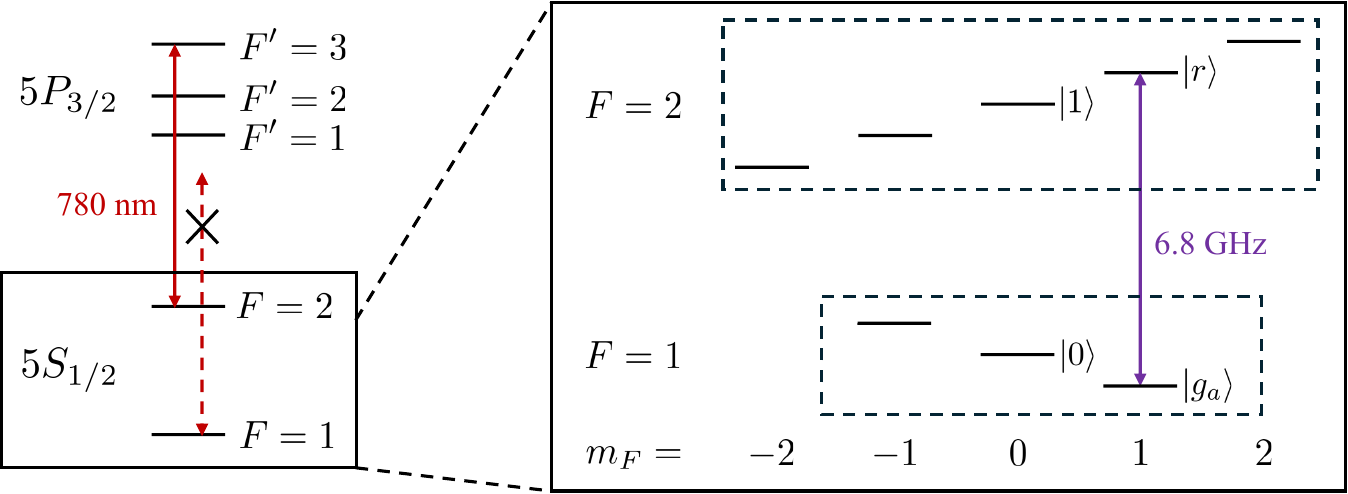}
\caption{Energy level diagram for optical (780 nm) and microwave (6.8 GHz) transition levels of $^{87}$Rb atom.\label{level-na-ensemble}}
\end{figure}

In this section, we describe the $^{87}\mathrm{Rb}$ atom ensemble protocol with possible level choices and its figures of merit. We choose the four relevant levels as
$\ket{g_a}=\ket{5S_{1/2},F=1,m_F=1}$,
$\ket{0}=\ket{5S_{1/2},F=1,m_F=0}$,
$\ket{1}=\ket{5S_{1/2},F=2,m_F=0}$, and
$\ket{r}=\ket{5S_{1/2},F=2,m_F=1}$.
To generate entanglement between an optical photon and the ensemble spin state, we consider a critically coupled optical cavity tuned to the $5S_{1/2},F=2\rightarrow 5P_{3/2},F'=3$ transition. While the $F=2$ manifold is resonant with the optical cavity, the $F=1$ manifold is far detuned by approximately $6.8~\mathrm{GHz}$ and is therefore effectively uncoupled from the cavity \cite{Tiecke2014,Dordevic2021}. We consider $\sim 10^6$ neutral atoms coupled to a traveling wave optical cavity with a single atom cooperativity of $C\sim 3$~\cite{Chen2022}. For the microwave coupling, a coplanar-waveguide resonator can be considered, yielding an ensemble spin-microwave coupling rate of $G_m/2\pi\sim 40~\mathrm{kHz}$ for the same number of atoms~\cite{Verd2009}. Atom-chip experiments have demonstrated seconds-scale coherence of hyperfine states~\cite{Bohi2009,Bernon2013}. We use a microwave Rabi frequency of $\Omega/2\pi=122~\mathrm{kHz}$, corresponding to a $\pi$-pulse time of $4.1~\mu\mathrm{s}$~\cite{Bohi2009}. This value is chosen to selectively address the desired hyperfine transition while suppressing off-resonant excitation separated by a few MHz.

Before analyzing the figures of merit, we first discuss several experimental challenges. Although the optical trap does not act as a pump in our protocol, i.e., the trapping field is not coupled directly into the atom-cavity system to enhance the interaction, stray trapping light incident on the superconducting chip may nevertheless cause Cooper-pair breaking and heating \cite{Beck2016,Wilde2025}. One possible sequence is to use an optical trap to transport or initially position the atoms near the superconducting chip and then switch off the trapping laser before running the protocol. After the optical trap is switched off, the atoms could in principle remain confined near the chip using a magnetic trap. However, the associated field inhomogeneity may produce broadening comparable to the collective coupling rate considered here and thereby reduce microwave-photon retrieval from the collective spin wave \cite{Verd2009,Kurucz2011,Hattermann2017}. We therefore consider, as a possible operating scenario, switching off the trapping fields during the protocol and restoring the optical trap afterward. In the absence of trapping, atomic displacement arises primarily from thermal ballistic motion and gravity \cite{Tuchendler2008}. For an atomic ensemble at a temperature of approximately 1 µK, these contributions result in a total displacement of less than 1 µm over a protocol runtime of approximately 50 µs. We therefore assume that the atoms remain sufficiently localized throughout the protocol. After the protocol is completed, the optical trap is restored, and the atoms are recooled in preparation for the next run. These preparation and re-cooling overheads are not included in the intrinsic heralding rates reported below.

Combining optical and microwave coupling in the same neutral-atom ensemble poses additional challenges. To entangle all participating atoms with the optical photon, it is important to avoid the wavelength-scale node structure of a standing-wave optical cavity. Without lattice registration, which typically requires additional optical access and trapping beams~\cite{Lee2014}, a Fabry-Perot cavity would couple atoms at different axial positions with strongly varying magnitude and phase~\cite{Brennecke2007}. A traveling-wave cavity removes this axial node problem and provides nonzero coupling along the propagation direction, at the cost of a smaller peak cooperativity than the corresponding standing-wave antinode~\cite{Chen2022}. On the microwave side, the wavelength is much longer than the typical $100$-$400~\mu\mathrm{m}$ extent of the neutral-atom cloud, so the coupling is not limited by wavelength-scale nodes~\cite{Hattermann2017}.

It is noteworthy that an atomic ensemble with simultaneous coupling to an optical and a millimeter-wave cavity has been demonstrated \cite{kumar_quantum-enabled_2023}. However, since the millimeter-wave transition in the experiments is encoded between the highly-excited Rydberg states, an optical pump is unavoidable, which is incompatible with the pump-free vision of our scheme. All these requirements make the experimental realization challenging, but they do not constitute a fundamental obstacle. We therefore estimate the potential performance of the protocol by combining parameters from the above components.

\subsection{Figures of Merit Analysis for Neutral Atom Ensemble}\label{ensemble_fom}

Due to the much longer microwave drive time compared to the ns-scale drive time of diamond color centers, the microwave photon detection probability should be described by Eq.~(\ref{mw_prob_four_level}), rather than by the fast drive limit expression used for the single spin protocol. In this setting, $\tau_w/\tau=0.5$ is no longer the optimal condition for maximizing the heralding probability and, consequently, the heralding rate. Moreover, decoherence starts once the superposition state is prepared. Thus, the long drive time limits the maximum fidelity achievable for the retrieved microwave photon.

The detection time becomes meaningful only once the protocol can be completed; that is, once the second time-bin microwave photon can be detected. Because the readout state can decay during the microwave drive, detection of the second time-bin photon can complete the protocol before the second $\pi$ pulse is finished. This defines the minimum detection time $\tau_\text{min}\equiv\tau_\pi+\tau_w$. Thus, $\tau>\tau_\text{min}$. Under this condition, the fidelity is bounded by $F_{M\text{,NA}}=\frac{1}{4}(1+e^{-\tau/T_{1,\text{t}}}+2e^{-\tau/(2T_{1,\text{t}})}e^{-(\tau+\tau_\pi)/T_{2,\text{s}}})$. In the instantaneous-pulse limit, for which the optimal waiting time approaches $\tau_w=\tau/2$, this constraint is automatically satisfied.

\begin{figure}[t!]
\centering
\includegraphics[width=\columnwidth]{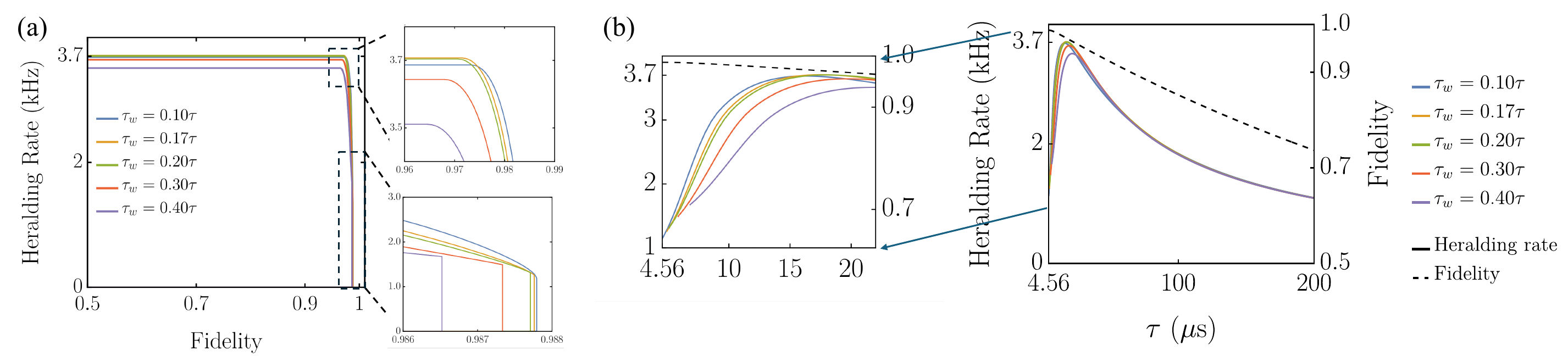}
\caption{Figures of merit analysis of the $^{87}\mathrm{Rb}$ ensemble for various values of $\tau_w/\tau$, assuming $T_{1,t}=300$ µs and $\epsilon=0.01$.\label{fom-na-ensemble}}
\end{figure}

To numerically find the optimal value of $\tau_w/\tau$, we vary $\tau_w/\tau$ while fixing $T_{1,t}=300~\mu\mathrm{s}$ and $\epsilon=0.01$, as shown in Fig.~\ref{fom-na-ensemble}(a). The maximum heralding rate is obtained at $\tau_w/\tau=0.17$, as shown in the upper inset of Fig.~\ref{fom-na-ensemble}(a). Fig.~\ref{fom-na-ensemble}(b) shows the heralding rate and fidelity as functions of the detection time $\tau$. Since $\tau_{\mathrm{min}}$ depends on $\tau_w/\tau$, the heralding rate and fidelity at $\tau_{\mathrm{min}}$ vary with the choice of $\tau_w/\tau$, producing different cutoffs in the lower branch of Fig.~\ref{fom-na-ensemble}(a). One can also see that the fidelity decreases monotonically after $\tau_{\mathrm{min}}$. Thus, the maximum fidelity is obtained only at $\tau=\tau_{\mathrm{min}}$. However, $\tau=\tau_{\mathrm{min}}$ corresponds to the moment when the second microwave drive has just started to emit the second time-bin microwave photon. In other words, since detection has proceeded only up to the first time-bin photon, the fidelity equation gives an upper bound for the total state, and such a short detection time could lead to an actual fidelity much lower than this bound. Considering the small variation of the maximum fidelity at different $\tau=\tau_{\mathrm{min}}$, as shown in the lower inset of Fig.~\ref{fom-na-ensemble}(a) and Fig.~\ref{fom-na-ensemble}(b), we choose $\tau_w/\tau=0.17$ as the optimized ratio. The plot in the main text is based on this choice. The maximum heralding rate of $3.71~\mathrm{kHz}$ is achieved at a detection time of $\tau=17.44~\mathrm{\mu s}$ with a fidelity of $0.97$.

\section{Phonon-Mediated MO-NODE}\label{phononappendix}

To achieve quantum state transfer between a solid-state spin and a microwave photon, we consider a three-mode hybrid system where the interaction is mediated by an intermediate mechanical phonon mode. The coherent exchange of a single excitation is described by the interaction Hamiltonian ($\hbar=1$):
\begin{equation}
    H_{int} = g_{sp}(b^\dagger\sigma_- + b\sigma_+) + g_{pe}(a^\dagger b + ab^\dagger),
\end{equation}
where $\sigma_-$ is the lowering operator for the spin, $b$ is the phonon annihilation operator, and $a$ is the microwave cavity photon annihilation operator. The parameters $g_{sp}$ and $g_{pe}$ denote the spin-phonon and phonon-microwave coupling strengths, respectively.

The system dynamics are heavily influenced by the intrinsic decay rates of each mode: the spin decoherence ($\gamma_s$), the mechanical phonon damping ($\kappa_m$), and the microwave cavity loss ($\kappa_e$). Depending on the hierarchy of these couplings and decay rates, the system can operate in distinctly different physical regimes, which can be used for either Purcell-enhanced microwave retrieval or coherent swap between the spin and the microwave state. 

\subsection{The Effective Purcell Rate}
If the objective is not to coherently swap a quantum state, but rather to induce a rapid, dissipative decay of the spin excitation directly into a microwave photon, the system can be engineered to operate in the effective Purcell regime.

Assuming the system is fully resonant, the Heisenberg-Langevin equations of motion for the single-excitation subspace are:
\begin{align}
    \dot{a} &= -i g_{pe} b - \frac{\kappa_e}{2} a, \\
    \dot{b} &= -i g_{sp} \sigma_- - i g_{pe} a - \frac{\kappa_m}{2} b, \\
    \dot{\sigma}_- &\approx -i g_{sp} b - \frac{\gamma_s}{2} \sigma_-.
\end{align}
To establish the Purcell enhancement, the microwave cavity must be the most heavily damped element ($\kappa_e \gg g_{pe}$). In this limit, the cavity field reaches a steady state almost instantaneously relative to the phonon dynamics. Setting $\dot{a} \approx 0$ allows us to adiabatically eliminate the cavity mode:
\begin{equation}
    a \approx -i \frac{2g_{pe}}{\kappa_e} b.
\end{equation}
Substituting this into the phonon equation reveals that the mechanical mode acquires a Purcell-enhanced decay rate, $\Gamma_m$, due to its coupling with the lossy cavity:
\begin{equation}
    \Gamma_m = \kappa_m + \frac{4g_{pe}^2}{\kappa_e} ,
\end{equation}

For the spin decay to be enhanced, this broadened phonon mode must act as a fast, memoryless bath for the spin. This requires a second adiabatic elimination ($\dot{b} \approx 0$), demanding that the enhanced phonon decay is much faster than the coherent spin-phonon exchange ($\Gamma_m \gg g_{sp}$). This condition can be satisfied given experimentally relevant parameters $(g_{sp},g_{pe},\kappa_m,\kappa_e)/2\pi =(0.3,2.4, 0.1, 12)$ MHz \cite{Meesala2024PRX, joe_purcell-enhanced_2026}. Applying this yields:
\begin{equation}
    b \approx -i \frac{2g_{sp}}{\Gamma_m} \sigma_-.
\end{equation}
Substituting back into the spin equation, we obtain the effective spin decay rate:
\begin{equation}
    \Gamma_{sp} = \frac{4g^2_{sp}}{\kappa_m + 4g^2_{pe}/\kappa_e}. 
\end{equation}
This derives the rate shown in the main text. While the expression for the effective spin decay rate, $\Gamma_{sp} \approx 4g_{sp}^2 / \Gamma_m$, suggests that the rate can be arbitrarily increased by minimizing the cavity-induced phonon decay, $\Gamma_m \approx 4g_{pe}^2/\kappa_e$, this minimization is strictly bounded by the physical requirements of the adiabatic hierarchy. Specifically, the second adiabatic elimination demands that the broadened phonon mode acts as a memoryless, Markovian bath for the spin, requiring $\Gamma_m \gg g_{sp}$. If we parameterize this operational boundary by a required margin $M \gg 1$ such that $\Gamma_m = M g_{sp}$, the effective spin decay rate simplifies to $\Gamma_{sp} \approx 4g_{sp} / M$. This reveals a fundamental tradeoff in the system architecture: enforcing a stricter validity of the adiabatic approximation (by increasing $M$) inherently suppresses the speed of the dissipation channel. Consequently, the maximally enhanced spin decay rate is physically bottlenecked by the bare spin-phonon coupling strength.

While the Purcell rate dictates the speed at which the spin excitation enters the dissipation channel, it does not guarantee that the excitation will successfully emerge as a usable microwave photon. The quantum excitation faces a competition between the desired forward-transfer process and non-radiative intrinsic losses. To determine the microwave emission probability ($p$)—defined as the likelihood of a single spin excitation successfully exiting into the external microwave transmission line, we must take into account the extraction efficiencies at each intermediate stage.

We partition the total microwave decay rate $\kappa_e$ into an external coupling rate $\kappa_{e,\text{ext}}$ and an internal loss rate $\kappa_{e,\text{int}}$, such that $\kappa_e = \kappa_{e,\text{ext}} + \kappa_{e,\text{int}}$. The probability that a microwave photon is successfully emitted into the transmission line is the microwave extraction ratio:
$\eta_e  = \kappa_{e,\text{ext}}/\kappa_e.$
Before reaching the microwave mode, the excitation must traverse the intermediate mechanical phonon mode. As established in the adiabatic elimination, the phonon mode acquires a cavity-induced decay rate $ 4g_{pe}^2/\kappa_e$. However, the phonon also suffers from intrinsic mechanical damping, $\kappa_m$. The probability that a phonon successfully decays into a microwave cavity photon, rather than being lost to thermal acoustic dissipation, is the mechanical extraction ratio:
$
\eta_m  = \frac{4g_{pe}^2}{\kappa_e}/\left(\frac{4g_{pe}^2}{\kappa_e} + \kappa_m\right).
$
To achieve $\eta_m \rightarrow 1$, the system requires strong cooperativity between the phonon and the microwave cavity, where the induced decay overwhelms the intrinsic mechanical loss ($4g_{pe}^2/\kappa_e \gg \kappa_m$).

The total probability of retrieving a microwave photon from a single spin excitation is formed from the product of the extraction ratios
\begin{equation}
    p_\text{mw} =  \eta_m \cdot \eta_e (1-e^{-\Gamma_{sp}\tau/2}) \approx \frac{C_{pe}}{C_{pe}+1} (1-e^{-\Gamma_{sp}\tau/2}),
\end{equation}
where the approximate sign means that we can write the extraction ratio in the form of cooperativity $C_{pe} = 4g_{pe}^2/\kappa_e\kappa_m$ if we assume a unit microwave extraction ratio $\eta_e \rightarrow 1$. 

\begin{figure}[t!]
\centering
\includegraphics[width=\columnwidth]{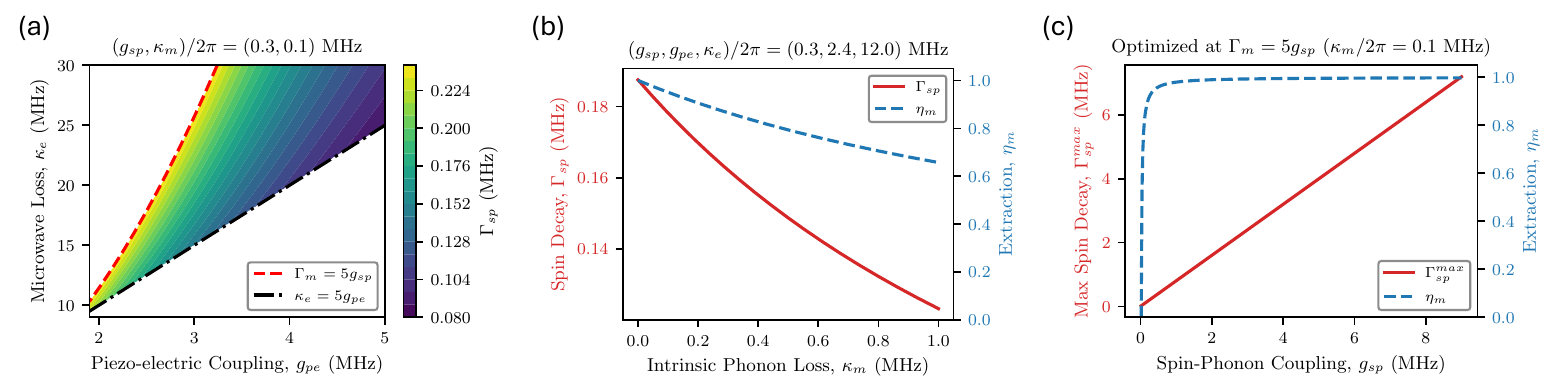}
\caption{(a) Behavior of the Purcell-enhanced spin decay under the two adiabatic conditions, $\Gamma_m \gg g_{sp}$ and $\kappa_e\gg g_{pe}$. (b) Performance of the spin-phonon scheme scaling with the loss of the phononic mode. (c) Performance of the spin-phonon scheme scaling with the spin-phonon coupling strength.} \label{sp_regime}
\end{figure}

To visualize the performance of the spin-phonon scheme under the two adiabatic elimination conditions $\Gamma_m \gg g_{sp}$ and $\kappa_e\gg g_{pe}$, we plot the two-dimensional design space in terms of $g_{pe}$ and $\kappa_e$ in Fig.~\ref{sp_regime}(a). For the purpose of demonstration, we set the two conditions to be $\Gamma_m >5 g_{sp}$ and $\kappa_e > 5g_{pe}$, which correspond to the red and black boundaries, respectively in the plot. In the parameter regime where the two conditions are satisfied, we show the color map of the enhanced spin decay rate, which achieves the maximal value exactly at the boundary $\Gamma_m = 5 g_{sp}$. In Fig.~\ref{sp_regime}(b), both the spin decay rate and the mechanical extraction ratio decrease with larger intrinsic phonon loss, which highlights the importance of a high-quality phononic cavity for the scheme. In Fig.~\ref{sp_regime}(c), both the maximal spin decay rate (achieved at the boundary of the adiabatic condition) and the extraction ratio increase with the spin-phonon coupling $g_{sp}$. Numerical simulation in \cite{joe_purcell-enhanced_2026} predicts a spin-phonon coupling rate of $9$ MHz, which corresponds to a $7.2$ MHz maximally achievable spin decay rate. Therefore, future experiments pursuing large spin-phonon interactions may pave the way for high-rate spin-phonon-based MO-NODE.

\subsection{The Coherent Swap Scheme}
\label{sec:coherent_swap}
To transfer a quantum state from a two-level spin to a microwave cavity via a mechanical phonon mode, we may consider a sequential resonant SWAP scheme. The sequential transfer is executed in two steps: first from the spin to the phonon, and second from the phonon to the microwave resonator.

To derive the fidelity of this process, we first derive the transfer fidelity for a single SWAP gate between two generic resonant modes (with annihilation operators $a$ and $b$, loss rates $\gamma_a$ and $\gamma_b$, and coupling $g$). In the single-excitation subspace, the non-Hermitian effective Hamiltonian governing the lossy dynamics is:
\begin{equation}
    H_{\text{eff}} = g(a^\dagger b + a b^\dagger) - i\frac{\gamma_a}{2}a^\dagger a - i\frac{\gamma_b}{2}b^\dagger b.
\end{equation}
Expressing the quantum state as $|\psi(t)\rangle = c_a(t)|1_a, 0_b\rangle + c_b(t)|0_a, 1_b\rangle$, the Schrödinger equation $i\partial_t |\psi\rangle = H_{\text{eff}}|\psi\rangle$ yields a system of coupled differential equations for the probability amplitudes:
\begin{equation}
    \frac{d}{dt} \begin{pmatrix} c_a \\ c_b \end{pmatrix} = \begin{pmatrix} -\gamma_a / 2 & -ig \\ -ig & -\gamma_b / 2 \end{pmatrix} \begin{pmatrix} c_a \\ c_b \end{pmatrix}.
\end{equation}
Assuming the strong coupling regime ($g \gg |\gamma_a - \gamma_b|$), the eigenvalues of this matrix simplify to $\lambda_{\pm} \approx -(\gamma_a + \gamma_b)/4 \pm ig$. For an initial state residing entirely in mode $a$ ($c_a(0)=1$, $c_b(0)=0$), the amplitude of the excitation in mode $b$ evolves as:
\begin{equation}
    c_b(t) = -i \exp\left(-\frac{\gamma_a + \gamma_b}{4} t\right) \sin(gt).
\end{equation}
A perfect SWAP operation requires a gate time $t_{\text{swap}} = \pi / 2g$. The transfer fidelity $F = |c_b(t_{\text{swap}})|^2$ evaluates to:
\begin{equation}
    F = \exp\left[-\frac{\pi}{2g}\left(\frac{\gamma_a + \gamma_b}{2}\right)\right].
\end{equation}
Because the exponent is small in the strong coupling regime, we can expand it to first order to obtain the standard linear approximation:
\begin{equation}
    F \approx 1 - \frac{\pi}{4g} (\gamma_a + \gamma_b).
\end{equation}

We now apply this result to the sequential three-mode transfer. The total fidelity $F_{\text{total}}$ is the product of the fidelities of the two sequential SWAP operations ($F_1$ for spin-to-phonon, and $F_2$ for phonon-to-microwave). Assuming the intrinsic spin loss is negligible ($\gamma_s \approx 0$) over the timescales of the transfer, the total fidelity is bounded by:
\begin{align}
    F_{\text{total}} &= F_1 \cdot F_2 \nonumber \\
    &\approx \left( 1 - \frac{\pi}{4} \frac{\kappa_m}{g_{sp}} \right) \left( 1 - \frac{\pi}{4} \frac{\kappa_m + \kappa_e}{g_{pe}} \right) \nonumber \\
    &\approx 1 - \frac{\pi}{4} \left( \frac{\kappa_m}{g_{sp}} + \frac{\kappa_m + \kappa_e}{g_{pe}} \right).
\end{align}

Given experimentally relevant parameters of $(g_{sp}, g_{pe}, \kappa_m,\kappa_e)/2\pi = (0.3, 2.4, 0.1,0.1)$~MHz, the transfer is bottlenecked by the relatively weak spin-phonon coupling. The time required to execute the full sequential transfer is $t_{\text{total}} = \pi/(2g_{sp})+\pi/(2g_{pe}) \approx 1~\mu$s. The estimated transfer fidelity is approximately $70\%$, limited primarily by phonon decay during the slower spin-to-phonon transfer.

\end{appendix}

\end{document}